\documentclass[aps,pre,reprint, amsmath, amssymb,superscriptaddress]{revtex4-1}

\usepackage{bm}
\usepackage[retainorgcmds]{IEEEtrantools}
\usepackage{graphicx}
\usepackage{mathrsfs}
\usepackage{amsmath}
\usepackage{amssymb}
\usepackage{color}
\usepackage{amsfonts}

\newcommand{\ud}{\,\mathrm{d}}

\DeclareMathOperator{\tr}{tr}

\newcommand{\trans}{^{\text{T}}}

\begin{document}

\title{Non-equilibrium quantum chains under multi-site Lindblad baths}
\date{\today}
\author{Pedro H. Guimar\~aes}
\affiliation{Instituto de F\'isica da Universidade de S\~ao Paulo,  05314-970 S\~ao Paulo, Brazil}
\author{Gabriel T. Landi}
\email{gtlandi@gmail.com}
\affiliation{Universidade Federal do ABC,  09210-580 Santo Andr\'e, Brazil}
\author{Mario J. de Oliveira}
\affiliation{Instituto de F\'isica da Universidade de S\~ao Paulo,  05314-970 S\~ao Paulo, Brazil}


\begin{abstract}

We study a quantum XX chain coupled to two heat reservoirs that act on multiple-sites and are kept at different temperatures and chemical potentials. 
The baths are described by Lindblad dissipators which are constructed by direct coupling to the fermionic normal modes of the chain.
Using a perturbative method, we are able to find analytical formulas for all steady-state properties of the system. 
We compute both the particle/magnetization current and the energy current, both of which are found to have the structure of Landauer's formula. 
We also obtain  exact formulas for the Onsager coefficients.
All properties are found to differ substantially from those of a single-site bath.
In particular, we find a strong dependence  on the intensity of the bath couplings.
In the weak coupling regime, we show that the Onsager reciprocal relations are satisfied.

\end{abstract}
\maketitle{}

%
%
%
%

\section{\label{sec:int}Introduction}

When a system is placed in contact with two reservoirs maintained at different temperatures or chemical potentials, it will usually tend to a non-equilibrium steady-state (NESS) characterized by the presence of finite currents between the two baths.
This physical scenario includes a wide range of important problems in condensed matter physics, with the most traditional examples being  measurements of  thermal conductivity \cite{Tritt2004} and electron transport in metals and semiconductors \cite{Ashcroft1976,Grosso2000}. 
But it also encompasses many other  problems, such as  ballistic  transport of electrons in mesoscopic devices \cite{Landauer1987,VanWees1988,Pastawski1991,Baringhaus2014,Datta1997}, hopping of ultra-cold atoms in optical lattices \cite{Bloch2005,Jedicke2004a,Diehl2008}, magnon transport in YIG crystals \cite{Demokritov2006,Rezende2009,Serga2010,*Chumak2014,*Chumak2015} and spintronics \cite{Fabian2004,Hogl2015}. 

Having a robust theoretical framework to study the properties of the NESS is therefore of importance to a broad range of areas. 
However, this is usually an extremely difficult endeavor, due mainly to two reasons.
First, the problem is heavily dependent on the possible scattering mechanisms that may affect the current-carrying excitations,
an effect which is usually modeled  using Boltzmann's equation \cite{Tritt2004,Grosso2000,Mahan}, Kubo's linear response formula \cite{Kubo1957,*Kubo1957a,Mahan,Jeon1995,*Jeon1995a} or the Landauer-B\"utiker formalism \cite{Landauer1987,VanWees1988,Pastawski1991,Baringhaus2014,Datta1997}.
Secondly, in contrast with thermal equilibrium, the NESS will in general be sensitive to the specific details of the coupling between the system and the baths.
For classical systems, this may be described using Nos\'e-Hover  \cite{Nose1984,*Hoover1985}  or  Langevin/Fokker-Planck equations \cite{Coffey2004,Risken1989}. 
These methods have been used extensively in the past to study heat flux and Fourier's law in classical chains of oscillators \cite{Rieder1967,Bolsterli1970, Casati1984,Prosen1999,Aoki2000,*Aoki2001,Pereira2004,*Pereira2006,*Pereira2013,Roy2012,Landi2013b,Landi2013a,*Landi2014a,*Guimaraes2015,Dhar2008}. 

The NESS of quantum systems, on the other hand, may be modeled using techniques from  open quantum systems \cite{Lindblad1976,Gardiner2004,Breuer2007,Caldeira1981a,Caldeira2014b}, such as the quantum Langevin equation  or the quantum master equation. 
One way to implement these methods is by starting with a microscopic derivation. 
That is, to start with a model for the system-bath interaction and then trace out the bath under suitable approximations. 
This approach was used,  for instance, in Refs.~\cite{Dhar2006,*Saito2007,Purkayastha2015,Bandyopadhyay2011} to study the heat flux through harmonic chains. 
However, in many cases the complexity of the model may easily render such approach unfeasible.

\begin{figure}
\centering
\includegraphics[width=0.45\textwidth]{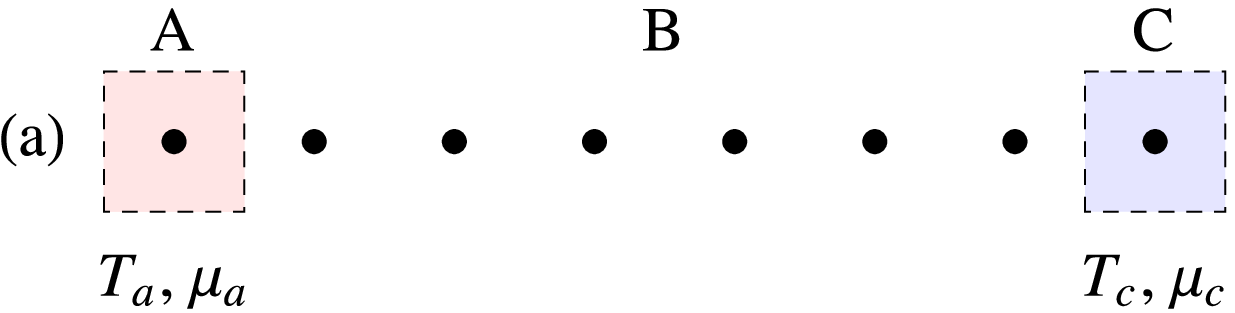}\\
\includegraphics[width=0.45\textwidth]{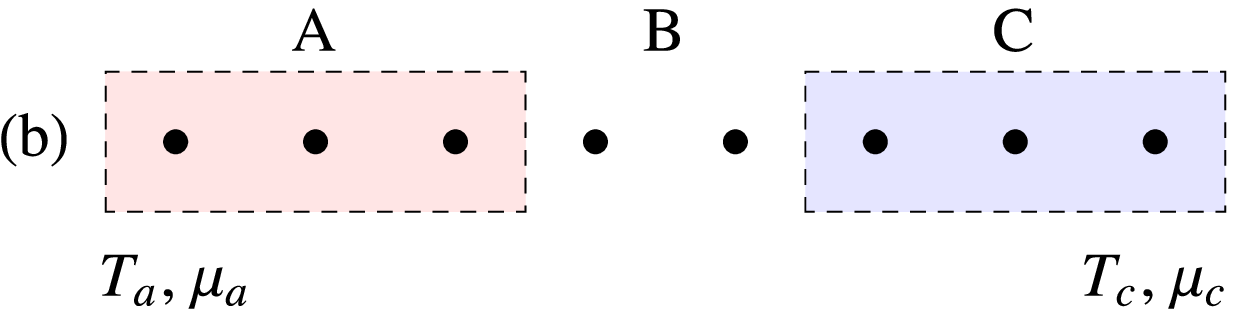}\\
\includegraphics[width=0.45\textwidth]{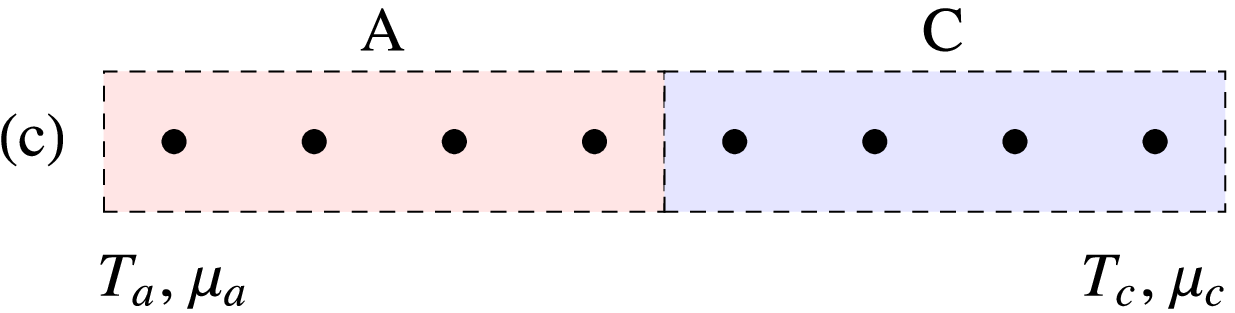}
\caption{\label{fig:drawing} 
Schematic illustration of multi-site baths. 
Each bath is characterized by a temperature $T$ and a chemical potential $\mu$. 
(a) A chain connected to two single-site baths.
(b) A chain connected to two multi-site baths.
(c) Multi-sites baths acting  precisely on each half of the chain.
}
\end{figure}

A more straightforward method is to implement Lindblad dissipators designed only for a specific part of the system. 
The idea is illustrated in Fig.~\ref{fig:drawing}(a), which depicts a chain of spins or harmonic oscillators  coupled  to two reservoirs kept at different temperatures and/or chemical potentials. 
The baths are then modeled by dissipators chosen such they  would correctly  thermalize the site in which they act, provided they were uncoupled from the rest of the system. 
Recently, this method has been used by a number of authors  to study the NESS of open quantum systems \cite{Manzano2012,*Asadian2013,Benenti2009,Karevski2009,*Platini2010,*Platini2008,Karevski2013,*Popkov2013a,Prosen2008,*Prosen2014,Prosen2015,*Prosen2011,*Prosen2011b,Prosen2009,*Prosen2012,*Prosen2013a,popkov1,*Popkov2013b,*popkov2,Mendoza-Arenas2013,*Mendoza-Arenas2014a,*Mendoza-Arenas2013a,Vorberg2015,Yan2009,*Zhang2009,Znidari2014,*Znidaric2015,*Znidaric2011,*Znidaric2013,Landi2014b,Landi2015a}.

In this paper we wish to consider alternatively  the case of a multi-site bath,  illustrated in Fig.~\ref{fig:drawing}(b). 
Now, the baths act on groups of particles and are such that they correctly thermalize the entire group in which they act. 
This idea was first considered using numerical simulations in Refs.~\cite{Mendoza-Arenas2014a,Prosen2009} for the case of two-spin baths (for a different approach to this idea, see Refs.~\cite{Platini2006a,Collura2014,Aschbacher2007,*Aschbacher2007a}). 
In this paper our goal is to implement these multi-site baths in a model which is analytically tractable and which allows the generalization to an arbitrary number of sites, including  the thermodynamic limit. 
As a working model, we will consider a quantum XX spin chain or, what is equivalent, the tight-binding model for electrons hopping in a lattice. 
Due to the quadratic nature of this system, the multi-site baths may be implemented for any chain size,   by coupling the  Lindblad operators directly to the normal modes of the chain. 
This will not only produce the correct target state, but will also produce the correct thermalization rates. 
Moreover, it is prone to analytical investigations, for any chain size.
As we will show, despite the simplicity of the model, the NESS shows a much more sophisticated structure than that of the single-site bath.

As a consequence of the exact duality between the quantum XX spin chain and the tight-binding model, all physical  results have two complementary interpretations.
In the case of the tight-binding model, the particle current will stand for the electric current in the system. 
As we will show, in this interpretation, our construction will resemble the ballistic conduction of electrons commonly studied in mesoscopic devices \cite{VanWees1988,Pastawski1991,Baringhaus2014}.
Indeed, we will show that the  steady-state particle current  may be written as a Landauer formula \cite{Landauer1987}. 
Moreover, since we have control of both temperature and chemical potential gradients, we are also able to study thermoelectric (Peltier-Seebeck) effects and  obtain exact formulas for the Onsager coefficients. 

Conversely, if our system is interpreted as a quantum XX spin chain, then the particle flux becomes the magnon flux.
The situation then approaches the experiments in \cite{Serga2010,*Chumak2014,*Chumak2015} involving the flux of magnons in engineered YIG crystals. 
In these experiments magnons are usually injected using a time-dependent local magnetic field generated by a microstrip antena.
These magnons are then parametrically converted due to natural 4-term interaction processes inside the system. 
The final result, as discussed in \cite{Demokritov2006}, is that the microstrip antena functions like an effective chemical potential for magnons.

%
%
%
%

\section{Multi-site Lindblad baths}

\subsection{The XX chain and Lindblad dissipators}

Consider a one-dimensional quantum XX chain with $L$ sites. The Hamiltonian of the system is 
\begin{equation}\label{H1}
H = - \frac{h}{2} \sum\limits_{n=1}^L \sigma_n^z - \frac{t}{2} \sum\limits_{n=1}^{L-1} (\sigma_n^x \sigma_{n+1}^x + \sigma_n^y \sigma_{n+1}^y)
\end{equation}
where the $\sigma_n^\alpha$ are the usual Pauli matrices. Here $h$ represents the magnetic field and $t$ represents the exchange constant between two neighboring spins.
This Hamiltonian may be converted to a fermionic representation through the Jordan-Wigner transformation \cite{Lieb1961,Lieb1964} by introducing a new set of operators $\eta_n$ according to
\begin{equation}\label{JW}
\eta_n = \left[ \prod\limits_{j = 1}^{n-1} e^{i \pi \sigma_j^+ \sigma_j^-} \right] \sigma_n^-
\end{equation}
where $\sigma_n^\pm  = (\sigma_n^x \pm i \sigma_n^y)/2$. 
These operators satisfy the usual fermionic algebra
\begin{equation}\label{comm}
\{\eta_n, \eta_{n'}^\dagger \} = \delta_{n,n'},\qquad \{\eta_n,\eta_{n'}\} = 0
\end{equation}
and, in terms of them,  Eq.~(\ref{H1}) is converted into
\begin{equation}\label{H0}
H = - h \sum\limits_{n=1}^L \eta_n^\dagger \eta_n - t\sum\limits_{n=1}^{L-1} (\eta_n^\dagger \eta_{n+1} + \eta_{n+1}^\dagger \eta_n)
\end{equation}
which is the fermionic representation of the XX chain.
Eq.~(\ref{H0}) also coincides exactly with the tight-binding Hamiltonian, describing the tunneling of electrons through a lattice (typical values of $t$ are in the order of 1 eV). 
In this case $t$ represents the probability amplitude for an electron to tunnel to a neighboring site, whereas $h$ represents the on-site energy of the electrons. 
The magnetization $\sigma_n^z$ and the site occupation numbers $\eta_n^\dagger \eta_n$ are related by $\sigma_n^z =  (2\eta_n^\dagger \eta_n - 1)$, so a fully occupied site is translated into a state fully magnetized  in the $+z$ direction, whereas a fully unoccupied site represents a fully magnetized state in the $-z$ direction.

Eq.~(\ref{H0}) is diagonalized trivially by moving to Fourier space.
But since we work with fixed boundary conditions, we must use a pseudo-momentum orthogonal transformation matrix
\begin{equation}\label{S}
S_{nk} = \sqrt{\frac{2}{L+1}} \sin(nk),\qquad k = \frac{\pi}{L+1}, \ldots, \frac{L\pi}{L+1}
\end{equation}
The pseudo-momenta $k$ take on $L$ distinct values in the interval $k\in [0,\pi]$. 
The $L\times L$ matrix $S$, with entries $S_{nk}$ is both orthogonal and symmetric. 
We now introduce a new set of fermionic operators according to 
\begin{equation}\label{ak_an}
\eta_k = \sum\limits_{n} S_{nk} \eta_n
\end{equation}
where we use the indices to distinguish between the two sets of operators, with momentum-like indices such as $k$ and $q$ referring to the Fourier transforms of the original operators, labeled with $n$. 
In terms of these new operators  the Hamiltonian~(\ref{H0}) is put in diagonal form:
\begin{equation}\label{H3}
H = \sum\limits_{k} \epsilon_k \; \eta_k^\dagger \eta_k,\qquad \epsilon_k = -h -2t \cos k
\end{equation}

Suppose now that we wish to couple the entire chain to a  reservoir at a temperature $T$ and a chemical potential $\mu$. 
This can be accomplished under the context of the Lindblad master equation,  by adding a dissipative term to the system's von Neumann equation, describing the time evolution of the density matrix $\rho$; viz, 
\begin{equation}\label{master}
\frac{\ud \rho}{\ud \tau} = -i [H,\rho] + \mathcal{D}(\rho)
\end{equation}
[we use $\tau$ for time in order to avoid confusion with the letter $t$, used for the tunneling rate.]
The choice of the dissipator $\mathcal{D}(\rho)$ is not unique.
Ideally, one should always attempt to derive it from an underlying microscopic theory describing the coupling between the system and the bath, in which case the final structure of $\mathcal{D}(\rho)$ will depend on the Hamiltonian $H$ of the system.
Of course, for models such as the one being studied here, this microscopic approach is unfeasible since we do not have any actual physical model for the bath. 
This problem is commonly avoided by using phenomenological dissipators (ie, dissipators which have not been derived from a microscopic theory). 
Although this may give physically reasonable results, it is well known that such choices of dissipators may also lead to physically wrong answers in certain cases. 
This is one of the main difficulties in using Lindblad master equations in the study of non-equilibrium phenomena.

Fortunately, for the particular Hamiltonian~(\ref{H3}), it is possible to contemplate a general structure for the Lindblad dissipators which correctly relaxes the chain toward the Gibbs thermal state while satisfying detailed balance. 
This can be done from a microscopic theory using a  bath   described by an infinite number of bosonic degrees of freedom, which is coupled linearly (in the $\eta_k$) to the system. 
Moreover, it assumes that the normal modes $\eta_k$ evolve independently of each other (as they must since the Hamiltonian~(\ref{H3}) factors into a sum of independent Hamiltonians for each mode).

The derivation of this dissipator is done the Appendix. 
The result is
\begin{IEEEeqnarray}{rCl}
\label{D}D(\rho) &=& \sum\limits_k 2\gamma_k \bar{n}_k \bigg[\eta_k^\dagger \rho \eta_k - \frac{1}{2} \{\eta_k \eta_k^\dagger,\rho\}\bigg] 
 	\\[0.2cm]
&+&  2\gamma_k (1-\bar{n}_k) \bigg[\eta_k \rho \eta_k^\dagger - \frac{1}{2} \{\eta_k^\dagger \eta_k,\rho\}\bigg] 
\nonumber
\end{IEEEeqnarray}
where
\begin{equation}\label{bar_n}
\bar{n}_k = \frac{1}{e^{(\epsilon_k-\mu)/T}+1}
\end{equation}
is the Fermi-Dirac distribution for mode $k$ and $\gamma_k$ are positive constants.
All information about the system-bath coupling, such as for instance, which particular sites are coupled to the bath, is contained within the $\gamma_k$.
Since we have no information about the system-bath coupling, we will leave our results as general functions of $\gamma_k$ and, eventually, we also assume for simplicity that  $\gamma_k = \gamma$ for all $k$.

The master Eq.~(\ref{master}), with $H$  given by Eq.~(\ref{H3}) and $\mathcal{D}(\rho)$ given by Eq.~(\ref{D}), will relax the system from \emph{any} initial 
density matrix $\rho(0)$ to the Grand Canonical Gibbs state $\rho(\infty) \propto e^{- (H - \mu N)/T}$. 
It can also be shown that this process satisfies detailed balance. 
That is, if we let $|i\rangle$ denote the eigenvectors of $H-\mu N$, with corresponding eigenvalues $\Omega_i$, then the time evolution of the diagonal entries $p_i = \langle i | \rho | i\rangle$ will evolve independently of the off-diagonal entries, according to the Pauli Master equation 
\[
\frac{\ud p_i}{\ud t} = \sum\limits_j \bigg\{ W_{i,j} p_j - W_{j,i} p_i\bigg\}
\]
where the transition rates $W_{i,j}$  satisfy the usual detailed balance relation
\[
\frac{W_{j,i}}{W_{i,j}} = e^{-(\Omega_j - \Omega_i)/T}
\]
In fact, this relation is a direct consequence of the Kubo-Martin-Schwinger condition of the bath degrees of freedom.


It is also interesting to look at the expectation values of the correlations $\langle \eta_k^\dagger \eta_{k'} \rangle = \tr(\eta_k^\dagger \eta_{k'} \rho)$. 
Using Eq.~(\ref{master}) we find
\begin{IEEEeqnarray}{rCl}
\frac{\ud }{\ud \tau} \langle \eta_k^\dagger \eta_k \rangle &=& 2\gamma_k (\bar{n}_k - \langle \eta_k^\dagger \eta_k \rangle)	\\[0.2cm]
\frac{\ud }{\ud \tau} \langle \eta_k^\dagger \eta_{k'} \rangle &=& -(\gamma_k + \gamma_{k'}) \langle \eta_k^\dagger \eta_{k'} \rangle,\quad k'\neq k
\end{IEEEeqnarray}
Hence, all cross correlations relax to zero, whereas the occupation numbers relax to the  equilibrium Fermi-Dirac occupations $\bar{n}_k$.

\subsection{Partial coupling to the  baths}

We now consider the situation depicted in Fig.~\ref{fig:drawing}, where our XX  chain of size $L$ is coupled to two heat baths kept at different temperatures and chemical potentials. 
We divide the chain into three parts, which we will henceforth refer to as A, B and C. 
The size of each part is $L_a$, $L_b$ and $L_c$, so the total size of the chain is $L = L_a + L_b + L_c$. 
For simplicity, we choose $L_c = L_a$. 
We will also be interested in the case that $L_b = 0$, which is illustrated in Fig.~\ref{fig:drawing}(c).

The Hamiltonian of the chain is given by Eq.~(\ref{H0}). 
For bookkeeping purposes, we rename the fermionic operators $\eta_n$ as follows:  $\eta_n = a_n$ with $n \in [1,L_a]$, $\eta_{L_a+n}  = b_n$, with $n \in [1,L_b]$ and $\eta_{L_a+L_b + n} = c_n$ with $n \in [L_a+L_b + 1, L]$.
The three set of operators $a_n$, $b_n$ and $c_n$ still satisfy the Fermionic algebra in Eq.~(\ref{comm}). 
We then divide the Hamiltonian~(\ref{H0}) as 
\begin{equation}
H = H_a + H_b + H_c + V_{ab} + V_{bc}
\end{equation}
where
\begin{equation}\label{Halpha}
H_\alpha = -h  \sum\limits_{n=1}^{L_\alpha} \alpha_n^\dagger \alpha_n   - t\sum\limits_{n=1}^{L_\alpha-1} (\alpha_n^\dagger \alpha_{n+1} + \alpha_{n+1}^\dagger \alpha_n)
\end{equation}
Here and henceforth $\alpha \in \{a,b,c\}$ will serve both as a label for each chain and to denote the corresponding creation and annihilation operators. 

As for the interactions between the chains, we now introduce a slight modification in the original model and write it as 
\begin{IEEEeqnarray}{rCl}
\label{VAB}V_{ab} &=& -g(a_{L_a}^\dagger b_{1} + b_{1}^\dagger a_{L_a})	\\[0.2cm]
\label{VBC}V_{bc} &=& -g(b_{L_b}^\dagger c_{1} + c_{1}^\dagger b_{L_b})	
\end{IEEEeqnarray}
That is, we use a different coupling constant $g$, instead of $t$. 
When $g = t$ we recover the uniform chain in Eq.~(\ref{H0}).
The assumption that $g \neq t$ means that the hopping rate inside the chains is different from the hopping rate between different chains. 
The reason for this choice is that, as  will be shown below, when $g\ll t$, the problem is amenable to analytical calculations using perturbation theory.

We now wish to couple   chains A and C to independent heat baths. 
To accomplish this we first diagonalize each chain individually by defining orthogonal transformation matrices exactly as in Eq.~(\ref{S}), but with the appropriate sizes, $L_a$, $L_b$ and $L_c$.
To avoid confusion, we will denote the corresponding matrices by $S^\alpha$, where $\alpha \in \{a,b,c\}$.
Notice also that for each matrix $S^\alpha$, the allowed values of $k$ may be different [cf. Eq.~(\ref{S})].

We then define new operators $a_k$, $b_k$ and $c_k$ exactly as in Eq.~(\ref{ak_an}), which  diagonalize the three chains individually:
\begin{IEEEeqnarray}{rCl}
\label{Halpha_diag}H_\alpha &=& \sum\limits_k \epsilon_{\alpha,k}\; \alpha_k^\dagger \alpha_k. \qquad \epsilon_{\alpha,k} = -h - 2t \cos k
\end{IEEEeqnarray}
In principle we could write $\epsilon_{k}$ instead of $\epsilon_{\alpha,k}$, but this notation is convenient for bookkeeping. 
It also emphasizes the fact that the allowed values of $k$ themselves depend on $\alpha$. 
In momentum space, the interaction terms in Eqs.~(\ref{VAB}) and (\ref{VBC}) become
\begin{IEEEeqnarray}{rCl}
\label{VAB2}V_{ab} &=& -g \sum\limits_{k,q} S_{L_a,k}^a S_{1,q}^b  (a_k^\dagger b_q + b_q^\dagger a_k)	\\[0.2cm]
\label{VBC2}V_{bc} &=& -g  \sum\limits_{q,k} S_{L_b,q}^b S_{1,k}^c  (b_q^\dagger c_k + c_k^\dagger b_q)
\end{IEEEeqnarray}

In order to couple chains A and C to heat reservoirs, we now write the quantum master equation for the system as 
\begin{equation}\label{master2}
\frac{\ud \rho}{\ud \tau} = -i [H,\rho] + D_a(\rho)+ D_c(\rho)
\end{equation}
where 
\begin{IEEEeqnarray}{rCl}
\label{D2}D_\alpha(\rho) &=& \sum\limits_{k} 2\gamma_{\alpha,k} \bar{n}_{\alpha,k} \bigg[\alpha_k^\dagger \rho \alpha_k - \frac{1}{2} \{\alpha_k \alpha_k^\dagger,\rho\}\bigg] 
 	\\[0.2cm]
&+&  2\gamma_{\alpha,k} (1-\bar{n}_{\alpha,k}) \bigg[\alpha_k \rho \alpha_k^\dagger - \frac{1}{2} \{\alpha_k^\dagger \alpha_k,\rho\}\bigg] 
\nonumber
\end{IEEEeqnarray}
and 
\begin{equation}\label{bar_n2}
\bar{n}_{\alpha,k} = \frac{1}{e^{(\epsilon_{\alpha,k}-\mu_\alpha)/T_\alpha}+1}
\end{equation}
is the Fermi-Dirac distribution for each individual chain.
For simplicity, we will usually assume that $\gamma_{\alpha,k} = \gamma$ but, again,  the notation $\gamma_{\alpha,k}$ may be useful for bookkeeping purposes.

As discussed above, the individual chains, with the their corresponding dissipators, will satisfy detailed balance. 
But when we couple them together, detailed balance is violated. 
An important question is therefore, whether or not one may recover detailed balance for certain parameter ranges. 
Below we will show that this happens when $\gamma$ is sufficiently small.
From a physical standpoint we indeed expect that $\gamma \ll t$, since $\gamma$ describes the rate at which particles are injected in the system, whereas the tunneling rate $t$ describes the typical propagation times of the excitations through the chains. 
Moreover, as shown in Ref.~\cite{Landi2015a}, if $\gamma \sim t$, particle-particle interactions become important and the non-interacting model in Eq.~(\ref{H0}) would no longer be valid. 
Notwithstanding, in this paper we will consider all values of $\gamma$, with the purpose of understanding exactly how it modifies the NESS and detailed balance.

\subsection{Lyapunov equation for the covariance matrix}

The quadratic nature of our model allows for the problem to be cast as a closed system of equations for the entries of the $L\times L$ covariance matrix: 
\begin{equation}\label{theta_def}
\theta_{\alpha k, \beta q} = \langle \alpha_k^\dagger \beta_q\rangle = \tr\bigg(\alpha_k^\dagger \beta_q \rho \bigg)
\end{equation}
It is convenient to divide $\theta$ into a $3\times 3$ block structure
\begin{equation}\label{theta_block}
\theta = \begin{pmatrix}
\theta_A 			&	\theta_{AB}			&	\theta_{AC}	\\[0.2cm]
\theta_{AB}^\dagger	&	\theta_B				&	\theta_{BC}	\\[0.2cm]
\theta_{AC}^\dagger &	\theta_{BC}^\dagger		&	\theta_C
\end{pmatrix}
\end{equation}
The time evolution of $\theta$ may be found directly from Eq.~(\ref{master2}) and reads:
\begin{equation}\label{theta}
\frac{\ud \theta}{\ud \tau} = i [W, \theta] - \{\Gamma,\theta\} + 2\mathcal{D}
\end{equation}
where $W$, $\Gamma$ and $\mathcal{D}$ are $L\times L$ matrices. The matrices $\Gamma$ and $\mathcal{D}$ stem from the dissipative part of the dynamics and read
\begin{IEEEeqnarray}{rCl}
\label{Gamma}\Gamma &=&  \text{diag}(\gamma_{a,k},0,\gamma_{c,k})	\\[0.2cm]
\label{mathD}\mathcal{D} &=& \;\text{diag}( \gamma_{a,k} \bar{n}_{a,k},0, \gamma_{c,k} \bar{n}_{c,k})
\end{IEEEeqnarray}
The matrix $W$, on the other hand, is a unitary contribution [stemming from the first term in Eq.~(\ref{master2})] and may be written as 
\begin{equation}\label{W}
W = W_0 - g W_1 
\end{equation}
where 
\begin{equation}\label{W0}
W_0 = \text{diag}(\epsilon_{a,k}, \epsilon_{b,k}, \epsilon_{c,k})
\end{equation}
and
\begin{equation}\label{Vmat}
W_1 = \begin{pmatrix} 
0 			&	 S_{ab}			&		0	\\[0.2cm]
S_{ab}\trans	&	0				&		S_{bc}	\\[0.2cm]
0			&	S_{bc}\trans		&		0
\end{pmatrix}
\end{equation}
Here $S_{ab}$ and $S_{bc}$ are rectangular matrices with entries 
\begin{equation}\label{Smats}
(S_{ab})_{k, q} = S_{L_a,k}^a S_{1,q}^b,\quad \text{ and } \quad (S_{bc})_{q, k} = S_{L_b,q}^b S_{1,k}^c 
\end{equation}

We are interested in the steady-state solution of Eq.~(\ref{theta}), which reads.
\begin{equation}\label{theta_ss0} 
\{\Gamma,\theta\} -i [W,\theta] = 2 \mathcal{D}
\end{equation}
This is a linear matrix equation for $\theta$. It can be solved numerically by writing it as a Lyapunov equation
\[
A \theta + \theta A^\dagger = 2 \mathcal{D}
\]
where $A = \Gamma - i W$. 
Efficient  Lyapunov solvers are nowadays implemented in most numerical libraries. 
The numerical solutions were used to check the correctness of all results shown in this paper.

%
%
%
%

\section{Perturbative solution}

The analytical solution of Eq.~(\ref{theta_ss0}) for arbitrary size is quite complicated. 
However, the problem may be treated analytically if we assume that $g\ll t$ and expand $\theta$ in a power series in $g$:
\begin{equation}\label{theta_series}
\theta = \theta_0 + g \theta_1 + g^2 \theta_2 + \ldots
\end{equation}
It is convenient to define the linear matrix operator 
\begin{equation}\label{Upsilon}
\Upsilon(\theta) = \{\Gamma,\theta\} - i [W_0,\theta]
\end{equation}
so that Eq.~(\ref{theta_ss0}) may be written as 
\begin{equation}\label{theta_ss} 
\Upsilon(\theta) = 2\mathcal{D} - i g [W_1, \theta]
\end{equation}
Inserting Eq.~(\ref{theta_series}) into this formula and collecting terms of the same order in $g$ then yields the following sequence of equations:
\begin{IEEEeqnarray}{rCl}
\label{theta0}\Upsilon(\theta_0) &=& 2 \mathcal{D}	\\[0.2cm]
\label{theta1}\Upsilon(\theta_1) &=& -i  [W_1,\theta_0]	\\[0.2cm]
\label{theta2}\Upsilon(\theta_2) &=& -i  [W_1,\theta_1]	
\end{IEEEeqnarray}
and etc. These equations may then be solved sequentially.
From extensive numerical analyses of Eq.~(\ref{theta_ss0}), we have concluded that high values of $g$ do not lead to any new physical effects. 
Hence, in this paper we will restrict the discussion up to linear order in $g$, Eqs.~(\ref{theta0}) and (\ref{theta1}).

Matrix equations are most easily handled using outer products, which we introduce through a vector basis $|\alpha,k\rangle$ (the use of Dirac's notation is not at all necessary, but simply convenient).
All matrices appearing in Eq.~(\ref{theta_ss}) may now be written in terms of outer products $|\alpha,k\rangle\langle \beta ,q|$. 
For instance, the matrices $\Gamma$, $\mathcal{D}$ and $W_0$ in Eqs.~(\ref{Gamma}), (\ref{mathD}) and (\ref{W0}) are all diagonal and read:
\begin{IEEEeqnarray}{rCl}
\label{Gamma2}\Gamma &=& \sum\limits_{\alpha,k} \; \gamma_{\alpha,k}|\alpha,k\rangle \langle \alpha, k|	\\[0.2cm]
\label{mathD2}\mathcal{D} &=& \sum\limits_{\alpha,k} \gamma_{\alpha,k}  \bar{n}_{\alpha,k}\; |\alpha,k\rangle \langle \alpha, k|	\\[0.2cm]
\label{W02} W_0 &=& \sum\limits_{\alpha,k} \epsilon_{\alpha,k}\; |\alpha,k\rangle \langle \alpha, k|
\end{IEEEeqnarray}
We similarly decompose the covariance matrix $\theta$ in Eq.~(\ref{theta_def}) by introducing two completeness relations: 
\begin{equation}\label{theta_def2}
\theta = \sum\limits_{\alpha,k,\beta,q} |\alpha,k\rangle\langle \alpha,k| \theta | \beta,q\rangle\langle \beta,q|
\end{equation}
The operator  $\Upsilon(\theta)$ in Eq.~(\ref{Upsilon}) may now be conveniently written as 
\begin{IEEEeqnarray}{rCl}
\label{Upsilon_theta}
\Upsilon(\theta) = \sum\limits_{\alpha,k,\beta,q} \bigg[\gamma_{\alpha,k}&& + \gamma_{\beta,q}-i(\epsilon_{\alpha,k}-\epsilon_{\beta,q})\bigg] 
\times\\[0.2cm]
&&\times \;|\alpha,k\rangle\langle \alpha,k| \theta | \beta,q\rangle\langle \beta,q|	\nonumber
\end{IEEEeqnarray}
where $\gamma_{a,k} = \gamma_{c,k} = \gamma$ and $\gamma_{b,k} = 0$.

With these results we may readily solve the zeroth-order Eq.~(\ref{theta0}). 
Since it represents the situation where the three chains are uncoupled, its solution will be a diagonal matrix whose entries are simply the equilibrium occupation numbers:
\begin{equation}\label{theta0_sol}
\theta_0 = \sum\limits_{\alpha,k} \bar{n}_{\alpha,k} |\alpha, k \rangle \langle \alpha, k| 
\end{equation}
where $\bar{n}_{\alpha,k}$ is given in Eq.~(\ref{bar_n2}).
However, since chain B is not coupled to any reservoirs, its zeroth-order occupation numbers $\bar{n}_{b,q}$ remain undetermined from this equation 
As we will show below, they can be fixed from the first order Eq.~(\ref{theta1}). 
[The off-diagonal elements of $\theta_B$ are zero; it is only the diagonal elements which remain undetermined.]

Next we turn to the first-order Eq.~(\ref{theta1}). 
In this case it is convenient to separate the cases $L_b \neq 0$ and  $L_b = 0$ [cf. Fig.~\ref{fig:drawing}(c)].
We begin with the latter.

\subsection{Solution when $L_b = 0$}

When $L_b = 0$ all formulas of the previous subsection remain valid, provided that the indices $\alpha$ be restricted to $\alpha\in \{a,c\}$.
Moreover, to solve Eq.~(\ref{theta1}) we need $[W_1,\theta_0]$ and the matrix $W_1$ in Eq.~(\ref{Vmat}) needs to be modified in this case.
It now becomes, in outer product notation,
\begin{equation}\label{Vmat2} 
W_1 = \sum\limits_{k,q} S_{L_a,k}^a S_{1,q}^c |a,k\rangle \langle c, q| + \text{trans}
\end{equation}
where ``trans'' stands for transpose.
Using this result together with Eq.~(\ref{theta0_sol}), we then find that 
\begin{equation}
-i [W_1,\theta_0] = - i  \sum\limits_{k,q} S_{L_a,k}^a S_{1,q}^c (\bar{n}_{c,q} - \bar{n}_{a,k}) |a,k\rangle \langle c,q| + \text{trans}
\end{equation}
Substituting this in Eq.~(\ref{theta1}) then allows us to conclude  that the only non-zero entries of $\theta_1$ will be 
\begin{equation}\label{theta1_sol}
\langle a,k| \theta_1 |c, q\rangle = i  \frac{S_{L_a,k}^a S_{1,q}^c (\bar{n}_{a,k} - \bar{n}_{c,q})}{2\gamma-i(\epsilon_{a,k}-\epsilon_{c,q})}
\end{equation}
In reference to the block structure in Eq.~(\ref{theta_block}), this corresponds to the elements $\theta_{AC}$. 
The complete covariance matrix, up to first order, is therefore $\theta = \theta_0 + g \theta_1$, where $\theta_0$ is given in Eq.~(\ref{theta0_sol}) and $\theta_1$ is given in Eq.~(\ref{theta1_sol}).

As can be seen in Eq.~(\ref{theta1_sol}), the result depends only on energy differences $\epsilon_{a,k} - \epsilon_{c,q}$, which are defined in Eq.~(\ref{Halpha_diag}). 
Hence, the constant $h$ cancels out in the denominator and remains only in the Fermi-Dirac occupation numbers. 
We will therefore  absorb $h$ into the definition of the chemical potentials $\mu_\alpha$, which is tantamount to setting $h = 0$.

\subsection{\label{ssec:sol_LbNEQ}Solution when $L_b \neq 0$}

Next we turn to the case $L_b \neq 0$, so once again $\alpha \in \{a,b,c\}$.
The commutator  $-i[W_1,\theta_0]$, using Eqs.~(\ref{Vmat}) and (\ref{theta0_sol}), becomes
\begin{IEEEeqnarray}{rCl}
-i [W_1,\theta_0] &=& - i  \sum\limits_{k,q} \bigg[S_{L_a,k}^a S_{1,q}^b (\bar{n}_{b,q} - \bar{n}_{a,k}) |a,k\rangle \langle b,q| 	\nonumber	\\[0.2cm]
&+& S_{L_b,q}^b S_{1,k}^c (\bar{n}_{c,k} - \bar{n}_{b,q}) |b,q\rangle\langle c,k| + \text{trans}\bigg]	\IEEEeqnarraynumspace
\end{IEEEeqnarray}
Combining this with Eq.~(\ref{Upsilon}) then gives us the non-zero entries of $\theta_1$:
\begin{IEEEeqnarray}{rCl}
\label{theta1_ab}\langle a,k | \theta_1 | b, q\rangle &=& i \frac{S_{L_a,k}^a S_{1,q}^b (\bar{n}_{a,k} - \bar{n}_{b,q})}{\gamma - i (\epsilon_{a,k} - \epsilon_{b,q})}	\\[0.2cm]
\label{theta1_bc}\langle b,q | \theta_1 | c, k\rangle &=& i \frac{S_{L_b,q}^b S_{1,k}^c (\bar{n}_{b,q} - \bar{n}_{c,k})}{\gamma - i (\epsilon_{b,q} - \epsilon_{c,k})}
\end{IEEEeqnarray}
Unlike Eq.~(\ref{theta1_sol}), in this formula the denominator depends on $\gamma$ and not $2\gamma$, which is a consequence of the fact that $\gamma_{b,q} = 0$.

Eqs.~(\ref{theta1_ab}) and (\ref{theta1_bc}) still depend on $\bar{n}_{b,q}$, which is not yet fixed.
That can be accomplished by imposing a symmetry conservation based on the time evolution of $\langle b_q^\dagger b_q \rangle$. 
Using Eq.~(\ref{master2}) we find that
\[
\frac{\ud \langle b_q^\dagger b_q \rangle}{\ud \tau} = i\langle [V_{ab},b_q^\dagger b_q] \rangle + i\langle [V_{bc}, b_q^\dagger b_q] \rangle
\]
and, using Eqs.~(\ref{VAB2}) and (\ref{VBC2}), we have 
\begin{IEEEeqnarray}{rCl}
i\langle [V_{ab},b_q^\dagger b_q] \rangle	 &=& 2 g \sum\limits_{k} S_{L_a,k}^a S_{1,q}^b \text{Im} \langle a,k | \theta | b,q\rangle 
\\[0.2cm]
-i\langle [V_{bc}, b_q^\dagger b_q] \rangle  &=& 2 g \sum\limits_{k} S_{L_b,q}^b S_{1,k}^c \text{Im} \langle b,q | \theta | c,k\rangle 
\end{IEEEeqnarray}
In the steady-state $\ud \langle b_q^\dagger b_q \rangle/\ud \tau = 0$ so these two quantities should be equal; ie, 
\begin{equation}
\sum\limits_{k} S_{L_a,k}^a S_{1,q}^b \text{Im} \langle a,k | \theta | b,q\rangle 
=
\sum\limits_{k} S_{L_b,q}^b S_{1,k}^c \text{Im} \langle b,q | \theta | c,k\rangle 
\end{equation}
The equality holds only for the sum as a whole and not for the individual elements. 
Inserting Eqs.~(\ref{theta1_ab}) and (\ref{theta1_bc}) into this result then determines $\bar{n}_{b,q}$ uniquely. 

To write down the final result we recall that  from Eq.~(\ref{Halpha_diag}),  $\epsilon_{c,k} = \epsilon_{a,k}$. 
Using also the explicit values of $S^\alpha$ in Eq.~(\ref{S}), we may then write
\begin{equation}\label{nbq}
\bar{n}_{b,q} = \frac{\sum\limits_{k} f(k,q) (\bar{n}_{a,k} + \bar{n}_{c,k})}{2\sum\limits_{k} f(k,q)}
\end{equation}
where
\begin{equation}\label{fkq}
f(k,q) = \frac{\sin^2k}{\gamma^2 + 4t^2(\cos k - \cos q)^2}
\end{equation}
This result is physically intuitive: $\bar{n}_{b,q}$ is given by a weighted average of the occupation numbers of chains A and C. 
Notice that this result makes no mention to the size of chain B, so that $q$ may be interpreted as a continuous function varying in the interval $q\in [0,\pi]$. 

An important particular case is that of $L_a = L_c = 1$, corresponding to Fig.~\ref{fig:drawing}(a). 
In this case, from Eq.~(\ref{S}), we find that $k$ will take on just a single value: $k = \pi /2$. 
Consequently, Eq.~(\ref{nbq}) is reduced to 
\begin{equation}\label{nbq_1}
\bar{n}_{b,q} = \frac{\bar{n}_a + \bar{n}_c}{2} 
\end{equation}
which is independent of $q$.
This is the typical behavior expected from a ballistic system \cite{Asadian2013,Karevski2009}: the occupation in the middle of the chain is the simple average of the occupation at the boundaries. 

Another particular case is that when $L_b = 1$, corresponding to a single spinless quantum dot in contact with two perfectly conducting leads. 
If we assume that $T_a = T_c$ and that the chemical potentials are inversely polarized ($\mu_a = - \mu_c = \mu$), then it follows that $\bar{n}_{b} = 1/2$ for any $\mu$.

\begin{figure}
\centering
\includegraphics[width=0.23\textwidth]{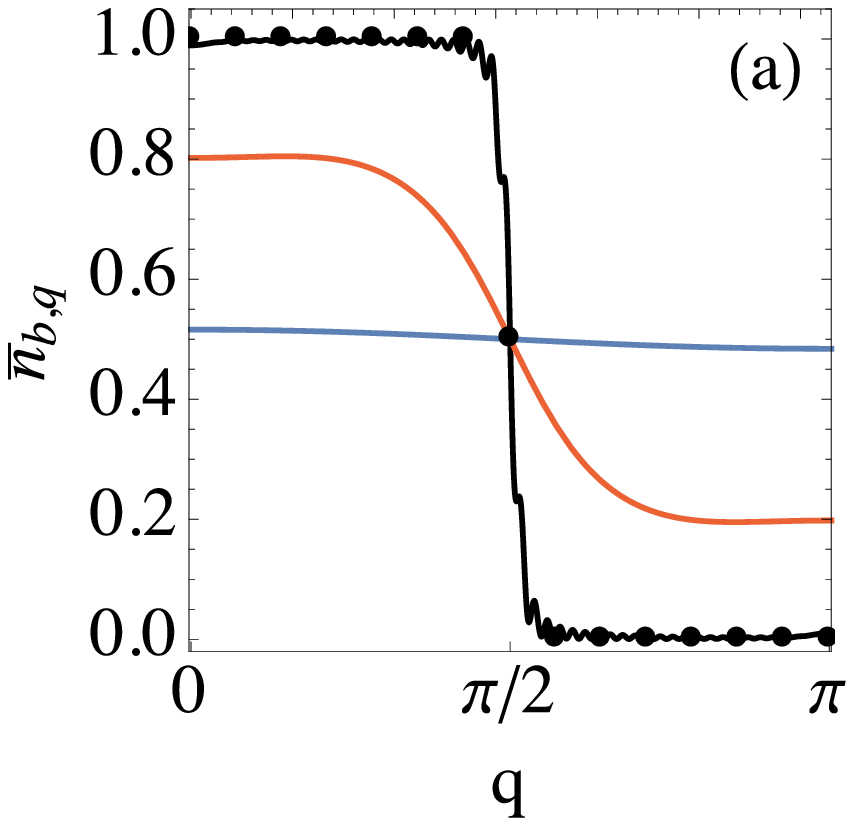}\quad
\includegraphics[width=0.23\textwidth]{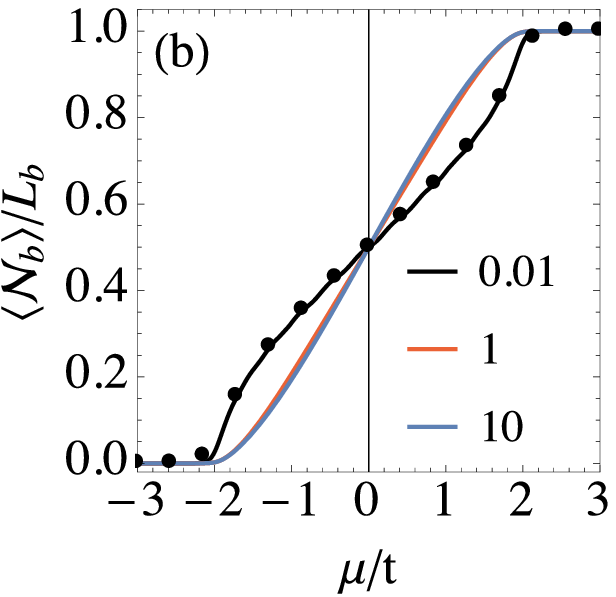}
\caption{\label{fig:occupation}
(a) The occupation numbers of chain B,  $\bar{n}_{b,q}$, computed from Eq.~(\ref{nbq}) with $T_a = T_c = 0.05 t$,  $\mu = 0$ and $L_a = L_c = 50$.
(b) Total number of excitations in chain B, $\langle \mathcal{N}_b \rangle/L_b$, as a function of $\mu= \mu_a = \mu_c$. 
The different curves correspond to different values of $\gamma/t$ and the solid points refer to the occupations of chains A and C. 
}
\end{figure}

In Fig.~\ref{fig:occupation} we illustrate the possible behaviors of $\bar{n}_{b,q}$ in Eq.~(\ref{nbq}).
The parameters used were   $T_a = T_c = 0.05 t$ and $L_a = L_c = 50$. The size $L_b$ does not need to be specified since $q$ may be trated as a continuous variable in Eq.~(\ref{nbq}). 
Fig.~\ref{fig:occupation}(a) shows the individual occupation numbers for $\mu_a = \mu_c = 0$ and Fig.~\ref{fig:occupation}(b) shows the total occupation of chain B, $\langle \mathcal{N}_b \rangle/L_b$ [cf. Eq.~(\ref{Nalpha}) below], as a function of $\mu_a = \mu_c = \mu$. 
Different curves correspond to different values of the bath coupling $\gamma$ and the solid points refer to the exact occupations of chains A or C.
As can be seen in both images, when $\gamma/t\ll 1$ the behavior of chain B mimics closely the behavior of chains A and C.
Conversely, when $\gamma/t \gg 1$ the normal modes are flattened out, leading to a distortion in the $\langle \mathcal{N}_b\rangle$~vs.~$\mu$ curve. 
The size $L_a = 50$ was chosen to illustrate some of the finite size effects that appear in the problem, in this case manifested by the ripples observed in the black curve ($\gamma = 0.01$) of Fig.~\ref{fig:occupation}(a). 
These ripples disappear quickly if  $L_a$ is increased further.

In order to better understand the results of Fig.~\ref{fig:occupation}, it is useful to study the case where chains A and C tend to the thermodynamic limit (while $L_b$ remains arbitrary). 
In this limit we may convert sums, such as Eq.~(\ref{fkq}), into integrals using the recipe $\sum_k = (L_a/\pi) \int_0^\pi \ud k$, which stems from the discrete values of $k$ in Eq.~(\ref{S}). 
The ensuing integral will have a shape that will be encountered often below. 
It is therefore best to consider first a general integral of the form:
\begin{equation}\label{Mdef}
\mathcal{M} = \frac{1}{\pi} \int\limits_0^\pi \ud k \frac{M(k)}{\gamma^2 + t^2 (\cos k - \cos q)^2 }
\end{equation}
where $M(k)$ is an arbitrary function.
This is the form of Eq.~(\ref{fkq}), with  $M(k) = \sin^2 k$ and $t\to 2t$  (up to an irrelevant constant factor). 
It is possible to determine the behavior of this function when $\gamma/t \ll 1$ and $\gamma/t \gg 1$. 
In the latter, we simply neglect the second term in the denominator, which gives 
\begin{equation}\label{M_high}
\mathcal{M} = \frac{1}{\pi \gamma^2} \int\limits_0^\pi \ud k  \; M(k)
\end{equation}
That is, the result becomes independent of $q$. 
Conversely, in the limit $\gamma/t \ll1$ we see from Eq.~(\ref{Mdef}) that the most important contribution to the integral will come from the region where $k \sim q$. 
In this case we may transform this integral into a contour problem and use the residue theorem to find  that
\begin{equation}\label{M_low}
\mathcal{M} = \frac{1}{\gamma t} \frac{M(q)}{\sin q}
\end{equation}
which is roughly the behavior of a $\delta$ function, up to a factor of $\gamma t \sin q$.

Using these results we may study the behavior of $\bar{n}_{b,q}$ in Eq.~(\ref{nbq}) in the limits of low and high $\gamma$. 
We find that 
\begin{IEEEeqnarray}{rCl}
\label{nbq_low}
(\gamma/t \ll 1) 		\quad	 \bar{n}_{b,q} &\simeq& \frac{\bar{n}_{a,q} + \bar{n}_{c,q}}{2}		\\[0.3cm]
\label{nbq_high}
(\gamma/t \gg 1)	\quad	\bar{n}_{b,q} &\simeq& \frac{2}{\pi} \int\limits_0^\pi \ud k\; \sin^2 k \frac{(\bar{n}_{a,k} + \bar{n}_{c,k})}{2}
\end{IEEEeqnarray}
We therefore see two very different physical results. 
When $\gamma/t\ll 1$,  $\bar{n}_{b,q}$ tends to the simple arithmetic average of the occupations of the chains A and C. 
But  when $\gamma/t \gg 1$,  $\bar{n}_{b,q}$ becomes \emph{independent} of $q$, tending to an average of all occupations numbers of chains A and C. 

We may also find an exact formula for Eq.~(\ref{nbq_high}) in the limit $T\to 0$ [the corresponding formula for Eq.~(\ref{nbq_low}) is trivial]. 
In this case we may write $\bar{n}_{\alpha,k} = \Theta(\mu - \epsilon_{\alpha,k})$, where $\Theta(x)$ is the Heaviside function. 
We also define the Fermi momentum from the relation $\epsilon_{k_F} = \mu$, which gives $k_F =  \arccos(-\mu/2t)$. 
Consequently, we find that when $\gamma/t \gg 1$, 
\begin{equation}
\bar{n}_{b,q} = \frac{\mathcal{C}(\mu_a) + \mathcal{C}(\mu_c)}{2}
\end{equation}
where
\begin{IEEEeqnarray}{rCl}
\label{Cmu}\mathcal{C}(\mu) &=& \frac{2}{\pi} \int\limits_0^\pi \ud k \sin^2 k \; \Theta(\mu-\epsilon_k) \nonumber	\\[0.2cm]
&=& \frac{1}{\pi}\bigg[ \mu\sqrt{4 t^2 - \mu^2} + \cos^{-1}(-\mu/2t)\bigg]
\end{IEEEeqnarray}
At $\mu = 0$ we get $\bar{n}_{b,q} = 1/2$, thence corresponding to the blue curve in Fig.~\ref{fig:occupation}(a). 
Moreover, since this result is independent of $q$, the  total occupation $\langle \mathcal{N}_{b}\rangle/L_b$ is then given by the same formula.That is, Eq.~(\ref{Cmu}) as a function of $\mu$   corresponds exactly  to the red and blue curves in Fig.~\ref{fig:occupation}(b).

%
%
%
%

\section{Particle  current}


We will now use the results from the previous section to study the steady-state particle current generated by the unbalance between the two baths.
Let
\begin{equation}\label{Nalpha}
\mathcal{N}_\alpha = \sum\limits_k \alpha_k^\dagger \alpha_k
\end{equation}
denote the total number of particles in  chain $\alpha$, and $\mathcal{N} = \mathcal{N}_a + \mathcal{N}_b + \mathcal{N}_c$ denote the total number of particles in the system. 
The equation for the time evolution of $\mathcal{N}$ may be obtained directly from Eq.~(\ref{master2}). 
Since $[H,\mathcal{N}] = 0$, it becomes simply
\begin{equation}\label{Ntot_evo}
\frac{\ud \langle \mathcal{N} \rangle }{\ud \tau}  = \tr\bigg[\mathcal{N}_a D_a(\rho)\bigg] +  \tr\bigg[\mathcal{N}_c D_c(\rho)\bigg]
\end{equation}
This equation shows that the reservoirs of A and C are the only two possible channels through which particles may flow into or out of the system. 
In the steady-state $\ud \langle \mathcal{N} \rangle/\ud \tau = 0$ and we therefore obtain
\begin{equation}\label{J1}
J := \tr\bigg[\mathcal{N}_a D_a(\rho)\bigg]  = - \tr\bigg[\mathcal{N}_c D_c(\rho)\bigg]
\end{equation}
The quantity $J$ represents the current of particles through the system. 
When $J>0$ it means particles are entering the system from reservoir A.
The electric current can be obtained from $J$ by multiplying by the electric charge $-e$.

Using Eqs.~(\ref{D}) and (\ref{Nalpha}) one may readily show  that
\begin{equation}\label{Jbad}
J = 2\gamma \sum\limits_{k}  (\bar{n}_{a,k} - \langle a_k^\dagger a_k \rangle)
\end{equation}
with a similar formula in terms of $\langle c_k^\dagger c_k \rangle$. 
It is also possible to obtain alternative formulas for the current, which coincide with Eq.~(\ref{Jbad}) in the steady-state, but  may be more convenient to work with. 
This is important because, as seen in Eq.~(\ref{theta0_sol}), deviations in the occupation numbers $\langle a_k^\dagger a_k \rangle$ will be of order $g^2$. Hence, to use Eq.~(\ref{Jbad}) we would need to continue the expansion of the covariance matrix up to terms $g^2$. 

Instead, we may look for an alternative formula starting from the equation describing the time-evolution of $\mathcal{N}_a$, also obtained from Eq.~(\ref{master2}). 
It reads 
\begin{equation}\label{NA}
\frac{\ud \langle \mathcal{N}_a \rangle}{\ud \tau}  = i \langle [V_{ab}, \mathcal{N}_a] \rangle + \tr\bigg[\mathcal{N}_a D_a(\rho)\bigg]
\end{equation}
Thus, we see that particles may flow away from chain A either to its reservoir or toward chain B (or chain C when $L_b = 0$).
Comparing with Eq.~(\ref{J1}) we see that in the steady-state we should have
\begin{equation}
J = - i \langle [V_{ab}, \mathcal{N}_a] \rangle
\end{equation}
and  using Eqs.~(\ref{VAB2}) and (\ref{Nalpha}) this finally becomes
\begin{equation}\label{J}
J = - i g\sum\limits_{k,q} S_{L_a,k}^a S_{1,q}^b  \langle a_k^\dagger b_q - b_q^\dagger a_k\rangle
\end{equation}
In the steady-state this formula is equivalent to  Eq.~(\ref{Jbad}). 
However, it has the advantage. but can be used together with the first order solution for the covariance matrix (an analogous formula could be defined for chain C).
When $L_b = 0$ it should be replaced by 
\begin{equation}\label{J_Lb0}
J = - i g\sum\limits_{k,q} S_{L_a,k}^a S_{1,q}^c  \langle a_k^\dagger c_q -c_q^\dagger a_k\rangle
\end{equation}

\subsection{Current when  $L_b  = 0$}

When $L_b = 0$ the relevant entries of the covariance matrix are given in Eq.~(\ref{theta1_sol}). 
Using the specific values of $\epsilon_{\alpha,k}$ in Eq.~(\ref{Halpha_diag}) and of $S^\alpha$ in Eq.~(\ref{S}), and exploiting the symmetry of Eq.~(\ref{J_Lb0}) with respect to $k$ and $q$, we may write  the particle current as
\begin{equation}\label{J_final}
J = \frac{4 g^2 \gamma}{(L_a+1)^2} \sum\limits_{k,q} \frac{\sin^2k \sin^2 q \;(\bar{n}_{a,k} - \bar{n}_{c,k})}{ \gamma^2 + t^2(\cos k-\cos q)^2}
\end{equation}
As expected, $J = 0$ if $g = 0$ or $\gamma = 0$. 
When $g=0$ we are uncoupling the two chains and when $\gamma = 0$ we are uncoupling the chains from their respective heat reservoirs. 
The current is also zero if $\bar{n}_{a,k} = \bar{n}_{c,k}$, as of course expected.

\begin{figure}
\centering
\includegraphics[width=0.42\textwidth]{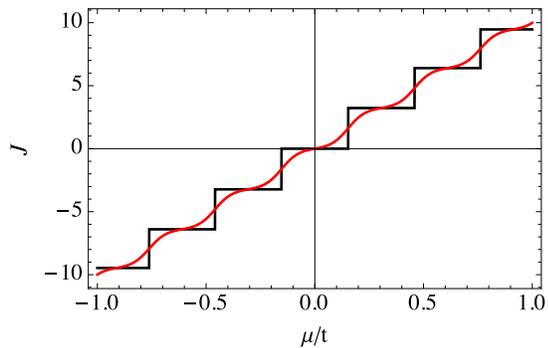}
\caption{\label{fig:voltage}(Color online) The particle current $J$ (in units of $g^2 \gamma/t^2$) as a function of the chemical potential difference (voltage bias) $\mu$  computed using Eq.~(\ref{J_final}) with $L_a = 40$ and $\gamma/t = 1$. 
The black curve corresponds to   $T_a = T_c = 0$  and the red curve to $T_a = T_c = 0.02 t$.
}
\end{figure}

It is convenient to define 
\begin{equation}\label{I_finite}
\mathcal{I}(k) = \frac{\sin^2k}{(L_a + 1)} \sum\limits_q \frac{\sin^2 q }{ \gamma^2 + t^2(\cos k-\cos q)^2}
\end{equation}
so that Eq.~(\ref{J_final}) may  be written as 
\begin{equation}\label{J_final2}
J = \frac{4  g^2 \gamma}{(L_a+1)} \sum\limits_{k} \mathcal{I}(k)(\bar{n}_{a,k} - \bar{n}_{c,k})
\end{equation}
This equation has the structure of Landauer's formula for the ballistic conduction of electrons through tunneling junctions \cite{Landauer1987,VanWees1988,Pastawski1991,Baringhaus2014}.
To illustrate this we present in Fig.~\ref{fig:voltage}  results for the current when $\mu_a = \mu/2$ and $\mu_c = -\mu/2$, so that the potential difference (voltagem bias) is $\mu$. 
Here and henceforth, all currents will be given in units of $g^2 \gamma/t^2$. 
As can be seen in the figure, the particle current shows a series of discrete jumps, as in electron tunneling experiments \cite{VanWees1988}. 
These jumps reflect the discreteness of the occupation numbers $\bar{n}_{\alpha,k}$ and are smoothed out as the temperature increases (illustrated in the image by the red curve). 

Next we turn to the case of infinitesimal unbalances. 
That is, we take $\mu_a = \mu+\delta\mu/2$, $\mu_c = \mu - \delta \mu/2$, $T_a = T+\delta T/2$ and $T_c = T - \delta T/2$, where $\delta \mu$ and $\delta T$ are assumed to be infinitesimal. 
In this case we may expand $\bar{n}_{a,k}$ and $\bar{n}_{c,k}$ in a power series. 
As a result, Eq.~(\ref{J_final}) may be written as 
\begin{equation}\label{Jinf}
J = \delta \mu \frac{\partial F}{\partial \mu} + \delta T \frac{\partial F}{\partial T}
\end{equation}
where
\begin{equation}\label{F}
F =  \frac{4  g^2 \gamma}{(L_a+1)} \sum\limits_{k} \mathcal{I}(k) \;\bar{n}_k
\end{equation}
We therefore see that $F$ plays the role of a non-equilibrium free energy, from which the different contributions to $J$ may be obtained by differentiation. 

\begin{figure}
\centering
\includegraphics[width=0.23\textwidth]{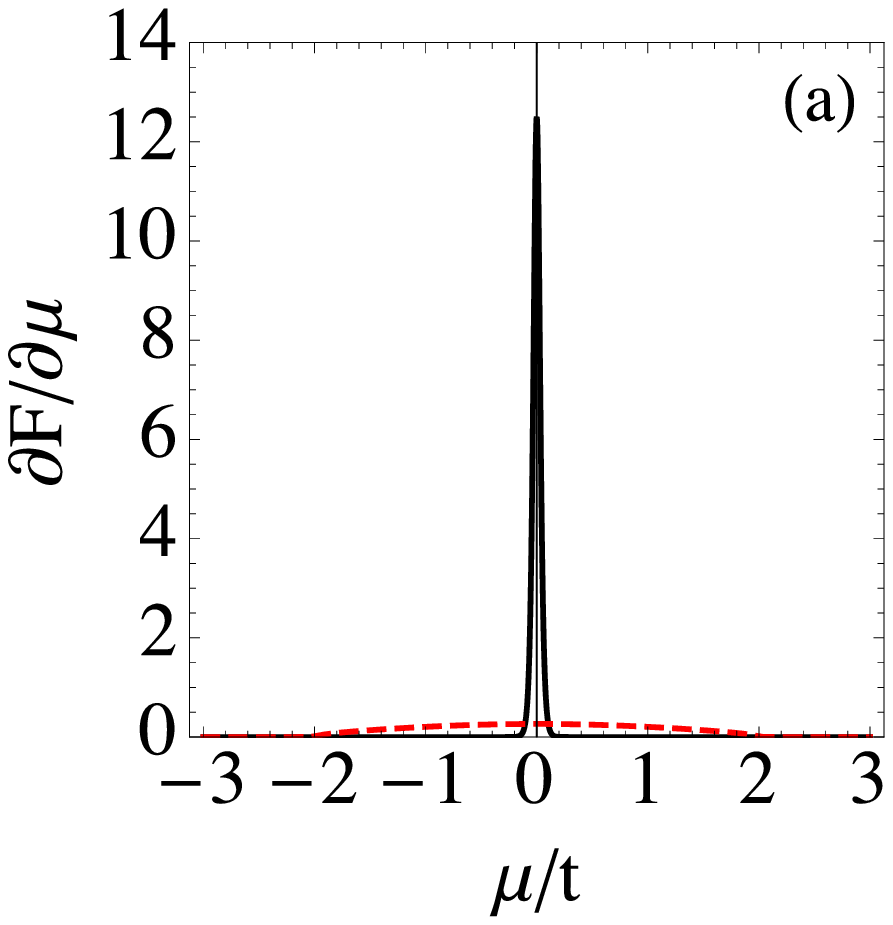}\quad
\includegraphics[width=0.23\textwidth]{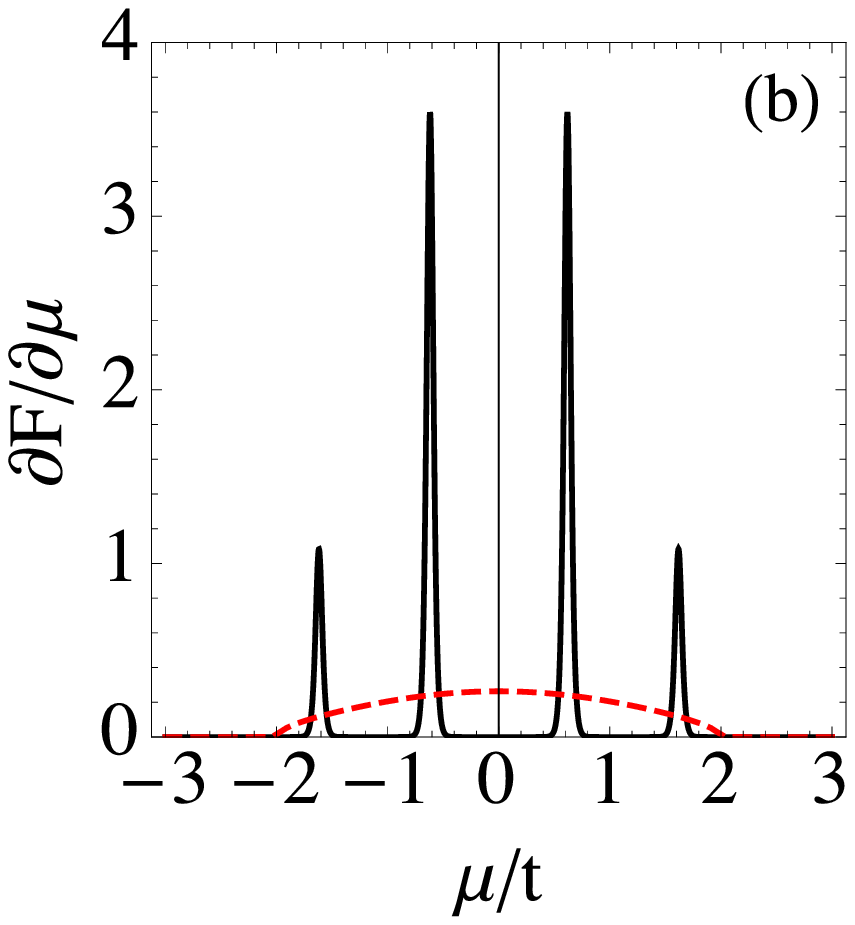}\\
\includegraphics[width=0.23\textwidth]{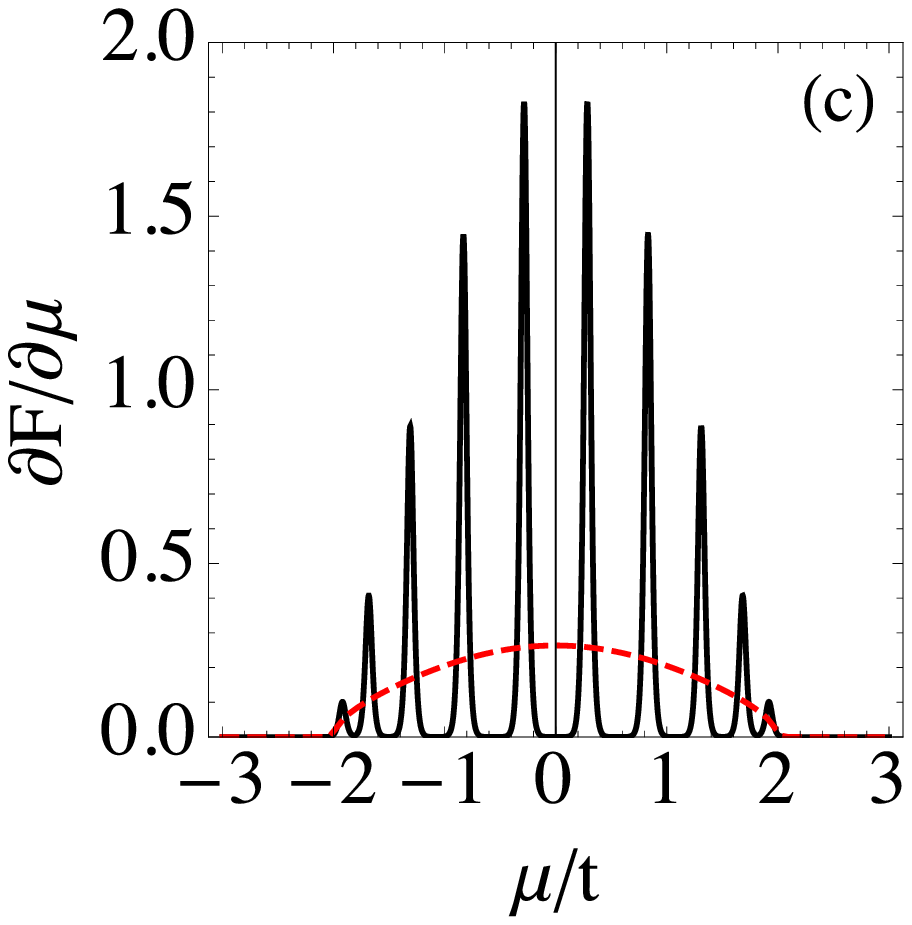}\quad
\includegraphics[width=0.23\textwidth]{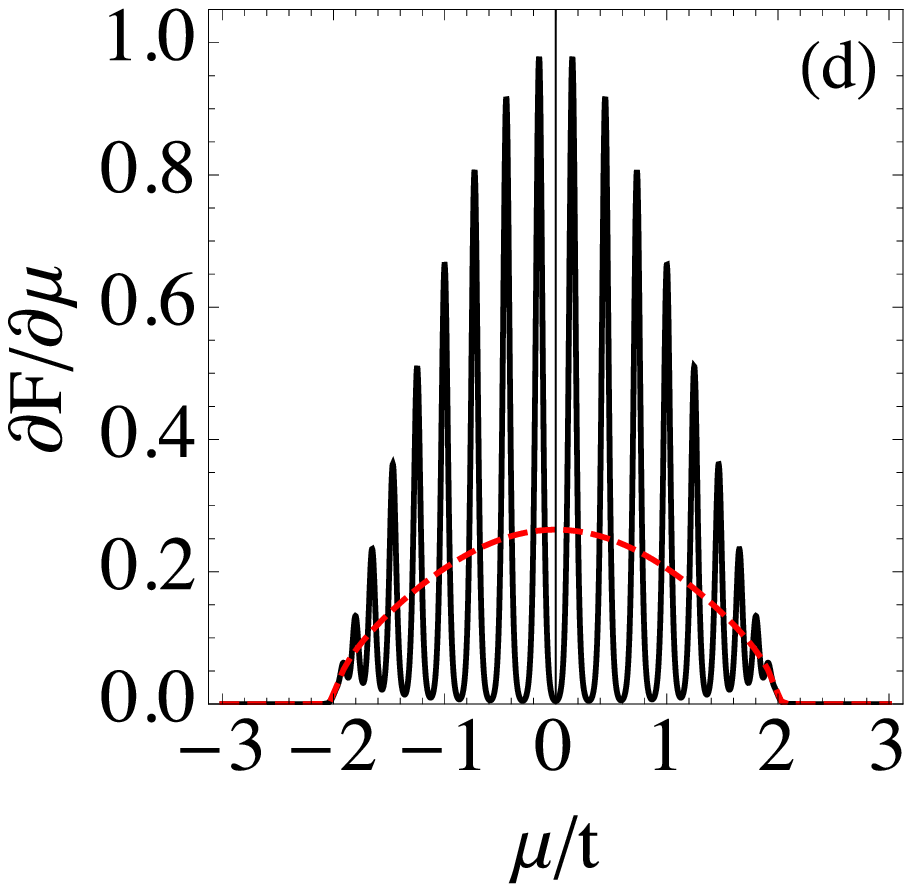}\\
\includegraphics[width=0.23\textwidth]{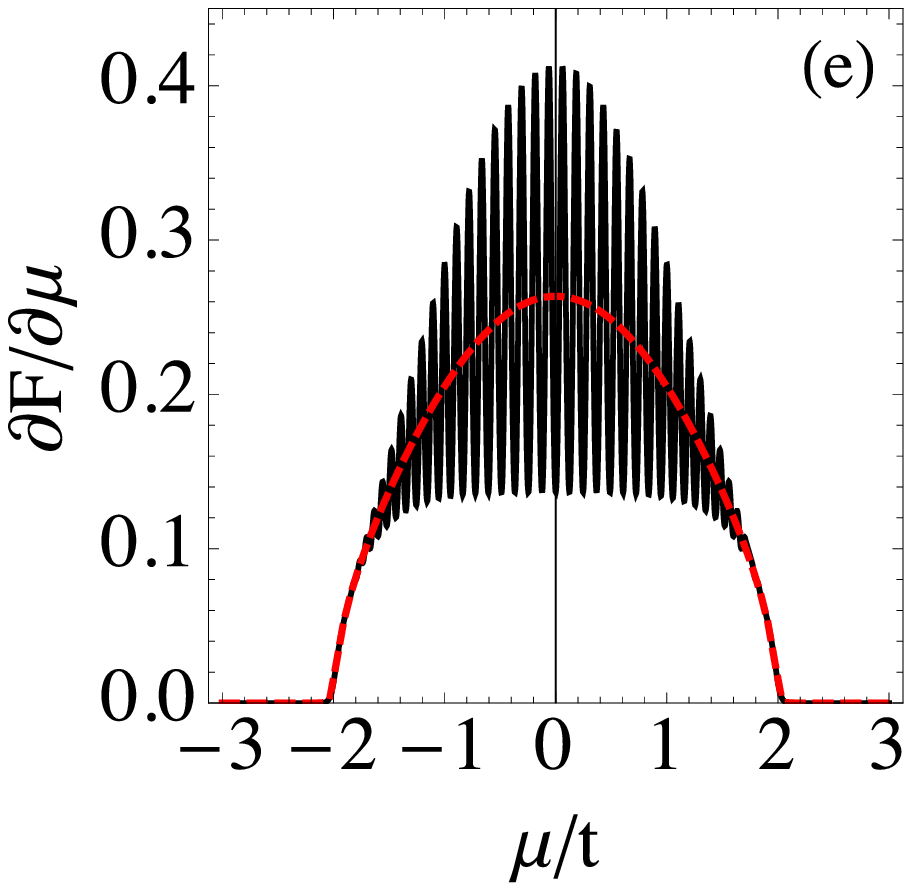}\quad
\includegraphics[width=0.23\textwidth]{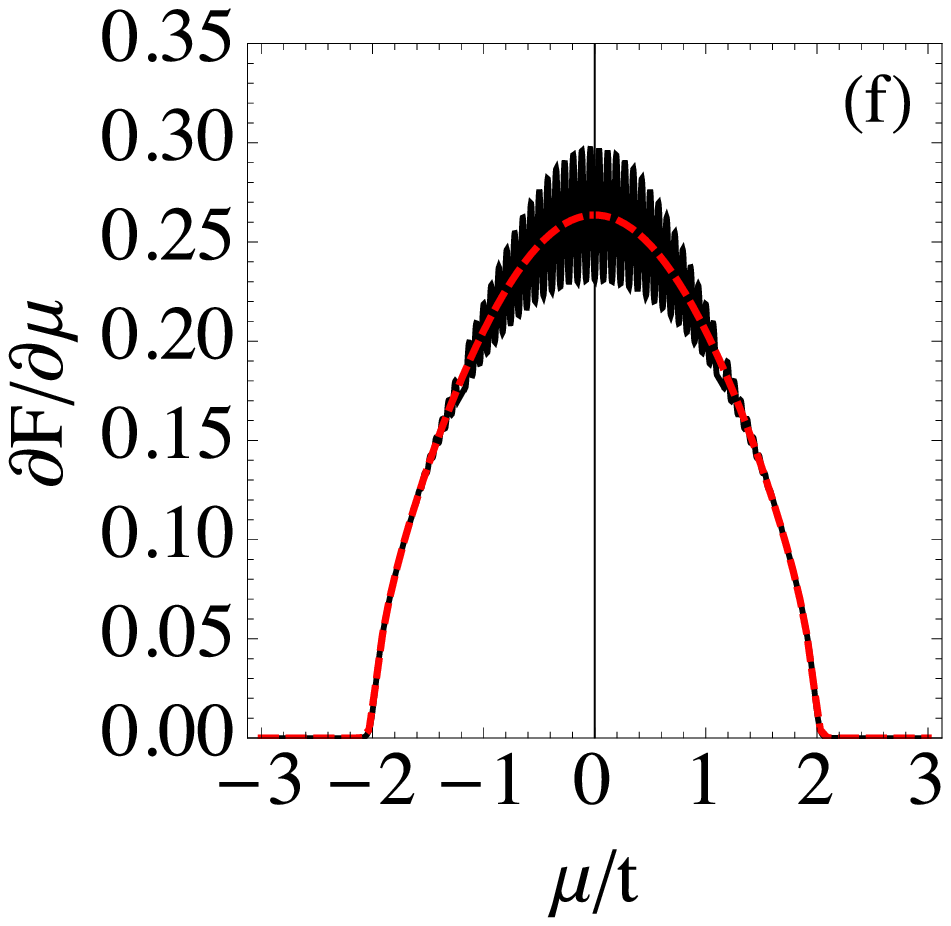}\\
\includegraphics[width=0.23\textwidth]{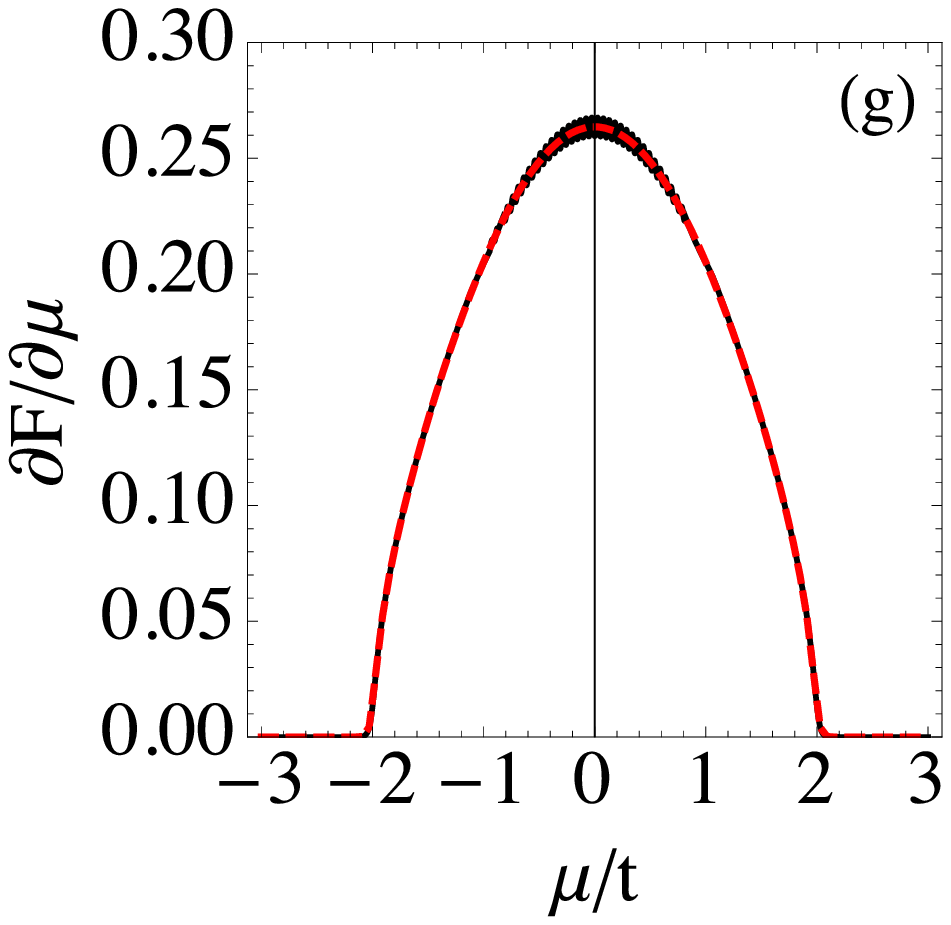}\quad
\includegraphics[width=0.23\textwidth]{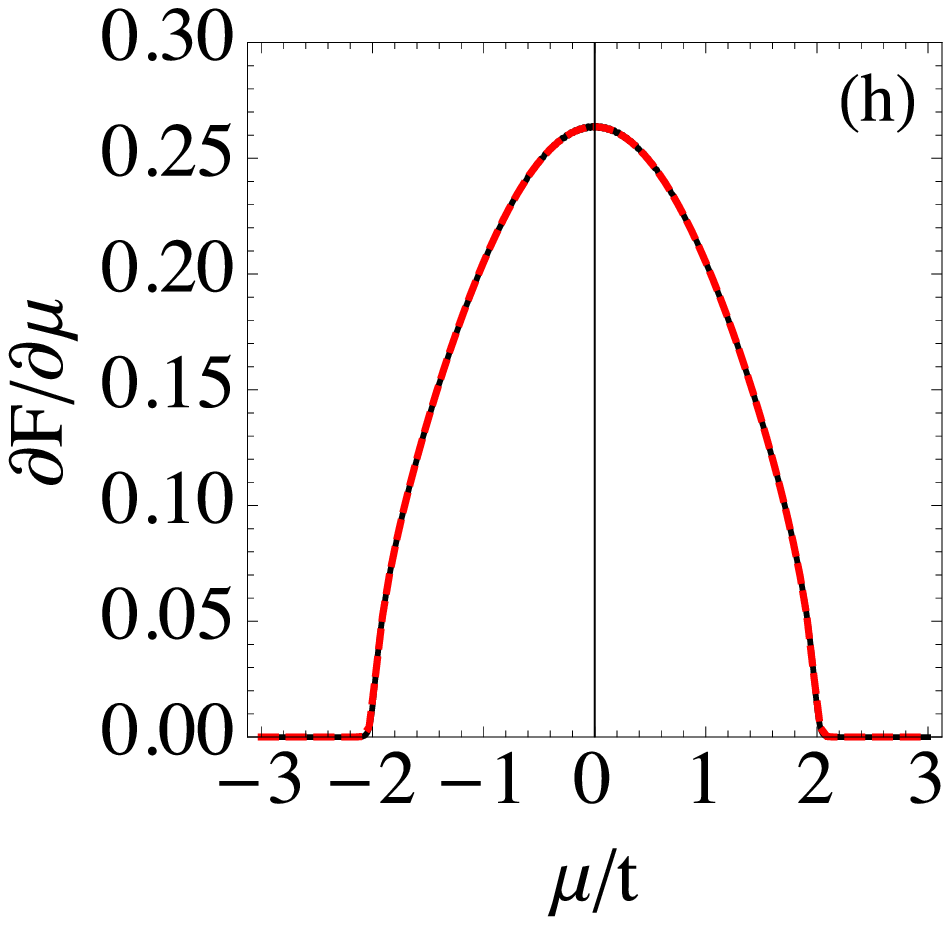}
\caption{\label{fig:J_mu_finite}(Color Online)
The particle current due to a gradient in the chemical potential,  $(\partial F/\partial \mu)$~vs.~$\mu$ plotted using Eq.~(\ref{F}) with $\gamma/t = 1$ and $T = 0.02 t$. 
Each curve correspond to a different value of $L_a = L_c$, respectively: (a) 1, (b) 4, (c) 10, (d) 20, (e) 50, (f) 80, (g) 120 and (h) 160. 
The red-dashed lines correspond to the thermodynamic limit, Eq.~(\ref{Jtl}). 
}
\end{figure}

\begin{figure}
\centering
\includegraphics[width=0.23\textwidth]{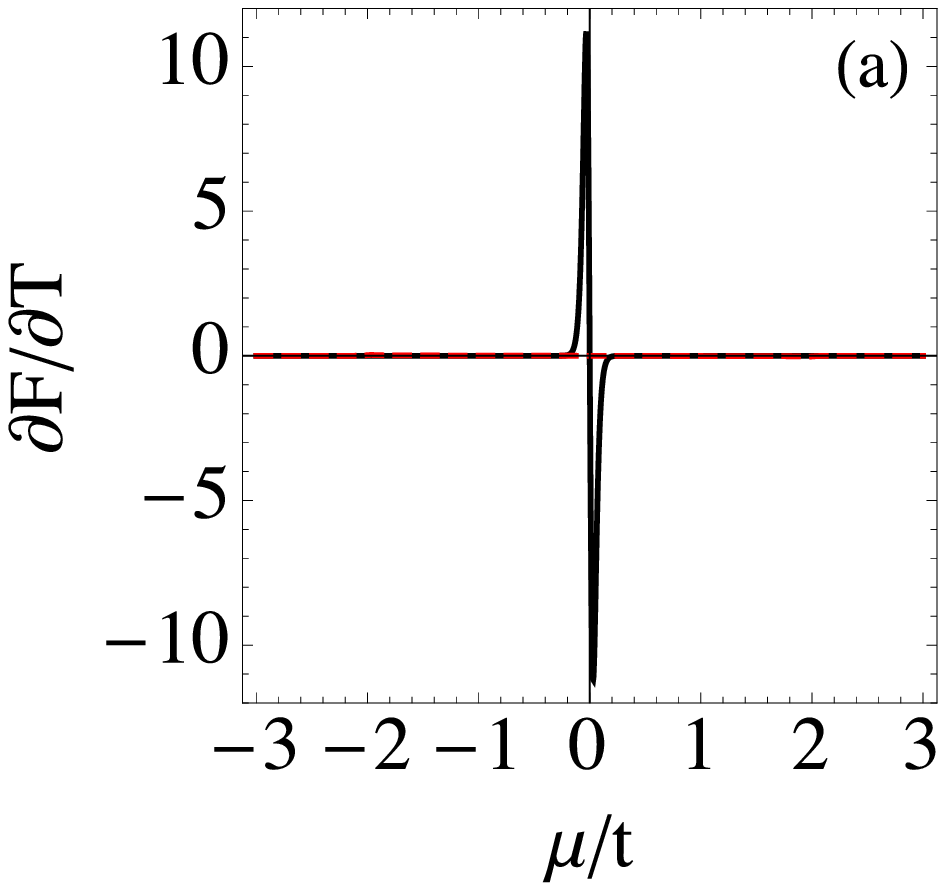}\quad
\includegraphics[width=0.23\textwidth]{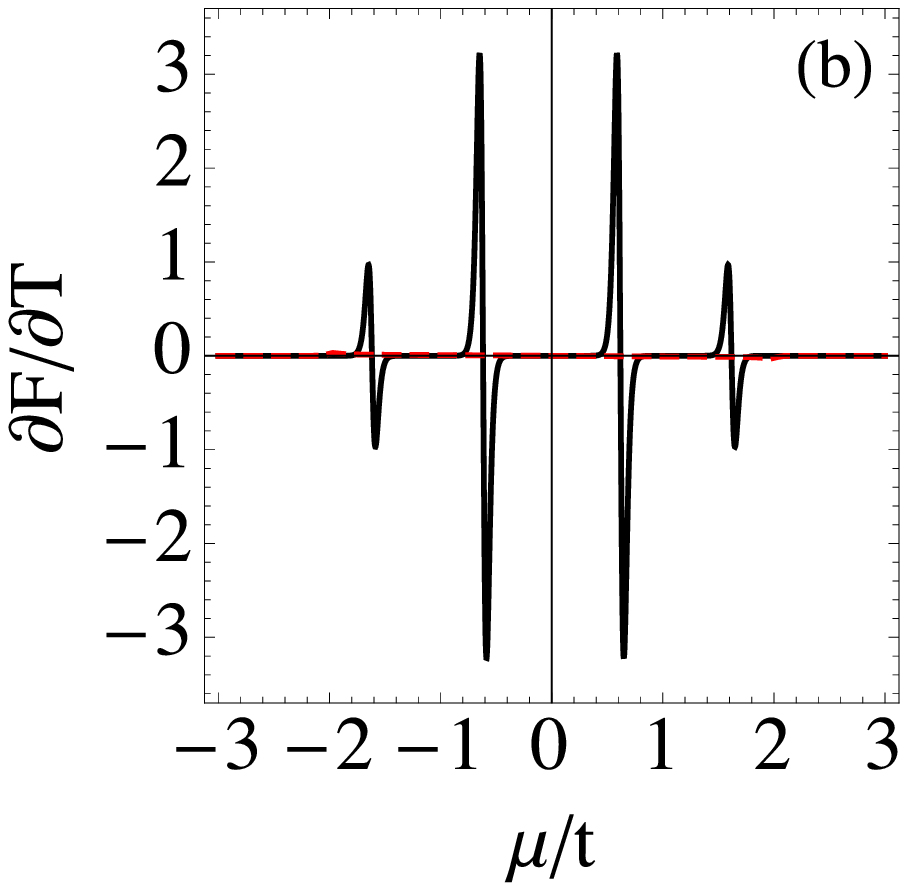}\\
\includegraphics[width=0.23\textwidth]{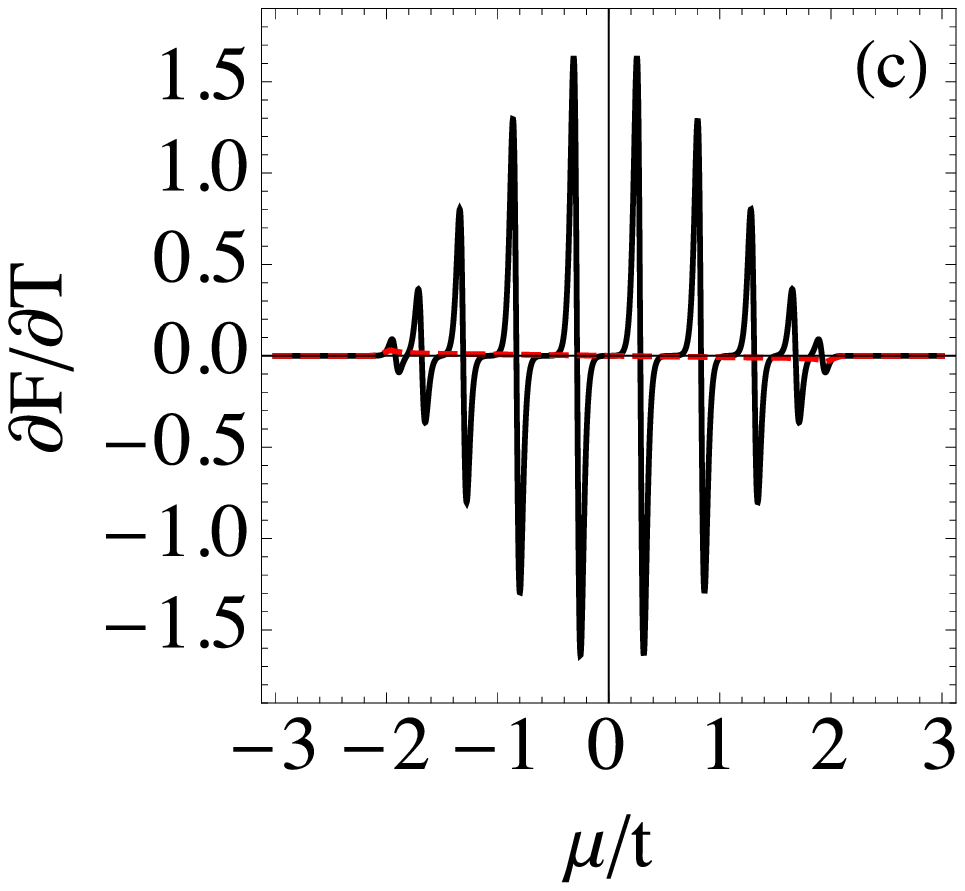}\quad
\includegraphics[width=0.23\textwidth]{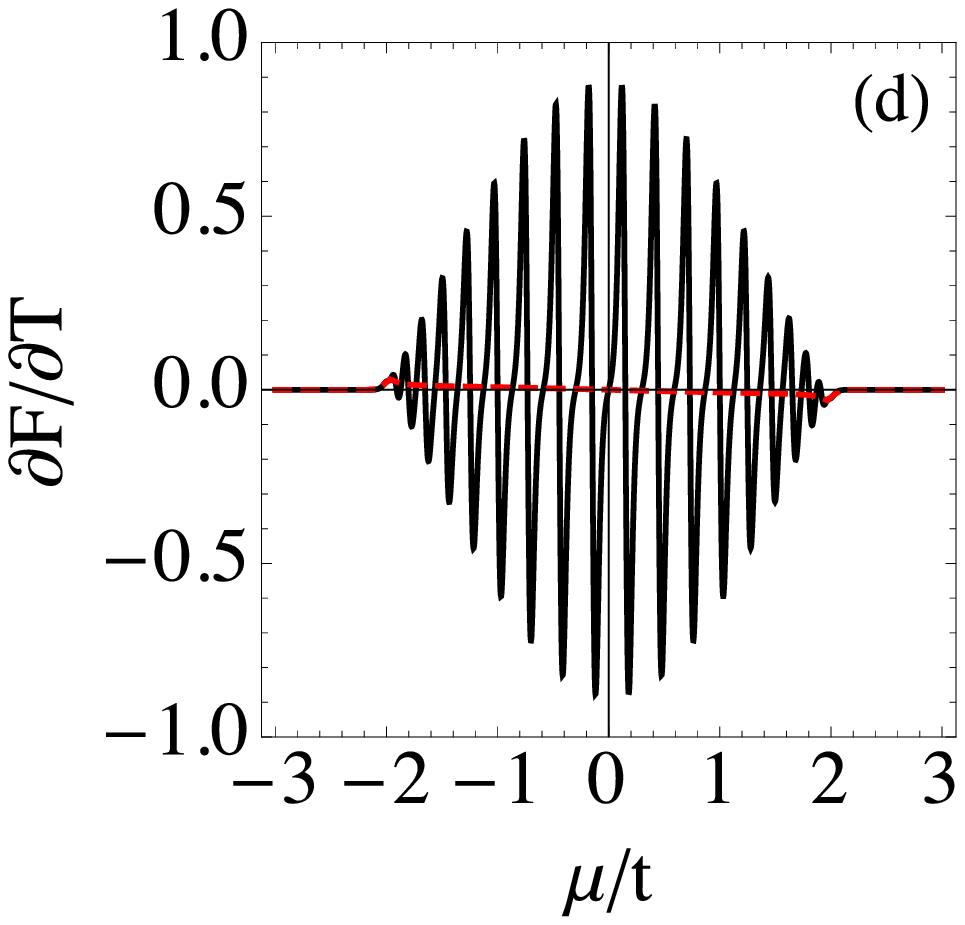}\\
\includegraphics[width=0.23\textwidth]{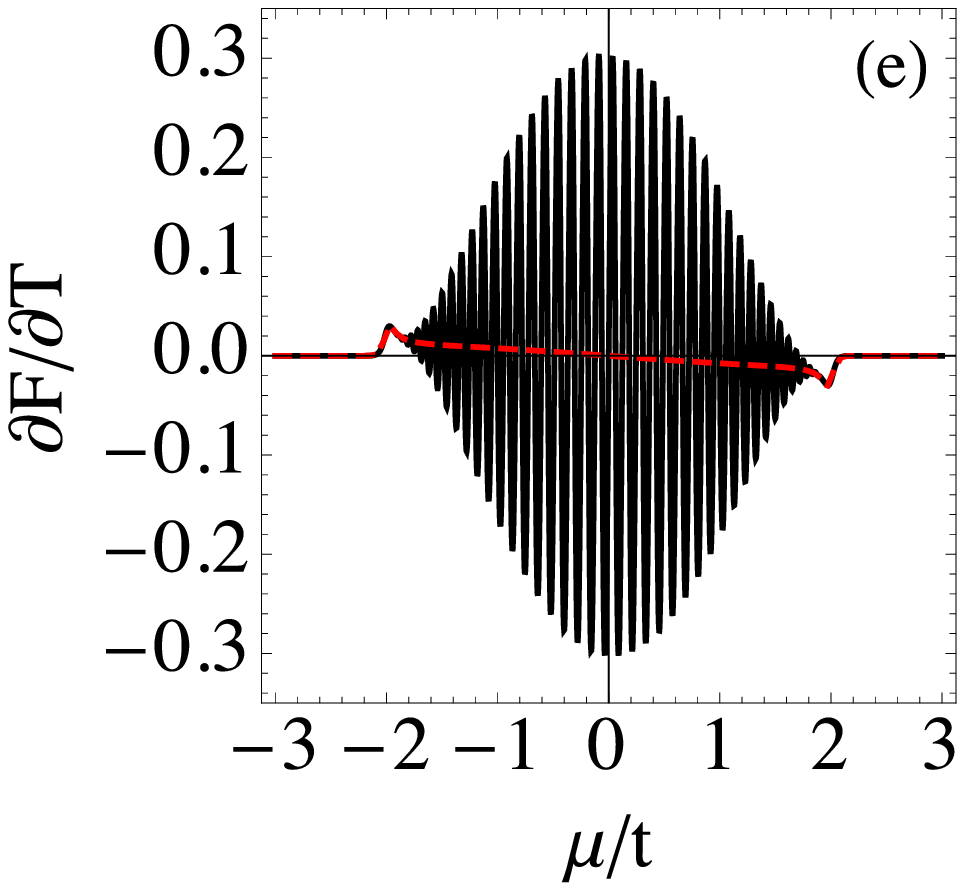}\quad
\includegraphics[width=0.23\textwidth]{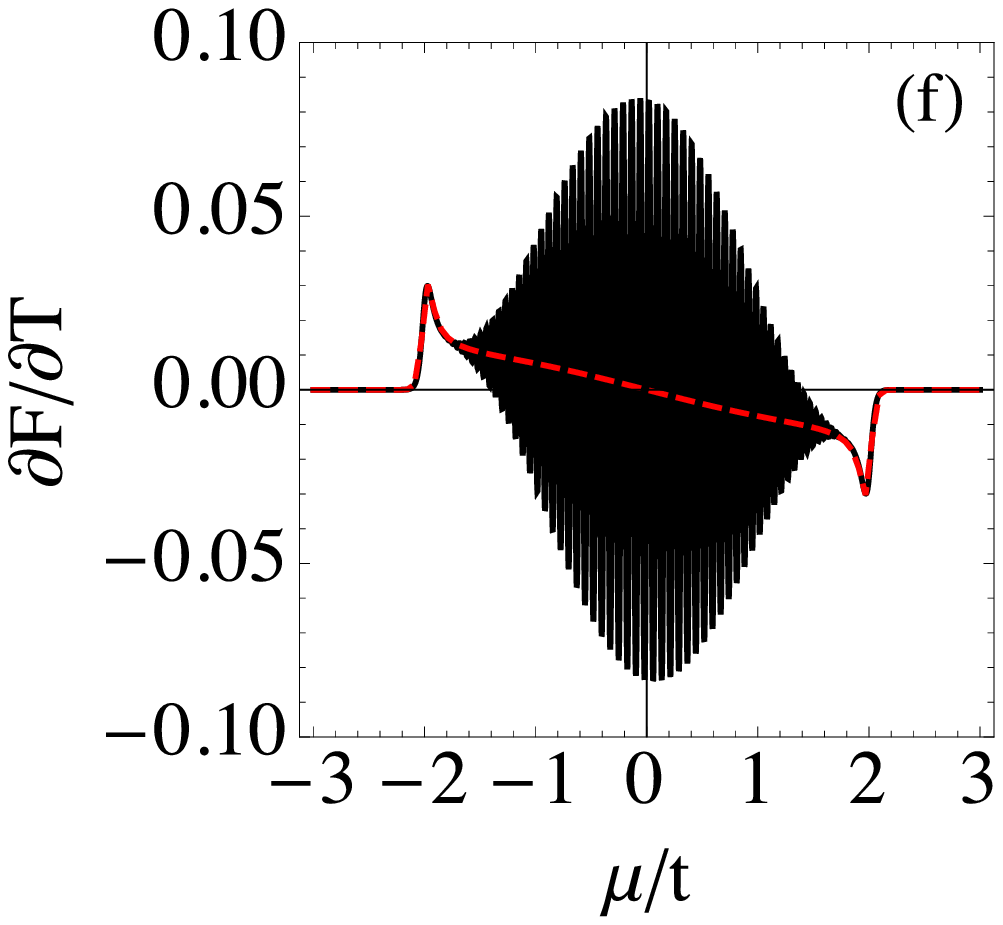}\\
\includegraphics[width=0.23\textwidth]{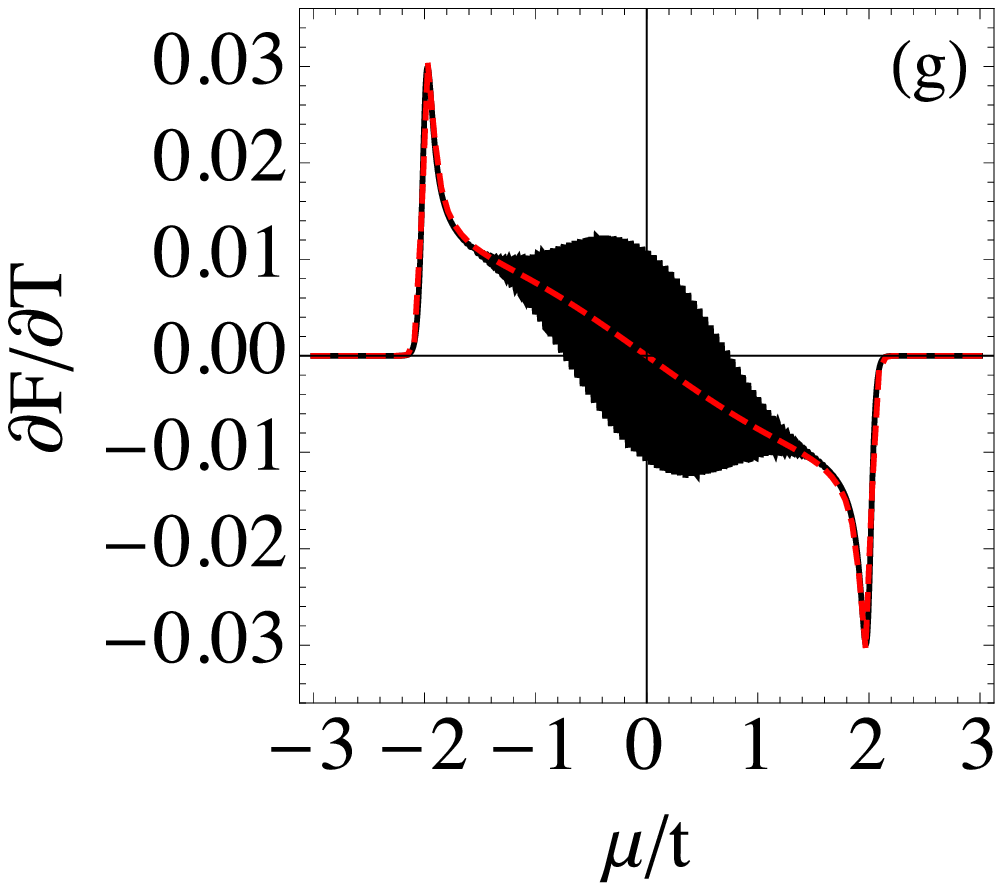}\quad
\includegraphics[width=0.23\textwidth]{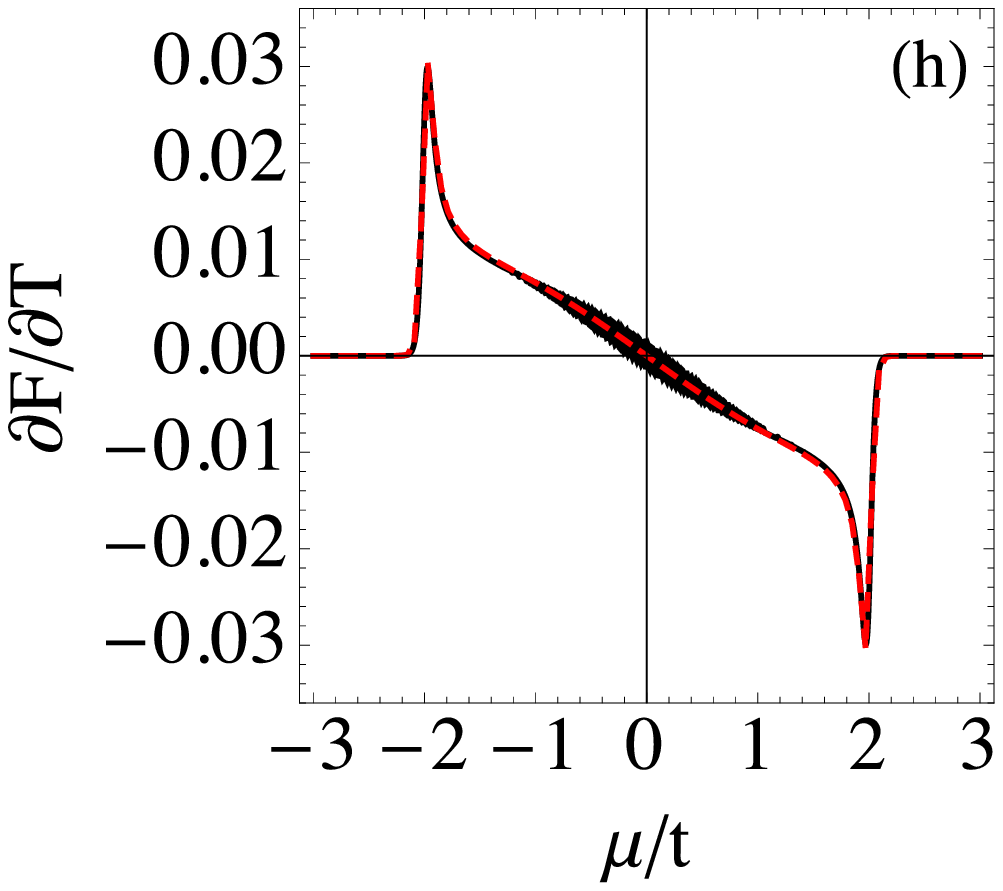}
\caption{\label{fig:J_T_finite}(Color Online)
The particle current due to a temperature gradient,  $(\partial F/\partial T)$~vs.~$\mu$ plotted using Eq.~(\ref{F}) with $\gamma/t = 1$ and $T = 0.02 t$. 
Each curve correspond to a different value of $L_a = L_c$, respectively: (a) 1, (b) 4, (c) 10, (d) 20, (e) 50, (f) 80, (g) 120 and (h) 160.
The red-dashed lines correspond to the thermodynamic limit, Eq.~(\ref{Jtl}). 
}
\end{figure}

Examples of the currents $\partial F/\partial \mu$ and $\partial F/\partial T$ are shown in Figs.~\ref{fig:J_mu_finite} and (\ref{fig:J_T_finite}) as a function of the chemical potential $\mu$, for $\gamma/t = 1$ and $T/t = 0.02$. 
The different images correspond to different sizes $L_a$ and the superimposed red-dashed curve corresponds to the thermodynamic limit [Eq.~(\ref{Jtl}) below]. 
The curves show the strong presence of finite size effects, which manifest themselves as sharp peaks  occurring when $\mu = - 2 t \cos k$ [recall the discrete structure of $k$ in Eq.~(\ref{S})].
As the size increases, these strong oscillations give place to a smooth curve, which gives a non-zero current only around the interval $\mu \in [-2t, 2t]$, corresponding to the bandwidth of $\epsilon_k$. 
It is also worth mentioning that these finite size oscillations are characteristic of low temperatures. 
If $T/t \sim 1$ they are replaced by smooth curves.

\subsection{Thermodynamic limit}

In the thermodynamic limit Eq.~(\ref{I_finite}) becomes 
\begin{equation}\label{q_int}
\mathcal{I}(k) = \frac{\sin^2 k}{\pi} \int\limits_0^\pi  \frac{\sin^2 q \; \ud q }{\gamma^2 + t^2 (\cos k - \cos q)^2}
\end{equation}
Similarly,  Eqs.~(\ref{J_final}) and (\ref{F})  are transformed to
\begin{equation}\label{Jtl}
J = \frac{4g^2 \gamma}{\pi} \int\limits_0^\pi\; \mathcal{I}(k) (\bar{n}_{a,k} - \bar{n}_{c,k})  \ud k
\end{equation}
and
\begin{equation}\label{Ftl}
F = \frac{4g^2 \gamma}{\pi} \int\limits_0^\pi  \; \mathcal{I}(k) \bar{n}_{k}\ud k
\end{equation}
These equations were used to plot the red curves in Figs.~\ref{fig:J_mu_finite} and \ref{fig:J_T_finite}.

\begin{figure}
\centering
\includegraphics[width=0.23\textwidth]{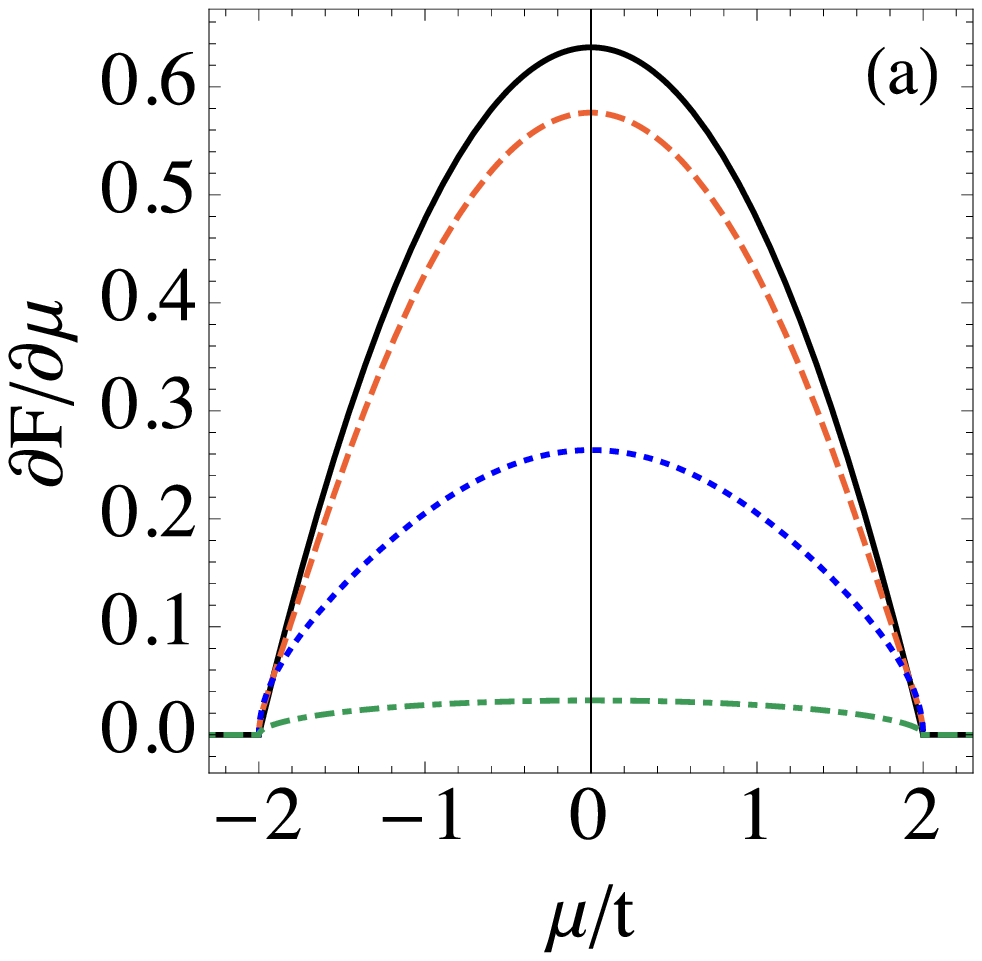}\quad
\includegraphics[width=0.23\textwidth]{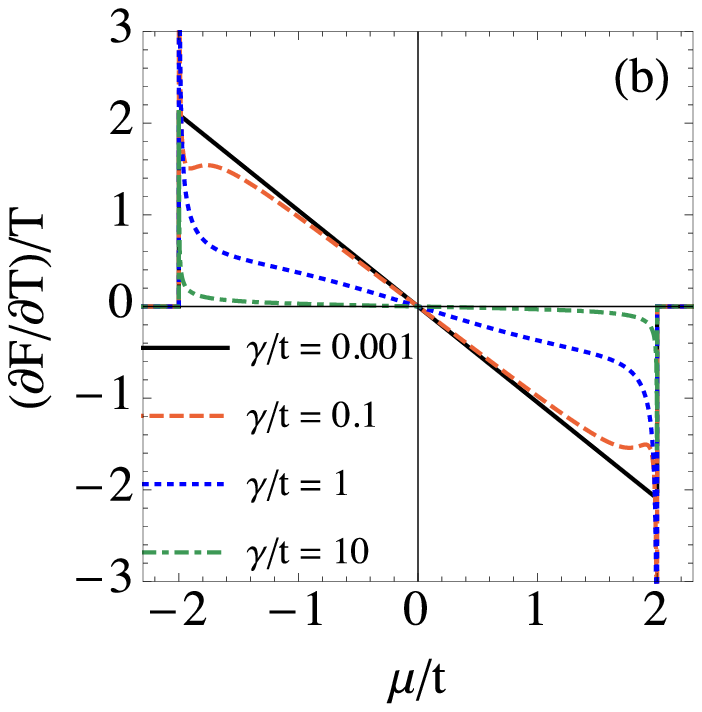}
\caption{\label{fig:J_gamma}
Influence of the coupling constant $\gamma$ in the particle current, at zero temperature and at the thermodynamic limit.
(a) $(\partial F/\partial \mu)$~vs.~$\mu$ and (b) $(\partial F/\partial T)$~vs.~$\mu$. 
The curves were computed using Eqs.~(\ref{Fmu_I}) and (\ref{FT_I}).
}
\end{figure}

The integral in Eq.~(\ref{q_int}) falls under the category of Eq.~(\ref{Mdef}). 
Hence the corresponding low and high $\gamma$ behaviors may be read off directly from Eqs.~(\ref{M_high}) and (\ref{M_low}):
\begin{equation}\label{I_limits}
\mathcal{I}(k) =  \begin{cases}
\displaystyle{ \frac{ |\sin k|^3}{\gamma t}}, & \text{ if } \gamma \ll t \\[0.4cm]
\displaystyle{\frac{ \sin^2k}{2\gamma^2}}, & \text{ if } \gamma \gg t
\end{cases}
\end{equation}
For completeness, we also mention that for intermediate values of $\gamma$, this integral may be computed analytically and reads
\begin{equation}\label{int}
\mathcal{I}(k) = \frac{\sin^2k}{t^2} \bigg\{\frac{[\lambda_1 + (\lambda_1^2 + \lambda_2^2)^{1/2}]^{1/2}}{\gamma\sqrt{2}} - 1\bigg\}
\end{equation}
where $\lambda_1 = \gamma^2 + t^2 \sin^2k$ and $\lambda_2 = 2 \gamma t \cos k$. 
The dependence of $\gamma$ on the particle current is investigated in Fig.~\ref{fig:J_gamma} for both a unbalance in the chemical potential and a unbalance in the temperature. 
We will now discuss how to obtain the analytical forms of these functions in the case that $T \to 0$.


At zero temperature we may again use that  $\bar{n}_{k} = \Theta(\mu - \epsilon_k)$, which implies that  $\partial \bar{n}_k/\partial \mu = \delta(\mu - \epsilon_k)$. Recalling also the definition of the Fermi momentum as  $k_F =  \arccos(-\mu/2t)$, we find for Eq.~(\ref{Ftl}) the following simple result:
\begin{equation}\label{Fmu_I}
\frac{\partial F}{\partial \mu} = \frac{4 g^2 \gamma}{\pi} \frac{\mathcal{I}(k_F(\mu))}{2 t \sin k_F(\mu)}
\end{equation}
where the factor in the denominator comes from transforming $\delta(\mu-\epsilon_k)$ into $\delta(k-k_F)$. 
Using the approximate results in Eq.~(\ref{I_limits}) we then obtain the explicit forms, valid for $\mu \in [-2t, 2t]$:
\begin{equation}\label{Fmu}
\frac{\partial F}{\partial \mu} = \begin{cases}
\displaystyle{\frac{g^2}{2 \pi t^4}(4 t^2 - \mu^2)} , 	&	  \gamma/t \ll 1 \\[0.4cm]
\displaystyle{\frac{g^2}{2 \pi t^2 \gamma} \sqrt{4 t^2 - \mu^2}}	& 	\gamma/t \gg 1
\end{cases}
\end{equation}
which match well the black ($\gamma/t = 0.001$) and green ($\gamma/t = 10$) curves  plotted in Fig.~\ref{fig:J_gamma}(a).

Similarly, we may analyze the behavior of $\partial F/\partial T$ as $T\to 0$. 
Of course, if $T = 0$  there can be no temperature unbalance, so we must look for the lowest contribution in $T$. 
To do that we perform a Sommerfeld expansion \cite{Ashcroft1976,Sommerfeld1927} by writing the integral in Eq.~(\ref{Ftl}) as 
\begin{IEEEeqnarray*}{rCl}
\int\limits_0^k \mathcal{I}(k) \bar{n}_k \ud k &=& \int\limits_{-2t}^{2t} \frac{\mathcal{I}(k(\epsilon))}{\ud \epsilon/\ud k} \frac{\ud \epsilon}{e^{(\epsilon - \mu)/T} + 1} \\[0.2cm]
&=& \mathcal{O}(T^0)  + \frac{\pi^2}{6} T^2 \frac{\partial}{\partial \epsilon} \bigg[ \frac{\mathcal{I}(k(\epsilon))}{\ud \epsilon/\ud k}\bigg]_{\epsilon = \mu}
\end{IEEEeqnarray*}
where the first term is independent of temperature. 
Consequently, we find that $\partial F/\partial T$ may be written as 
\begin{equation}\label{FT_I}
\frac{\partial F}{\partial T} = \frac{4 \pi T g^2 \gamma}{3} \frac{\partial}{\partial \mu} \bigg[ \frac{\mathcal{I}(k_F(\mu))}{2t \sin k_F(\mu)}\bigg]
\end{equation}
Using Eq.~(\ref{I_limits}) for the low and high $\gamma$ behavior of $\mathcal{I}$, we finally conclude that 
\begin{equation}\label{FT}
\frac{\partial F}{\partial T} = \begin{cases}
\displaystyle{-\frac{\pi g^2 T \mu}{3 t^4}} , 	&	  \gamma/t \ll 1 \\[0.4cm]
\displaystyle{-\frac{\pi g^2 T \mu}{6 t^2 \gamma \sqrt{4 t^2 - \mu^2}}}	& 	\gamma/t \gg 1
\end{cases}
\end{equation}
which, again, hold only for $\mu \in [-2t, 2t]$. 
These two formulas match precisely the black ($\gamma/t = 0.001$) and green ($\gamma/t = 10$) curves  plotted in Fig.~\ref{fig:J_gamma}(b).
It shows that when $\gamma/t \ll 1$ the current due to a temperature gradient is linear in $\mu$, but when $\gamma/t \gg 1$, it acquires sharp peaks near the band edges. 

%
%

\subsection{Current when  $L_b  \neq 0$}

We now turn to the particle current when the size of chain B is non-zero. 
The definition of $J$  in this case is given in Eq.~(\ref{J}), with the relevant matrix elements given in Eq.~(\ref{theta1_ab}).
The current then comes
\begin{equation}\label{JLb_final}
J = \frac{8 g^2 \gamma}{(L_a+1) (L_b+1)}\sum\limits_{k,q}\; \frac{\sin^2 k \sin^2 q (\bar{n}_{a,k} - \bar{n}_{b,q})}{\gamma^2 + 4t^2 (\cos k - \cos q)^2}
\end{equation}
where, recall, the allowed values of $k$ and $q$ are different since $L_a$ and $L_b$ are arbitrary.
This equation also depends on the occupation numbers $\bar{n}_{b,q}$, which are given in Eq.~(\ref{nbq}). 

When $L_a = L_c = 1$ Eq.~(\ref{JLb_final}) reduces to 
\begin{equation}\label{JLa1_a}
J = \frac{4 g^2 \gamma}{(L_b+1)} (\bar{n}_a - \bar{n}_c) \sum\limits_{q}\; \frac{\sin^2 q}{\gamma^2 + 4t^2 \cos^2 q}
\end{equation}
which is simply a constant coefficient multiplied by the occupation difference $(\bar{n}_a - \bar{n}_c)$. 
A homogeneous ($g = t$) XX spin chain under a single spin bath was studied in Ref.~\cite{Karevski2009}, which found for the current the exact formula 
\begin{equation}\label{JLa1_b}
J = \frac{ \gamma}{t^2 + \gamma^2} \frac{\bar{n}_a - \bar{n}_c}{2} 
\end{equation}
The difference between this result and Eq.~(\ref{JLa1_a}) is due to the fact that we assumed a non-homogeneous chain ($g \neq t$). 
If we take $g = t$ \emph{and} if we continue the series expansion of $\theta$ up to higher orders, we recover exactly Eq.~(\ref{JLa1_b}), 
as easily verified  from  numerical simulations.

\begin{figure}
\centering
\includegraphics[width=0.23\textwidth]{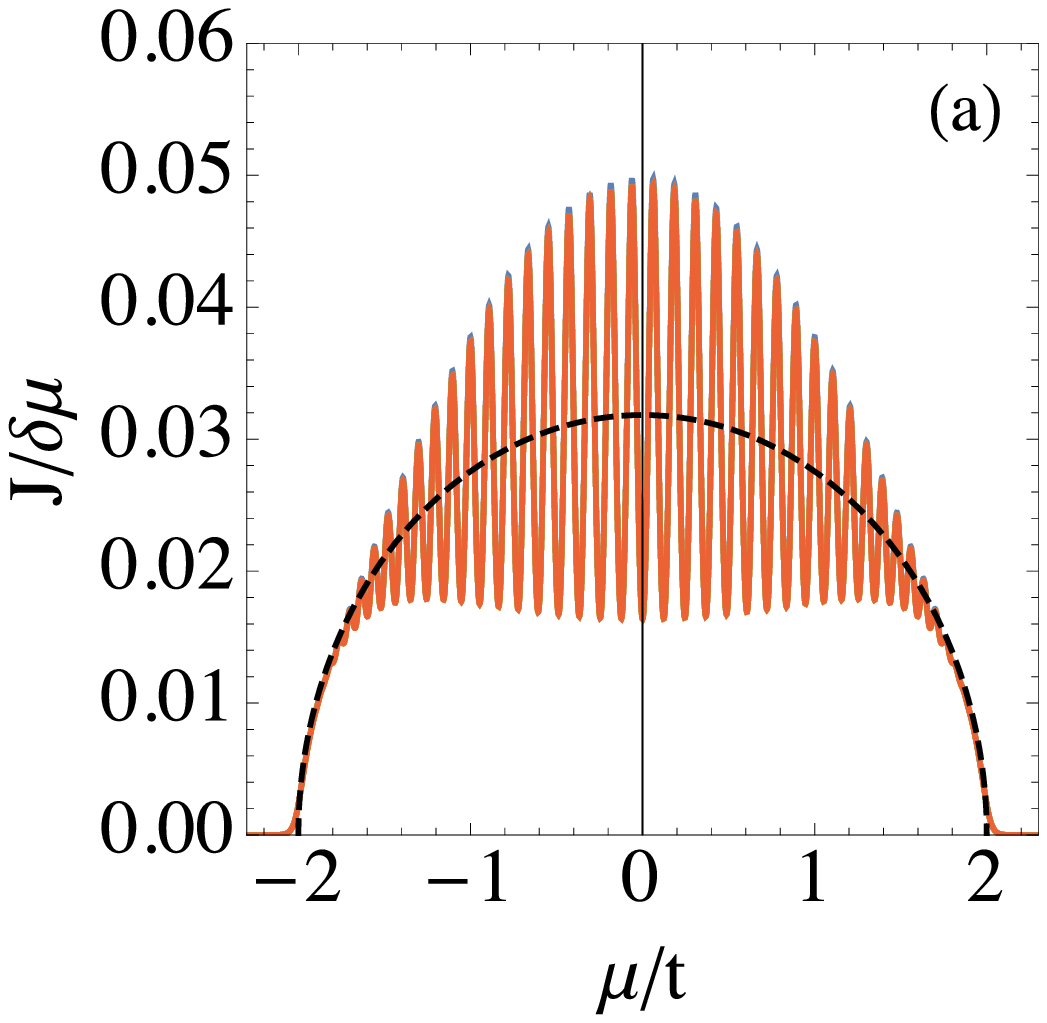}\quad
\includegraphics[width=0.23\textwidth]{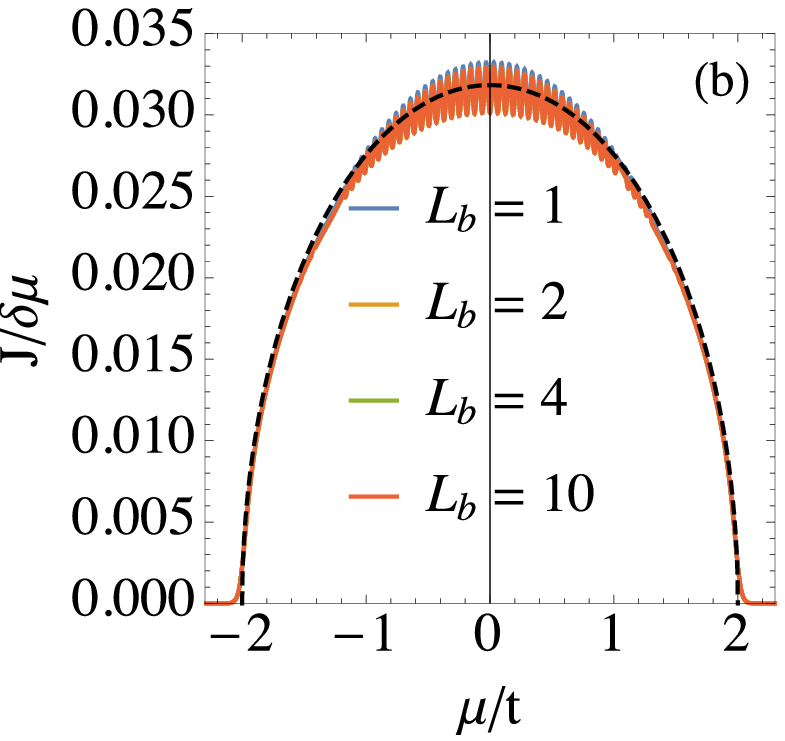}\\
\includegraphics[width=0.23\textwidth]{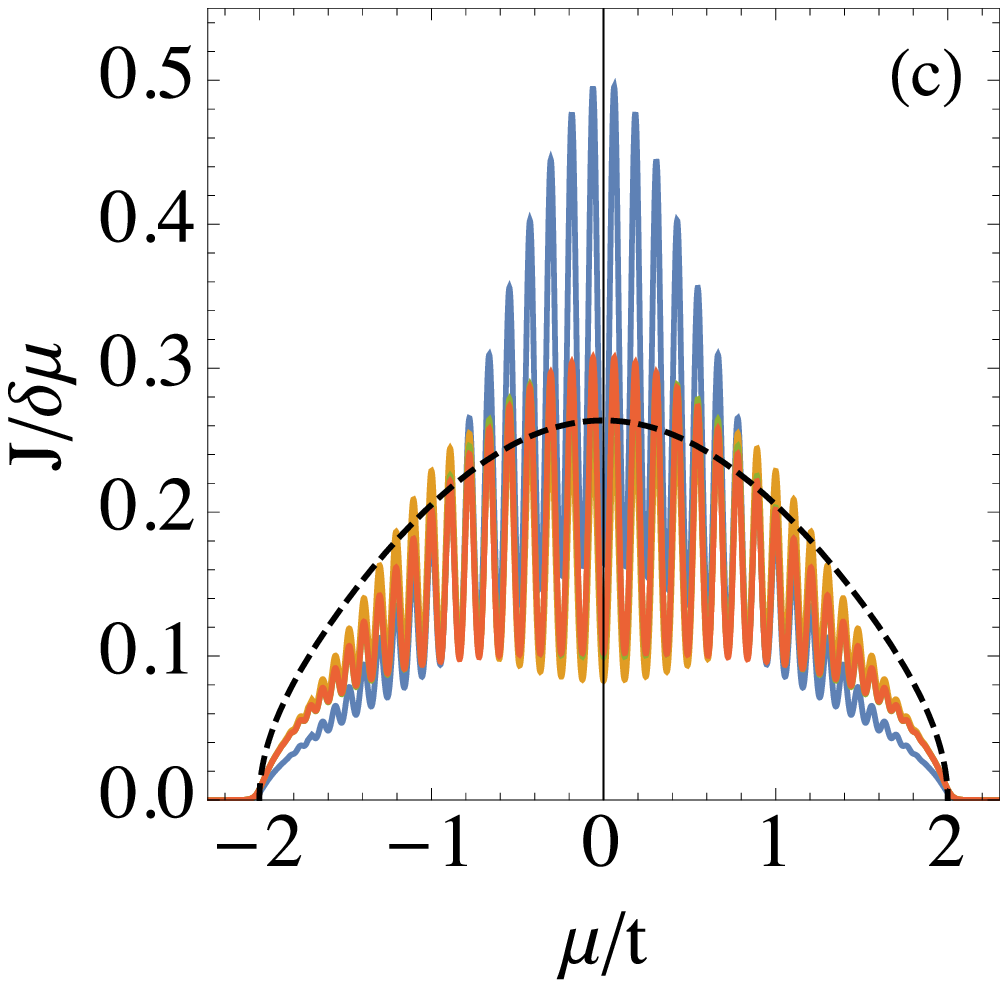}\quad
\includegraphics[width=0.23\textwidth]{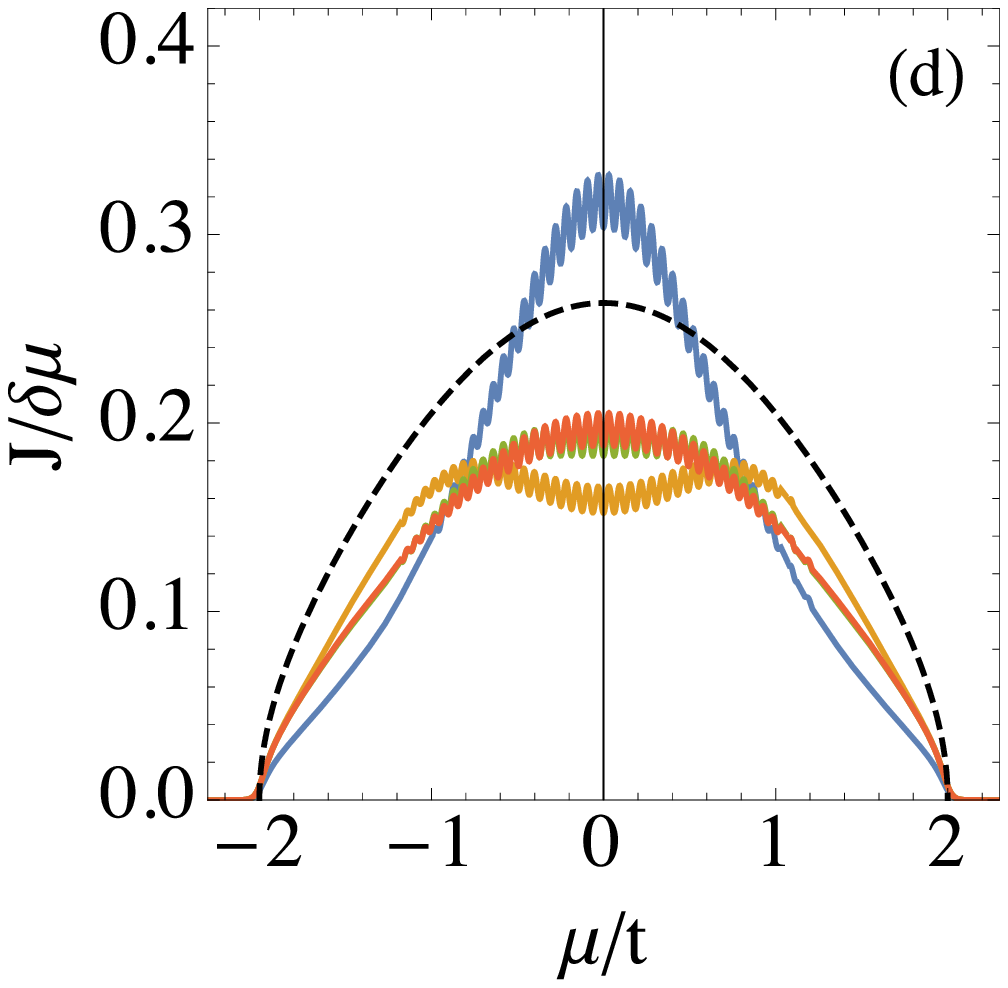}
\caption{\label{fig:JLb_gam}
Particle current $J/\delta\mu$~vs.~$\mu$ for different sizes of the middle chain (see legend in image (b)), 
computed using Eq.~(\ref{JLb_final}) with $T = 0.02 t$ for different 
combinations of  $L_a$ and $\gamma/t$:  (a) $L_a=50$, $\gamma/t = 10$, (b) $L_a = 100$, $\gamma/t = 10$, (c) $L_a = 50$, $\gamma/t = 1$ and (d) $L_a = 100$, $\gamma/t = 1$. 
The dashed black curve corresponds to Eq.~(\ref{Fmu_I}). 
}
\end{figure}

Examples of Eq.~(\ref{JLb_final}) are shown in Fig.~\ref{fig:JLb_gam} for $\gamma/t = 10$ and $\gamma/t = 1$, with different choices of $L_a$ and $L_b$. 
When $\gamma/t \gg 1$, as shown in Eq.~(\ref{nbq_high}), the occupation numbers $\bar{n}_{b,q}$ become independent of $q$. 
Consequently, in this case the current $J$ in Eq.~(\ref{JLb_final}) becomes independent of the size $L_b$ of chain B.  
This is visible in Figs.~\ref{fig:JLb_gam}(a) and (b), which correspond to $\gamma/t = 10$, where we see that the curves for different values of $L_b$ practically coincide. Moreover, we also see that these curves mimic the behavior of the current when $L_b = 0$, represented here by the dashed black curves plotted from Eq.~(\ref{Fmu_I}).
Thus, we conclude that when $\gamma/t \gg 1$, the presence of chain B does not affect in any way the current through the system. 
When $\gamma/t = 1$ [Figs.~\ref{fig:JLb_gam}(c) and (d)], on the other hand, a dependence  in $L_b$ becomes visible. 
However, even though the changes are substantial when moving from $L_b = 1$ to $L_b = 2$, the curves for $L_b = 4$ and $L_b = 10$ already practically coincide. 
Notwithstanding, none of the curves coincide with that from $L_b = 0$, thus showing that when $\gamma/t = 1$, the presence of chain B does have an effect on the properties of the current.

The behavior of Eq.~(\ref{JLb_final}) when $\gamma/t \ll 1$, on the other hand, is much more intricate since it will depend sensibly on the sizes $L_a$ and $L_b$. 
The reason for this is that the flux will have substantial contributions whenever $(\cos k - \cos q) \sim 0$. 
But  $k$ and $q$ take on a mesh of discrete values, as denoted in Eq.~(\ref{S}) (with $L_a$ and $L_b$ respectively). 
Consequently, the behavior of $J$ will change substantially for different combinations of $L_a$ and $L_b$. 

\begin{figure}
\centering
\includegraphics[width=0.47\textwidth]{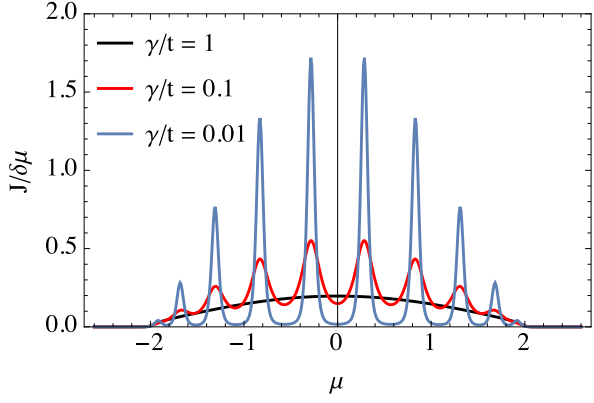}
\caption{\label{fig:JLb_gam_small}$J/\delta\mu$~vs.~$\mu$ for different values of $\gamma/t$, with $L_b = 10$ and $T/t = 0.02$. 
When $\gamma/t \ll 1$, the curve approaches Fig.~\ref{fig:J_mu_finite}(c). 
}
\end{figure}

Instead, let us suppose for simplicity that chains A and C tend to the thermodynamic limit, whereas the size of chain B remains arbitrary. 
In this case we may convert the sum over $k$ in Eq.~(\ref{JLb_final}) to an integral, to find
\begin{equation}\label{JLb_final_tl}
J = \frac{8 g^2 \gamma}{\pi (L_b+1)}\sum\limits_{q}\int\limits_0^\pi \ud k \; \frac{\sin^2 k \sin^2 q (\bar{n}_{a,k} - \bar{n}_{b,q})}{\gamma^2 + 4t^2 (\cos k - \cos q)^2}
\end{equation}
Next we may use Eq.~(\ref{M_low}) to approximate the result for $\gamma/t \ll 1$. 
Using also Eq.~(\ref{nbq_low}) we then find that 
\begin{equation}
J \simeq \frac{2 g^2\gamma}{L_b + 1} \sum\limits_q \sin^3 q (\bar{n}_{a,q} - \bar{n}_{c,q})
\end{equation}
Comparing this with Eq.~(\ref{J_final}), and noticing also Eq.~(\ref{I_limits}), we conclude that when $\gamma/t \ll 1$ the flux through chain B will behave somewhat like the flux  for $L_b = 0$ studied in the previous subsection. 
However, it will be governed by  $L_b$, instead of $L_a$. 
This idea is illustrated in Fig.~\ref{fig:JLb_gam_small}, where we plot Eq.~(\ref{JLb_final_tl}) for $L_b = 10$ and several values of $\gamma/t$. 
As can be seen, when $\gamma/t$ decreases the current approaches the behavior of Fig.~\ref{fig:J_mu_finite}(c), which was computed with $L_b = 0$, and $L_a = L_c = 10$. 

%
%
%
%

\section{Heat current and Onsager Coefficients}

We now discuss the energy and heat currents through the system, and also compute the Onsager coefficients for this problem. 
To find a formula for the energy current we may repeat the procedure that led us to Eq.~(\ref{J}), but with the Hamiltonian $H$ instead of $\mathcal{N}$.
Starting with Eq.~(\ref{master2}) we find the following equation for the time-evolution of $\langle H\rangle$: 
\begin{equation}
\frac{\ud \langle H \rangle}{\ud t}  = \tr\bigg[H D_a(\rho)\bigg] +  \tr\bigg[H D_c(\rho)\bigg]
\end{equation}
Thus, the flux of energy in the steady-state will be 
\begin{equation}\label{JE1}
J_E = \tr\bigg[H D_a(\rho)\bigg]  = \tr\bigg[H_a D_a(\rho)\bigg]  + \tr\bigg[V_{ab} D_a(\rho)\bigg] 
\end{equation}
where
\begin{IEEEeqnarray*}{rCl}
\label{fluxAA} \tr\bigg[H_a D_a(\rho)\bigg]   &=&2\gamma \sum\limits_k  \epsilon_{a,k} (\bar{n}_{a,k} - \langle a_k^\dagger a_k \rangle)
\\[0.2cm]
\label{fluxVA} \tr\bigg[V_{ab} D_a(\rho)\bigg] &=& g\gamma \sum\limits_{k,q}  S_{L_a,k}^a S_{1,q}^b \bigg[\langle a_k^\dagger b_q \rangle + \langle b_q^\dagger a_k \rangle\bigg]
\end{IEEEeqnarray*}
The first equation requires knowledge of the second-order expansion of Eq.~(\ref{theta_series}). 
A formula based on the first order solution  may be found by looking at the time evolution of $\langle H_a \rangle$,  again obtained from Eq.~(\ref{master2}):
\begin{equation}
\frac{\ud \langle H_a \rangle}{\ud t}  =  i \langle [V_{ab}, H_a] \rangle + \tr\bigg[H_a D_a(\rho)\bigg] 
\end{equation}
Thus, we see that 
\begin{IEEEeqnarray*}{rCl}
\tr\bigg[H_a D_a(\rho)\bigg]  &=& - i \langle [V_{ab},H_a] \rangle \\[0.2cm]
&=& - i g \sum\limits_{k,q} \epsilon_{a,k} S_{L_a,k}^a S_{1,q}^b  \langle a_k^\dagger b_q - b_q^\dagger a_k\rangle
\end{IEEEeqnarray*}
Combining the results we conclude that the energy flux in Eq.~(\ref{JE1}) may therefore be written as 
\begin{IEEEeqnarray}{rCl}\label{JE}
J_E =  g \sum\limits_{k,q} &S_{L_a,k}^a S_{1,q}^b& \bigg[  \langle a_k^\dagger b_q \rangle  \left(\gamma - i \epsilon_{a,k}\right) 
+\langle  b_q^\dagger a_k\rangle \left(\gamma + i \epsilon_{a,k}\right)\bigg]	\nonumber
\end{IEEEeqnarray}
If $L_b = 0$, we should write instead
\begin{equation}\label{JE_AC}
J_E =  g \sum\limits_{k,q} S_{L_a,k}^a S_{1,q}^c \bigg[  \langle a_k^\dagger c_q \rangle  \left(\gamma - i \epsilon_{a,k}\right) 
+\langle  c_q^\dagger a_k\rangle \left(\gamma + i \epsilon_{a,k}\right)\bigg]	
\end{equation}

\subsection{Energy current when $L_b = 0$}

For simplicity, we will restrict the discussion of the energy current to the case $L_b = 0$. 
In this case, similarly to Eq.~(\ref{J_final}), we obtain for the energy current (\ref{JE_AC}) the following result:
\begin{equation}\label{JE_sum}
J_E=\frac{4  g^2 \gamma }{(L_a+1)^2} \sum\limits_{k,q} \frac{ \sin^2 k \sin^2 q\; (\bar{n}_{a,k}- \bar{n}_{c,k})(\epsilon_k + \epsilon_q)/2}{\gamma^2 + t^2 (\cos k - \cos q)^2} 
\end{equation}
It is also convenient to  define 
\begin{equation}\label{IE_def}
\mathcal{I}_E(k) = \frac{\sin^2 k}{L_a + 1} \sum\limits_q \frac{\sin^2 q (\epsilon_k + \epsilon_q)/2}{\gamma^2 + t^2 (\cos k - \cos q)^2}
\end{equation}
so that Eq.~(\ref{JE_sum}) becomes
\begin{equation}\label{JE_final}
J_E = \frac{4 g^2 \gamma}{L_a + 1} \sum\limits_k \mathcal{I}_E(k) \; (\bar{n}_{a,k} - \bar{n}_{c,k})
\end{equation}

\begin{figure}
\centering
\includegraphics[width=0.23\textwidth]{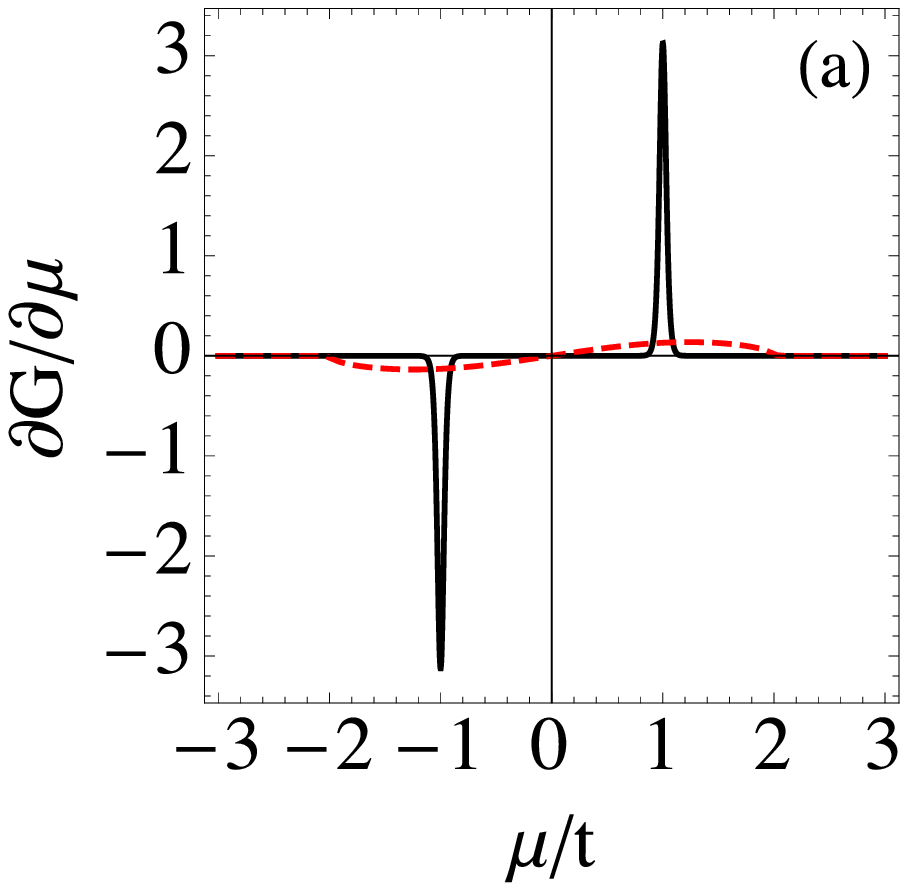}\quad
\includegraphics[width=0.23\textwidth]{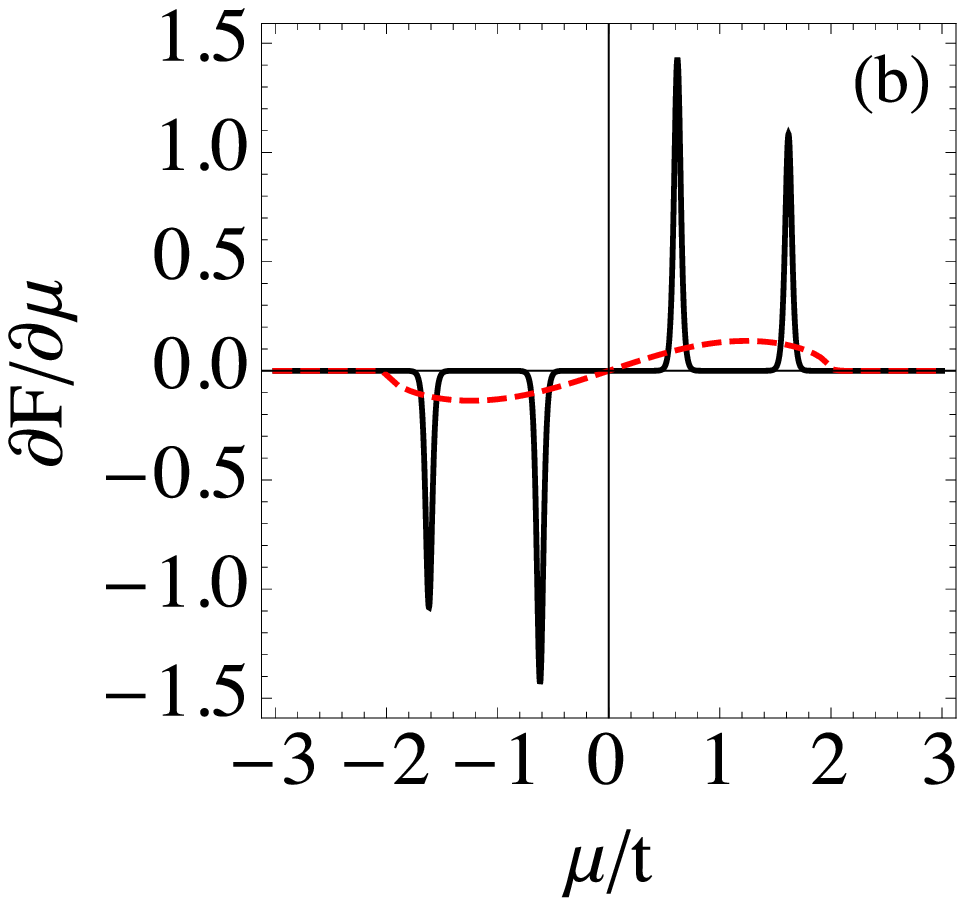}\\
\includegraphics[width=0.23\textwidth]{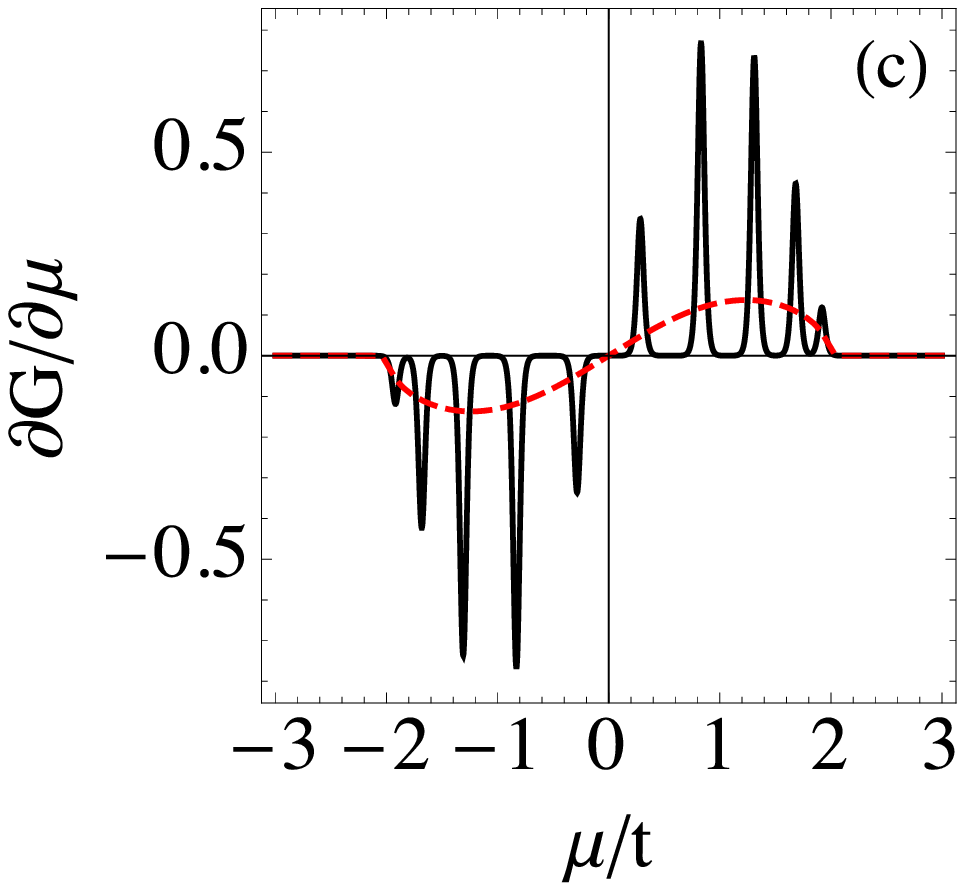}\quad
\includegraphics[width=0.23\textwidth]{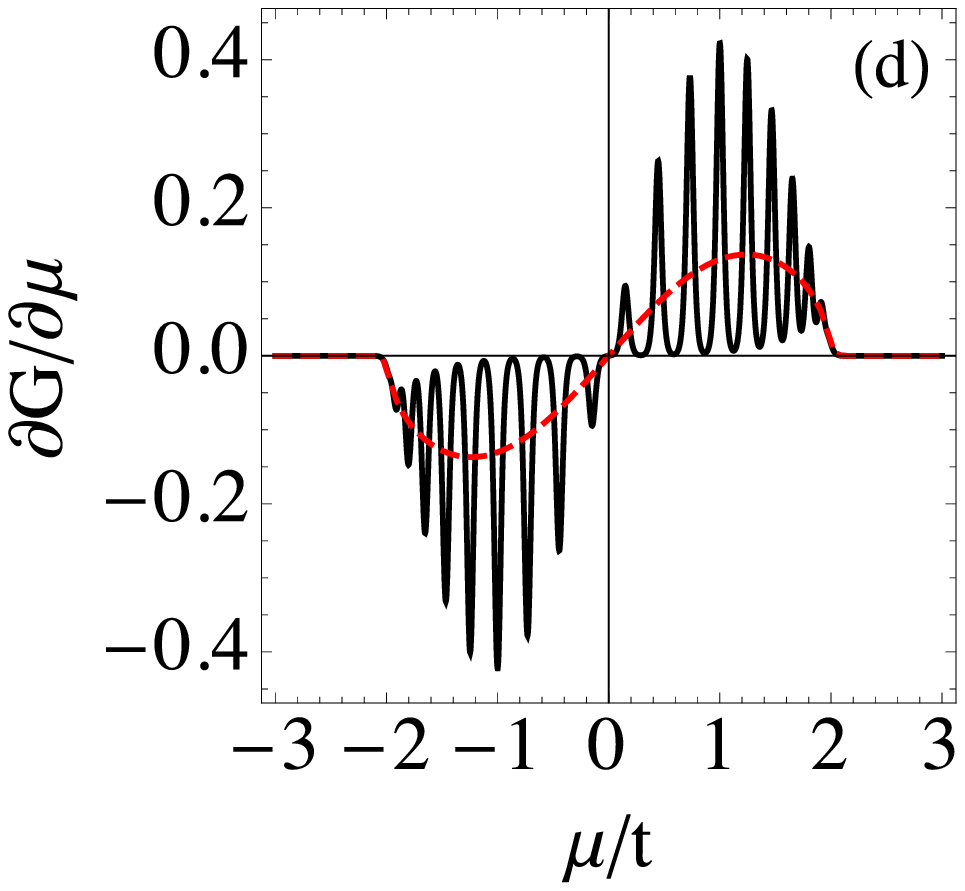}\\
\includegraphics[width=0.23\textwidth]{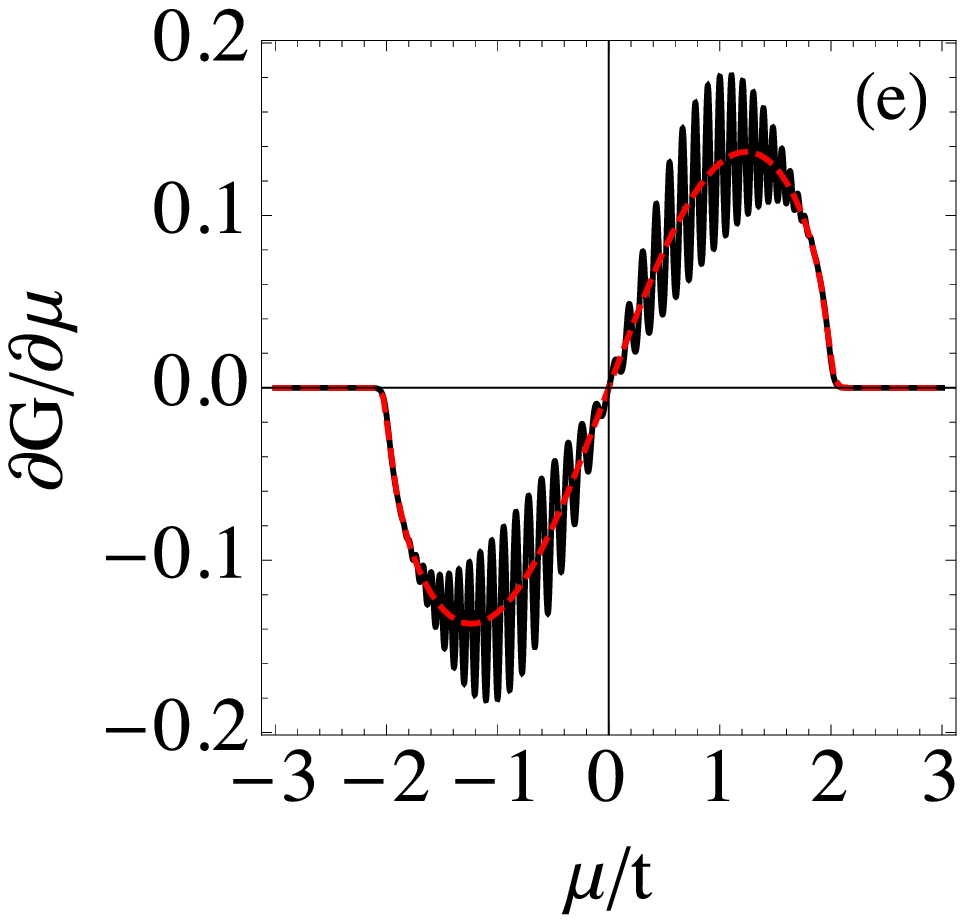}\quad
\includegraphics[width=0.23\textwidth]{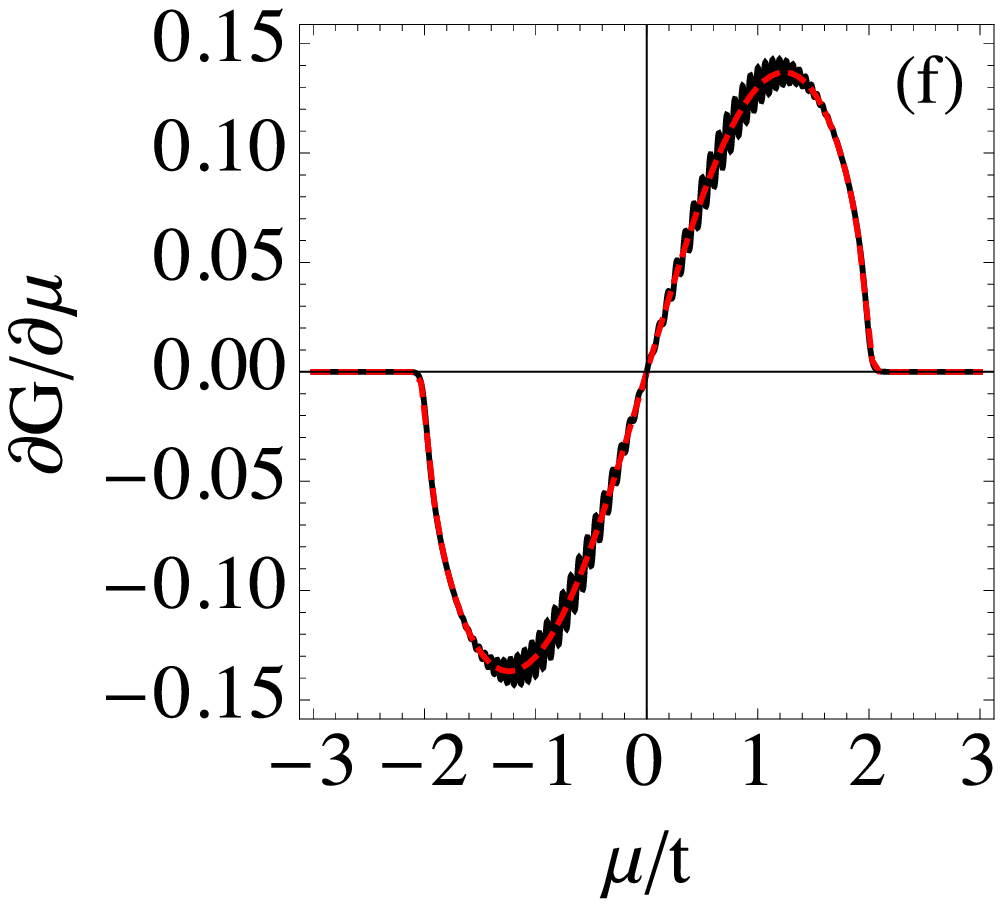}\\
\includegraphics[width=0.23\textwidth]{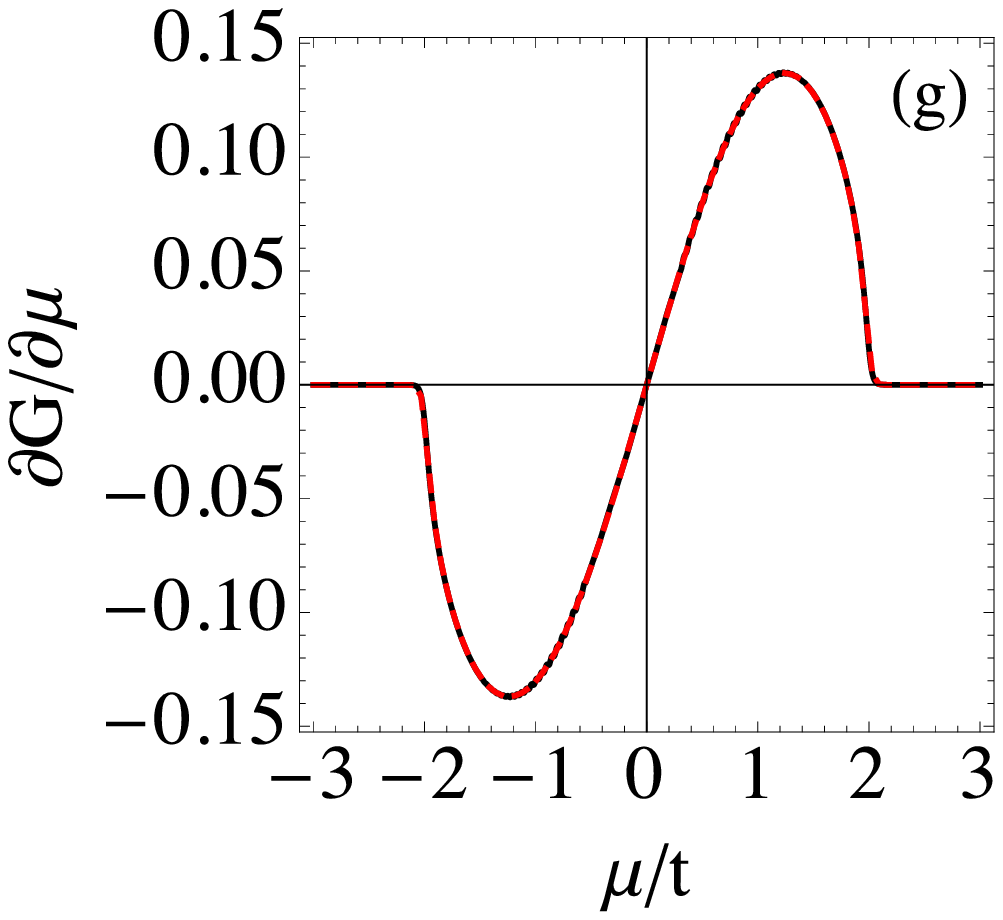}\quad
\includegraphics[width=0.23\textwidth]{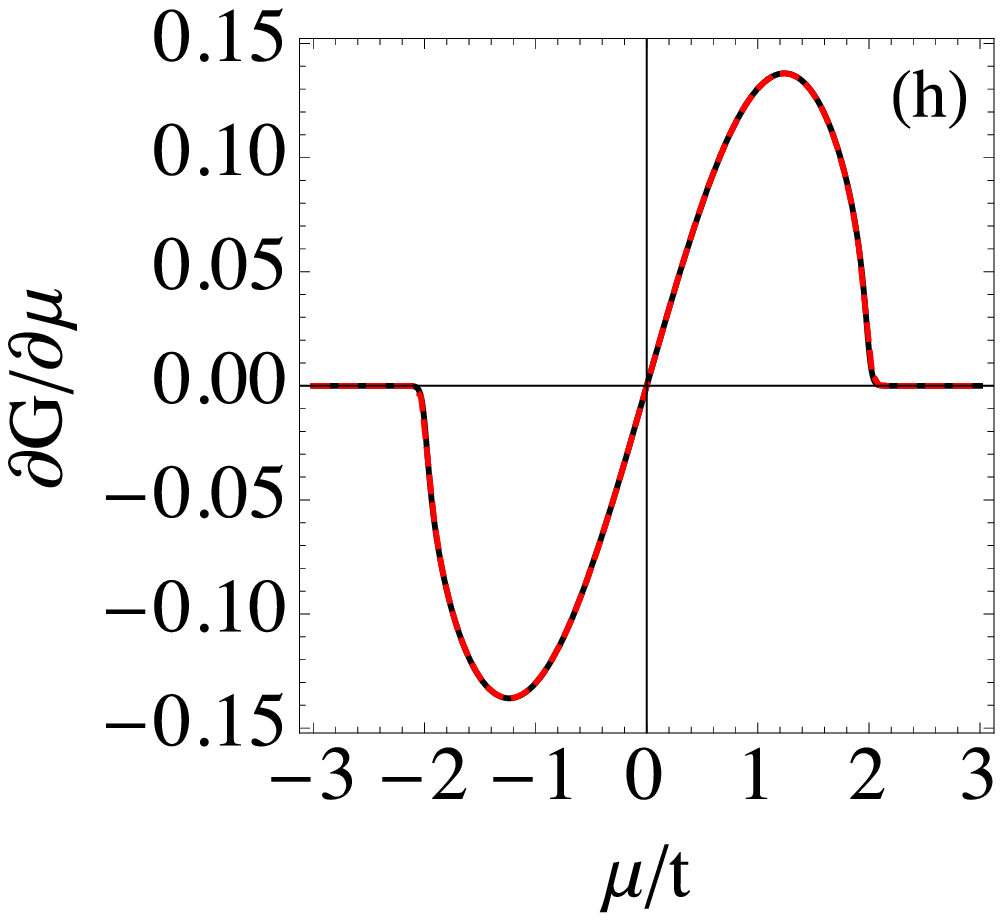}
\caption{\label{fig:JE_mu_finite}(Color Online)
The energy  current due to a gradient in the chemical potential, $(\partial G/\partial \mu)$~vs.~$\mu$ plotted using Eq.~(\ref{G}) with $\gamma/t = 1$ and $T = 0.02 t$. 
Each curve correspond to a different value of $L_a = L_c$, respectively: (a) 2, (b) 4, (c) 10, (d) 20, (e) 50, (f) 80, (g) 120 and (h) 160.
The red-dashed lines correspond to the thermodynamic limit.
}
\end{figure}

\begin{figure}
\centering
\includegraphics[width=0.23\textwidth]{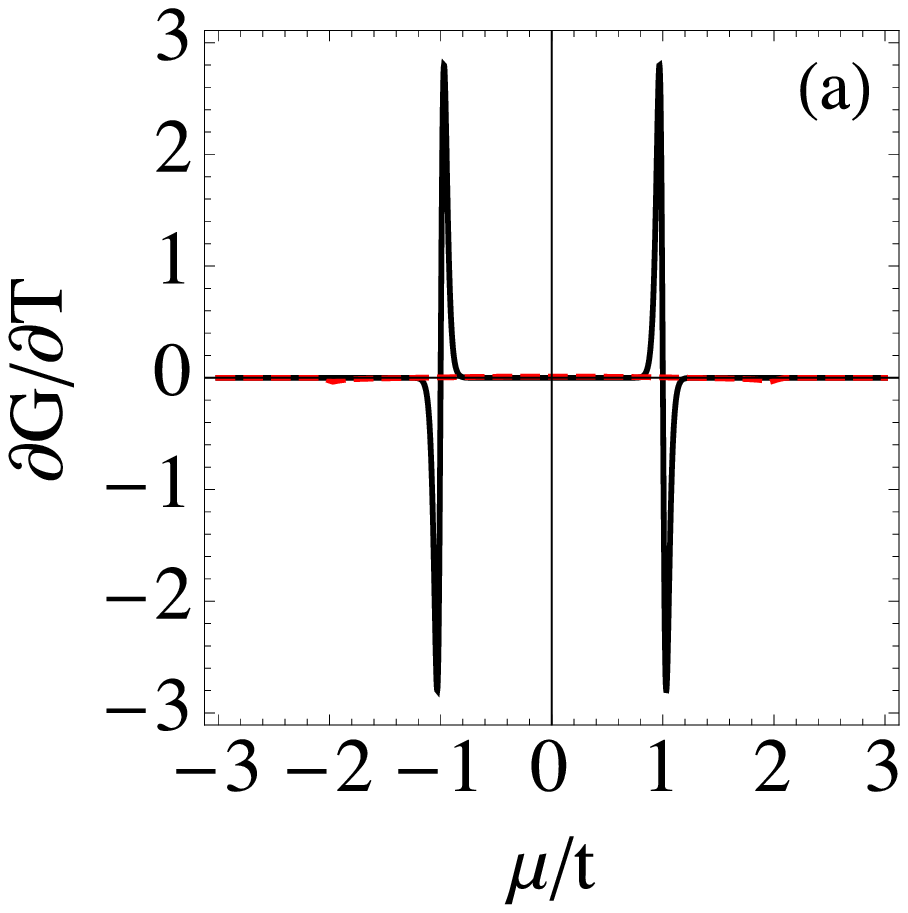}\quad
\includegraphics[width=0.23\textwidth]{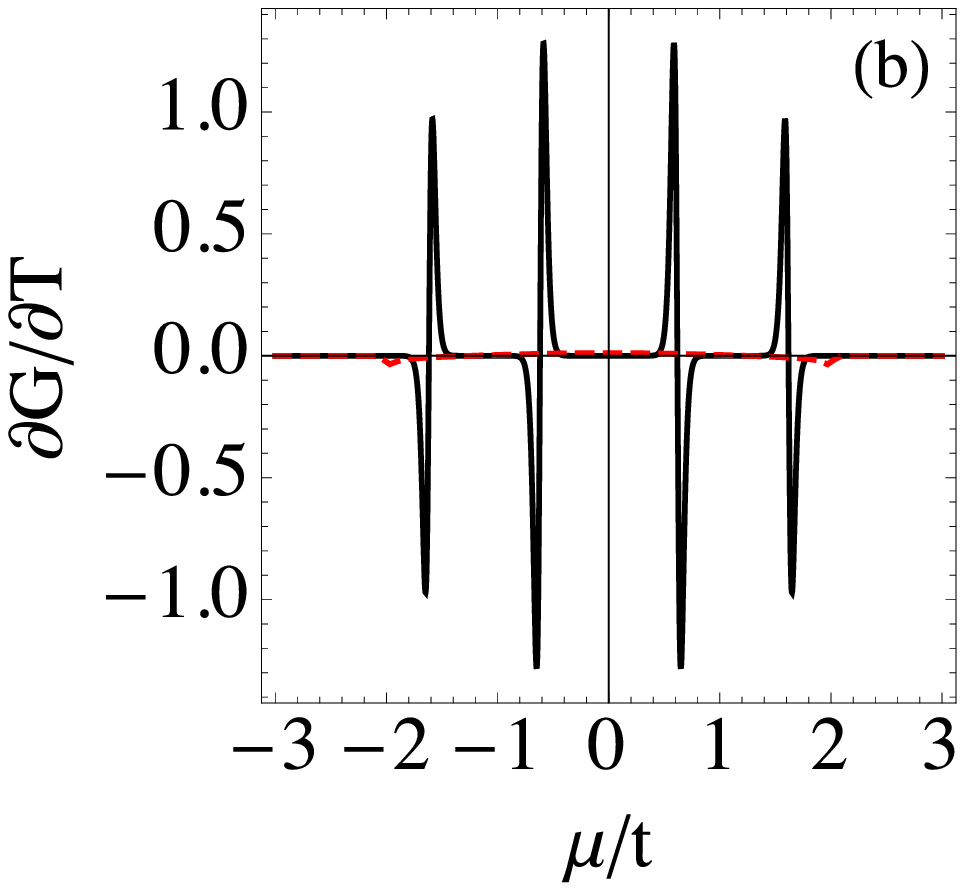}\\
\includegraphics[width=0.23\textwidth]{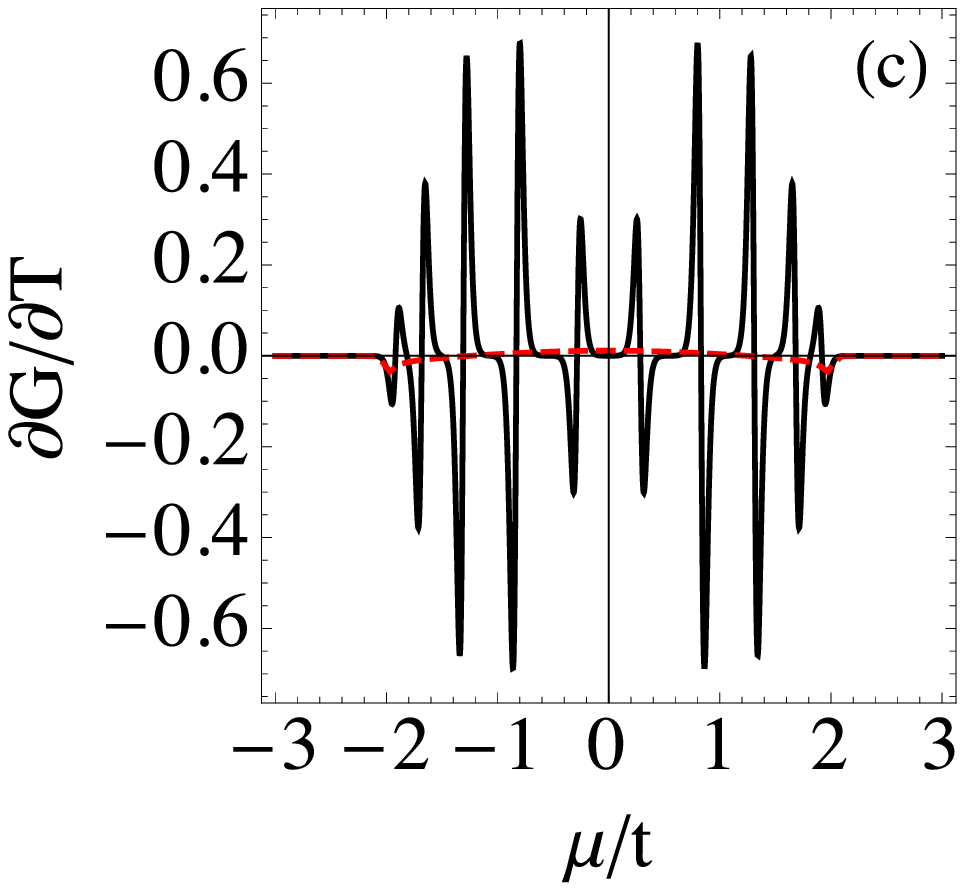}\quad
\includegraphics[width=0.23\textwidth]{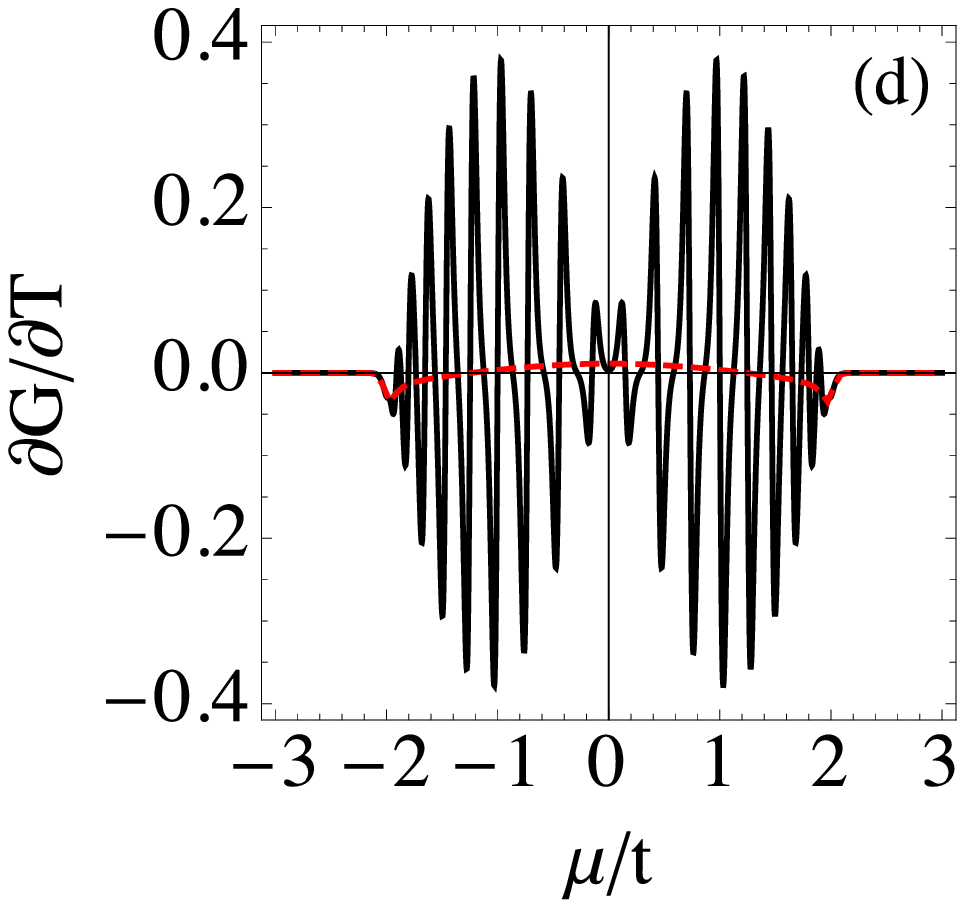}\\
\includegraphics[width=0.23\textwidth]{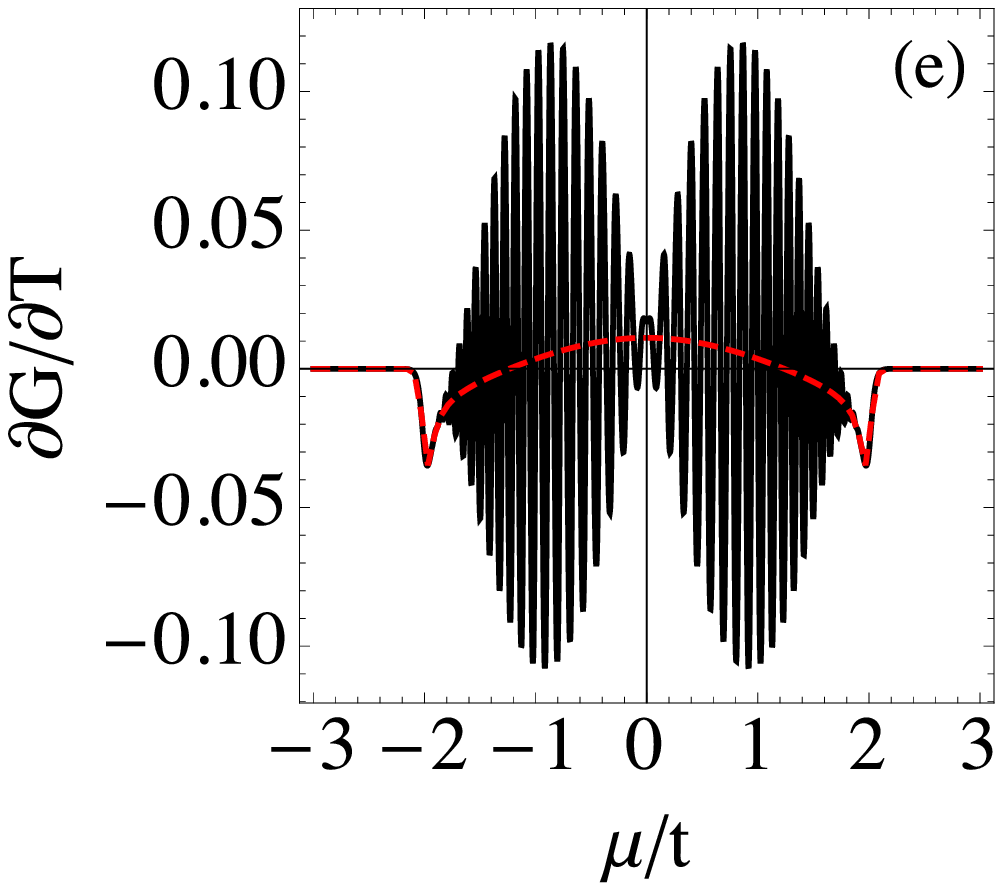}\quad
\includegraphics[width=0.23\textwidth]{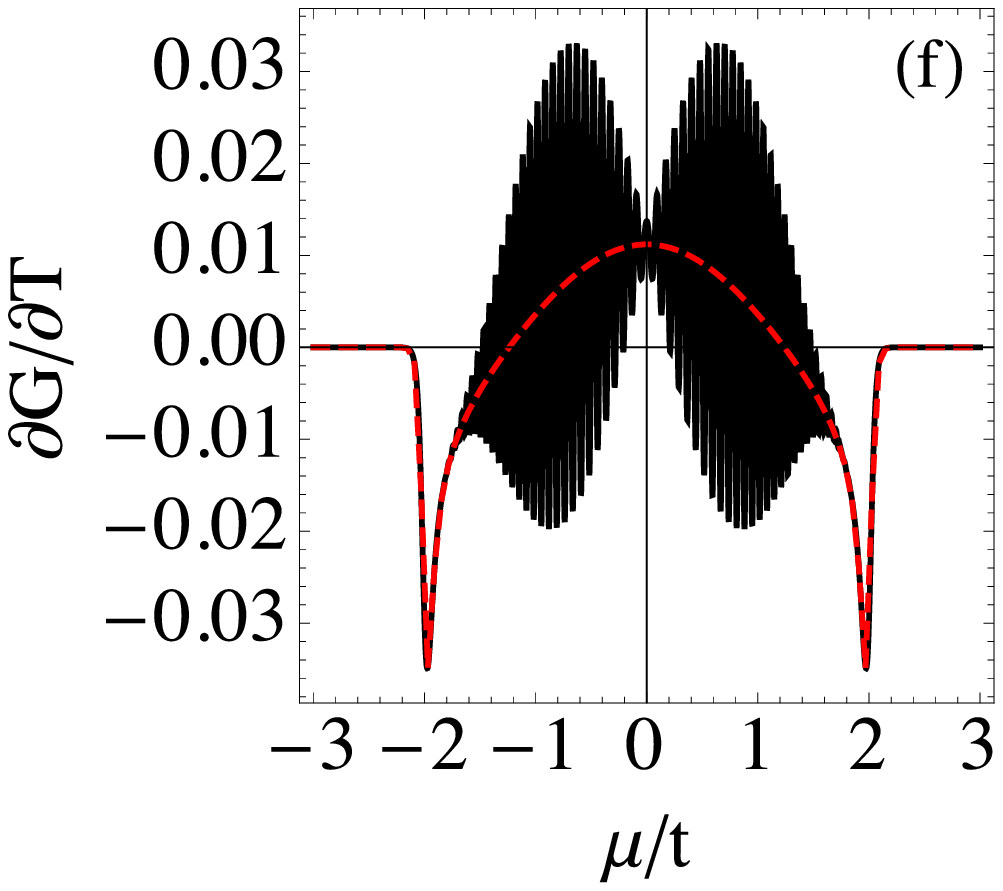}\\
\includegraphics[width=0.23\textwidth]{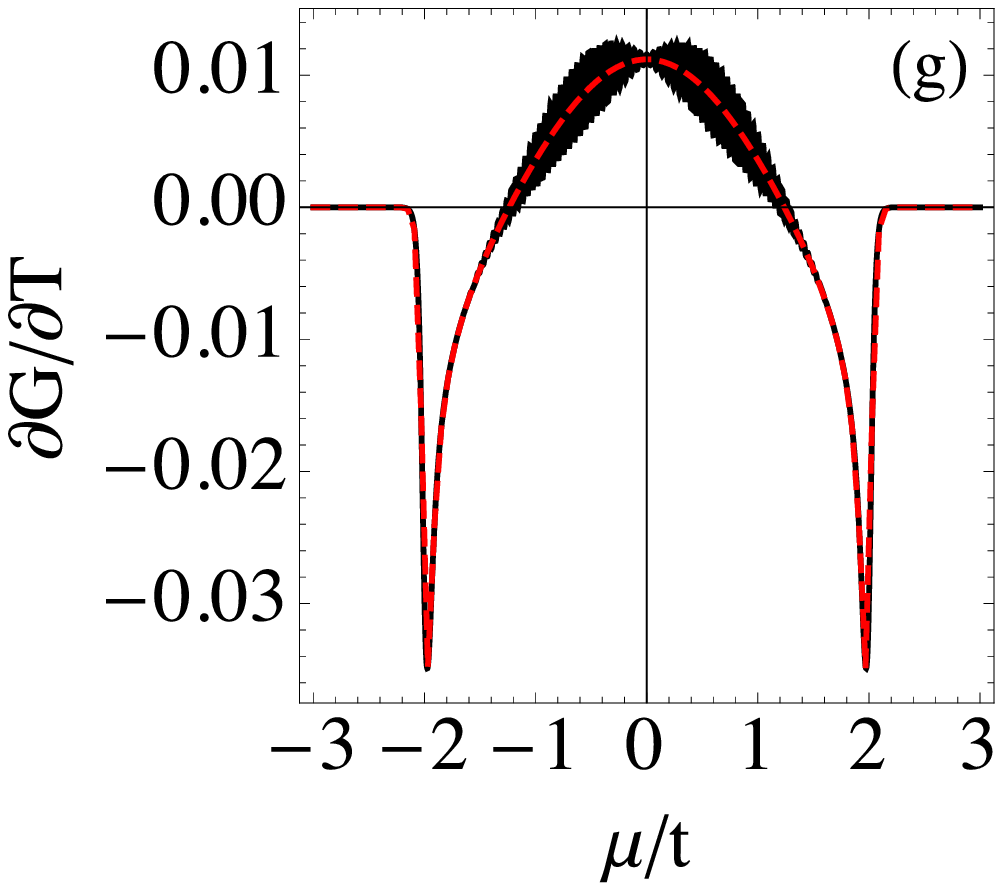}\quad
\includegraphics[width=0.23\textwidth]{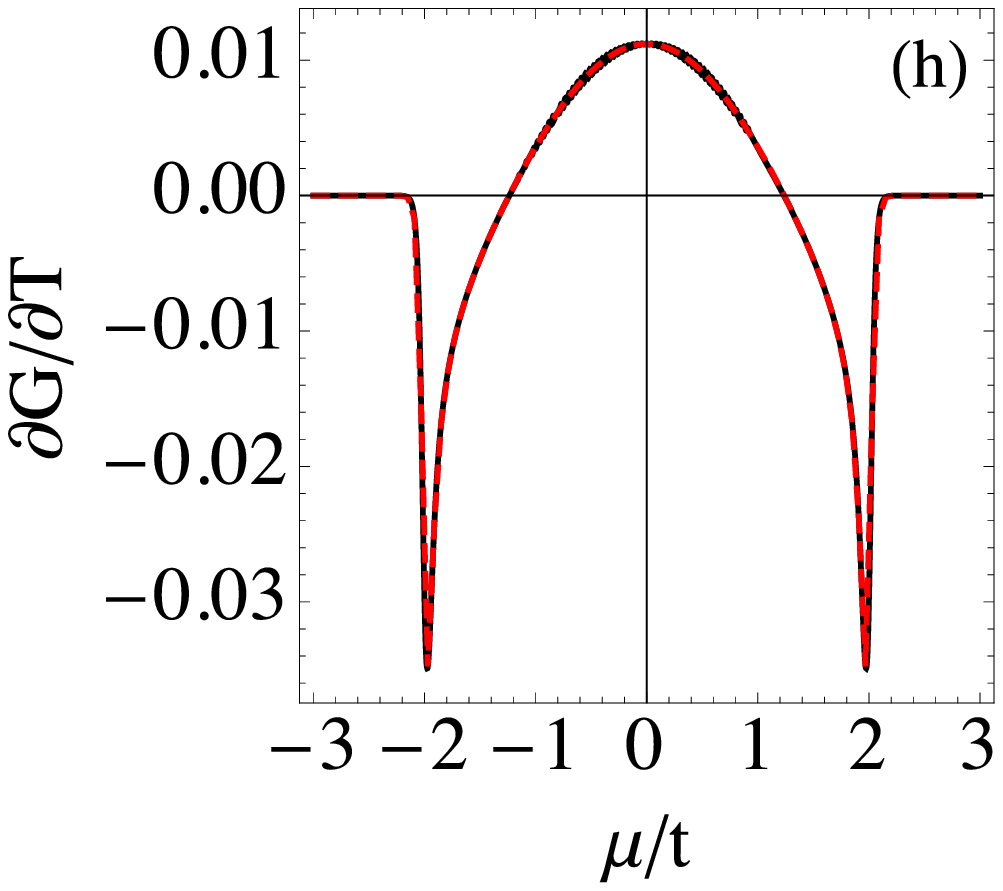}
\caption{\label{fig:JE_T_finite}(Color Online)
The energy current due to a temperature gradient, $(\partial G/\partial T)$~vs.~$\mu$ plotted using Eq.~(\ref{G}) with $\gamma/t = 1$ and $T = 0.02 t$. 
Each curve correspond to a different value of $L_a = L_c$, respectively: (a) 2, (b) 4, (c) 10, (d) 20, (e) 50, (f) 80, (g) 120 and (h) 160.
The red-dashed lines correspond to the thermodynamic limit.
}
\end{figure}

In the case of infinitesimal unbalances the energy current becomes
\begin{equation}\label{JEinf}
J_E = \delta\mu \frac{\partial G}{\partial \mu} + \delta T \frac{\partial G}{\partial T}
\end{equation}
where
\begin{equation}\label{G}
G = \frac{4 g^2 \gamma}{L_a + 1} \sum\limits_k \mathcal{I}_E (k) \bar{n}_k
\end{equation}
Examples of $\partial G/\partial \mu$ and $\partial G/\partial T$ are shown in Figs.~\ref{fig:JE_mu_finite} and \ref{fig:JE_T_finite} for conditions similar to those used in Figs.~\ref{fig:J_mu_finite} and \ref{fig:J_T_finite}. 
As can be seen, the role of finite size effects is similar to the previous case.

In the thermodynamic limit we may convert $\mathcal{I}_E(k)$ in Eq.~(\ref{IE_def}) into an integral and then use Eq.~(\ref{Mdef}) to explore the limits where $\gamma/t \ll 1$ and $\gamma/t \gg 1$. 
We then get 
\begin{equation}\label{IE_limits}
\mathcal{I}_E(k) \simeq  \begin{cases}
\displaystyle{ \frac{\epsilon_k |\sin k|^3}{\gamma t } =  \epsilon_k \mathcal{I}(k)} , & \text{ if } \gamma \ll t \\[0.4cm]
\displaystyle{\frac{ \epsilon_k \sin^2k}{4\gamma^2}= \frac{\epsilon_k \mathcal{I}(k)}{2}}, & \text{ if } \gamma \gg t
\end{cases}
\end{equation}
The presence of the factor of 1/2 in the second equation has, as we will show below, important consequences to the behavior of the system. 
For intermediate values of $\gamma/t$, the integral may also be computed analytically but the result is cumbersome and will not be presented. 

Using these results we find that the flux in Eq.~(\ref{JE_final}) may be written as 
\begin{equation}\label{JE_TL}
J_E \simeq  \begin{cases}
\displaystyle{ \frac{4 g^2 \gamma}{\pi} \int\limits_0^\pi \ud k \; \mathcal{I}(k) \epsilon_k (\bar{n}_{a,k} - \bar{n}_{c,k})} , & \text{ if } \gamma \ll t \\[0.8cm]
\displaystyle{ \frac{4 g^2 \gamma}{\pi} \int\limits_0^\pi \ud k \; \frac{\mathcal{I}(k)}{2} \epsilon_k (\bar{n}_{a,k} - \bar{n}_{c,k})}, & \text{ if } \gamma \gg t
\end{cases}
\end{equation}
which may be compared directly with Eq.~(\ref{Jtl}).
Similarly, Eq.~(\ref{G}) becomes
\begin{equation}\label{G_TL}
G \simeq  \begin{cases}
\displaystyle{ \frac{4 g^2 \gamma}{\pi} \int\limits_0^\pi \ud k \; \mathcal{I}(k) \epsilon_k \bar{n}_{k}} , & \text{ if } \gamma \ll t \\[0.8cm]
\displaystyle{ \frac{4 g^2 \gamma}{\pi} \int\limits_0^\pi \ud k \; \frac{\mathcal{I}(k)}{2} \epsilon_k \bar{n}_{k}}, & \text{ if } \gamma \gg t
\end{cases}
\end{equation}
which may be compared with Eq.~(\ref{Ftl}).

\begin{figure}
\centering
\includegraphics[width=0.23\textwidth]{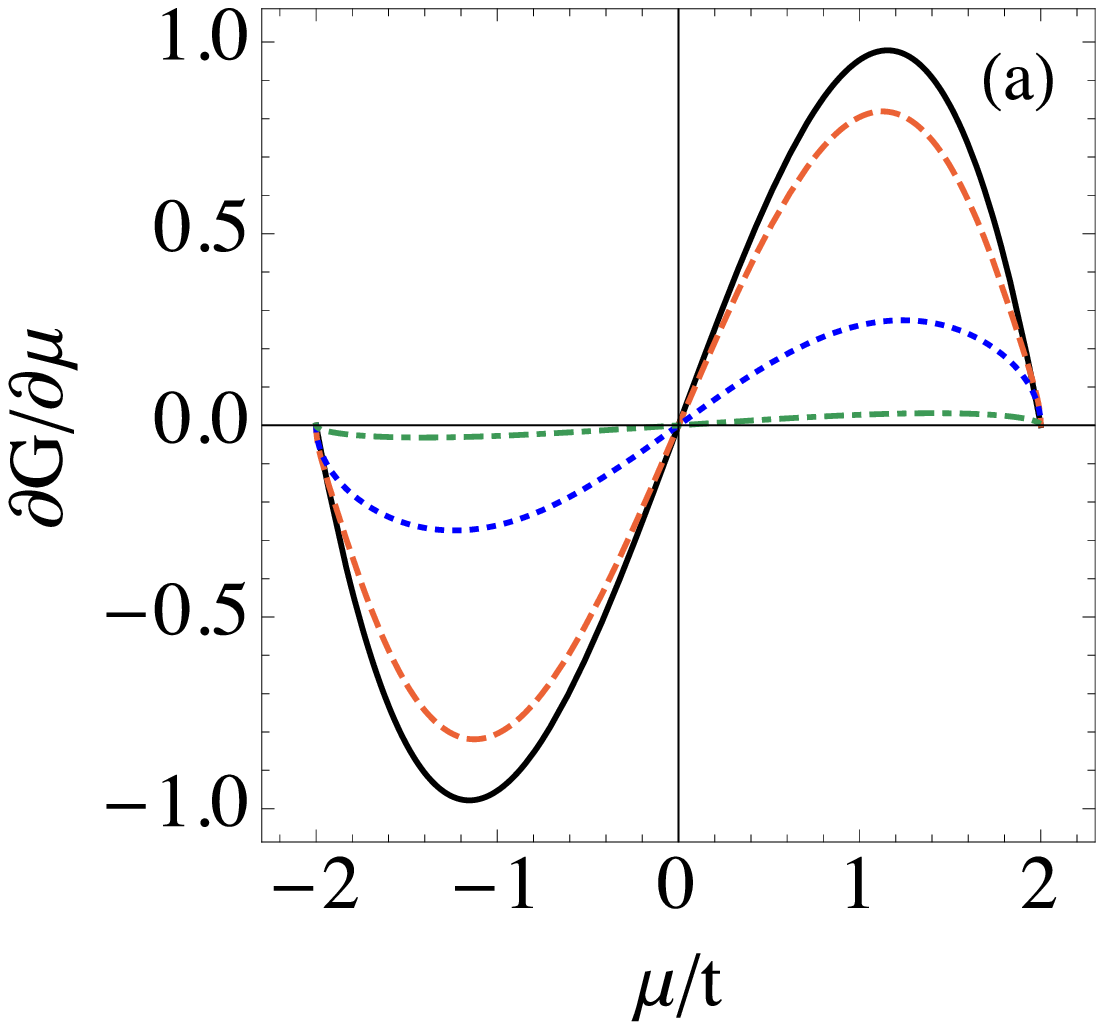}\quad
\includegraphics[width=0.23\textwidth]{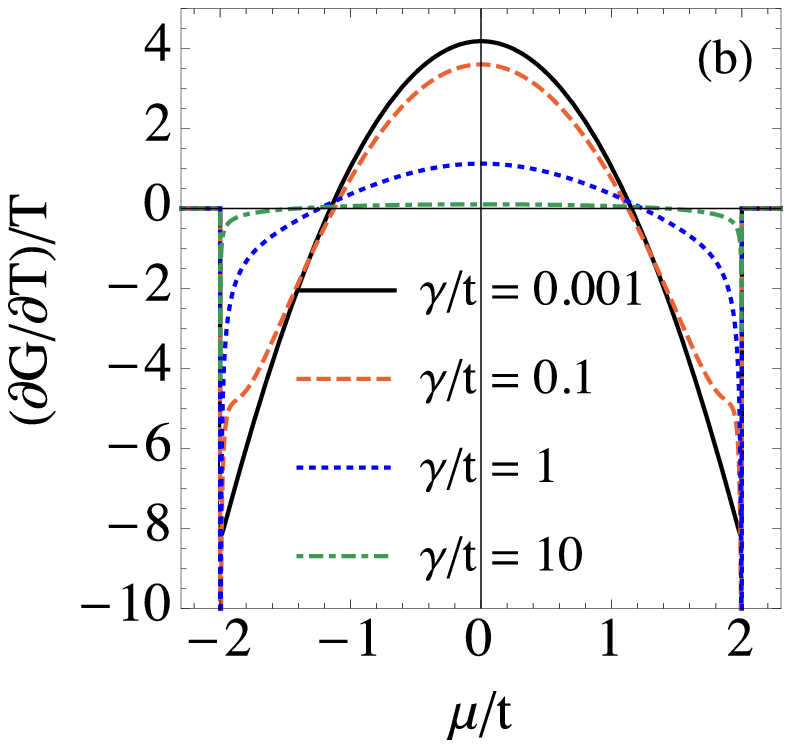}
\caption{\label{fig:JE_gamma}
(a)  $\partial G/\partial \mu$~vs.~$\mu$ and (b) $(\partial G/\partial T)$~vs.~$\mu$ for $T = 0$ in the thermodynamic limit. 
}
\end{figure}

At zero temperatures, Eqs.~(\ref{Fmu_I}) and (\ref{FT_I}) remain valid for the energy current, provided we replace $\mathcal{I}$ with $\mathcal{I}_E$.
We therefore find that
\begin{equation}\label{Gmu}
\frac{\partial G}{\partial \mu} = \begin{cases}
\displaystyle{\frac{g^2 \mu (4 t^2 - \mu^2)}{2 \pi t^4}} , 	&	  \gamma/t \ll 1 \\[0.4cm]
\displaystyle{\frac{g^2 \mu \sqrt{4 t^2 - \mu^2}}{4\pi t^2 \gamma}}	& 	\gamma/t \gg 1
\end{cases}
\end{equation}
and
\begin{equation}\label{GT}
\frac{\partial G}{\partial T} = \begin{cases}
\displaystyle{\frac{\pi g^2 T (4 t^2 - 3 \mu^2)}{6 t^4}} , 	&	  \gamma/t \ll 1 \\[0.4cm]
\displaystyle{\frac{\pi g^2 T}{6\gamma t^2}\frac{(2t^2-\mu^2)}{\sqrt{4 t^2 - \mu^2}}}	& 	\gamma/t \gg 1
\end{cases}
\end{equation}
These results, together with the general dependence when $T = 0$, are shown in Fig.~\ref{fig:JE_gamma}.

\subsection*{\label{ssec:onsager}Onsager coefficients}

From the first law of thermodynamics, the current of energy should have a term due to the current of heat and another due to the current of particles. 
That is, we should have  $J_E = J_Q + \mu J$, where $J_Q$ is the heat current through the system.
Since we know $J$ and $J_E$, we may then use this to compute the heat current:
\begin{equation}
J_Q = J_E - \mu J
\end{equation}
The currents  $J$ and $J_Q$ may be cast in the language of Onsager's coefficients \cite{Onsager1931,*Onsager1931a} by defining the unbalances as $(\delta \mu)/T$ and $-\delta(1/T)$.
Then the fluxes $J$ and $J_Q$ should satisfy 
\begin{equation}
\begin{pmatrix} J \\[0.2cm] J_Q \end{pmatrix}
=
\begin{pmatrix} \ell_{11}	&	\ell_{12} \\[0.2cm] \ell_{21} 	&	\ell_{22} \end{pmatrix}
\begin{pmatrix} (\delta\mu)/T \\[0.2cm] -\delta(1/T)\end{pmatrix}
\end{equation}
where the $\ell_{ij}$ are the Onsager coefficients. 
According to Onsager's reciprocity relations \cite{Onsager1931,*Onsager1931a}, we expect that $\ell_{12} = \ell_{21}$.
Moreover, the entropy production rate in the NESS is defined as 
\begin{equation}
\Pi = J (\delta\mu)/T - J_Q \delta(1/T)	
\end{equation}
and it should be a non-negative quantity. 
This will be satisfied for any infinitesimal unbalance provided the determinant of the Onsager matrix, $\ell_{11} \ell_{22} - \ell_{12} \ell_{21}$, is non-negative. 

We now use all our previous results to obtain the Onsager coefficients. 
Using Eq.~(\ref{Jinf}) we find that 
\begin{equation}\label{onsager_1}
\ell_{11} = T \frac{\partial F}{\partial \mu},\qquad \ell_{12} =  T^2 \frac{\partial F}{\partial T}
\end{equation}
Similarly, using Eqs.~(\ref{Jinf}) and (\ref{JEinf}) we may write 
\begin{equation}
J_Q = \delta_\mu \bigg[\frac{\partial G}{\partial \mu} - \mu \frac{\partial F}{\partial \mu} \bigg] + \delta T \bigg[\frac{\partial G}{\partial T} - \mu \frac{\partial F}{\partial T} \bigg]
\end{equation}
Thus, the other Onsager coefficients are 
\begin{equation}\label{onsager_2}
\ell_{21} = T \bigg[\frac{\partial G}{\partial \mu} - \mu \frac{\partial F}{\partial \mu} \bigg], \quad \ell_{22} = T^2 \bigg[\frac{\partial G}{\partial T} - \mu \frac{\partial F}{\partial T} \bigg]
\end{equation}

\begin{figure}
\centering
\includegraphics[width=0.23\textwidth]{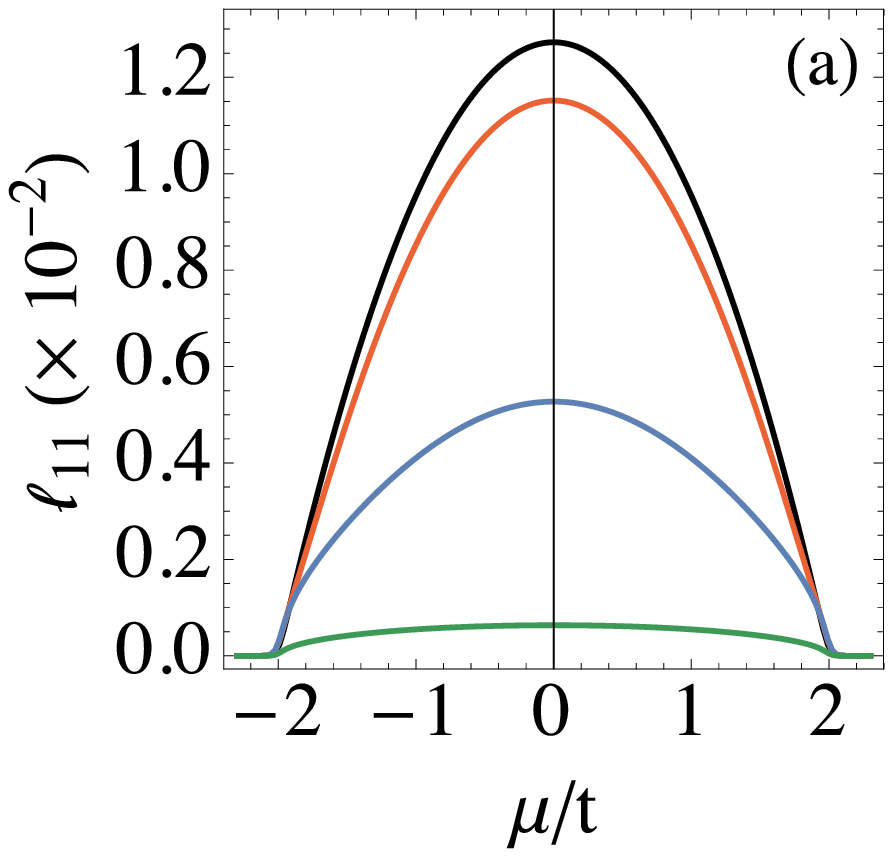}\quad
\includegraphics[width=0.23\textwidth]{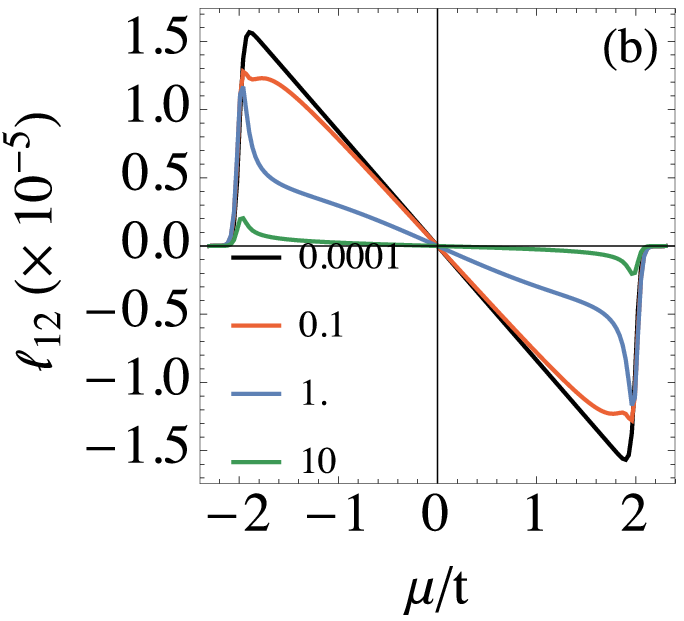}\\
\includegraphics[width=0.23\textwidth]{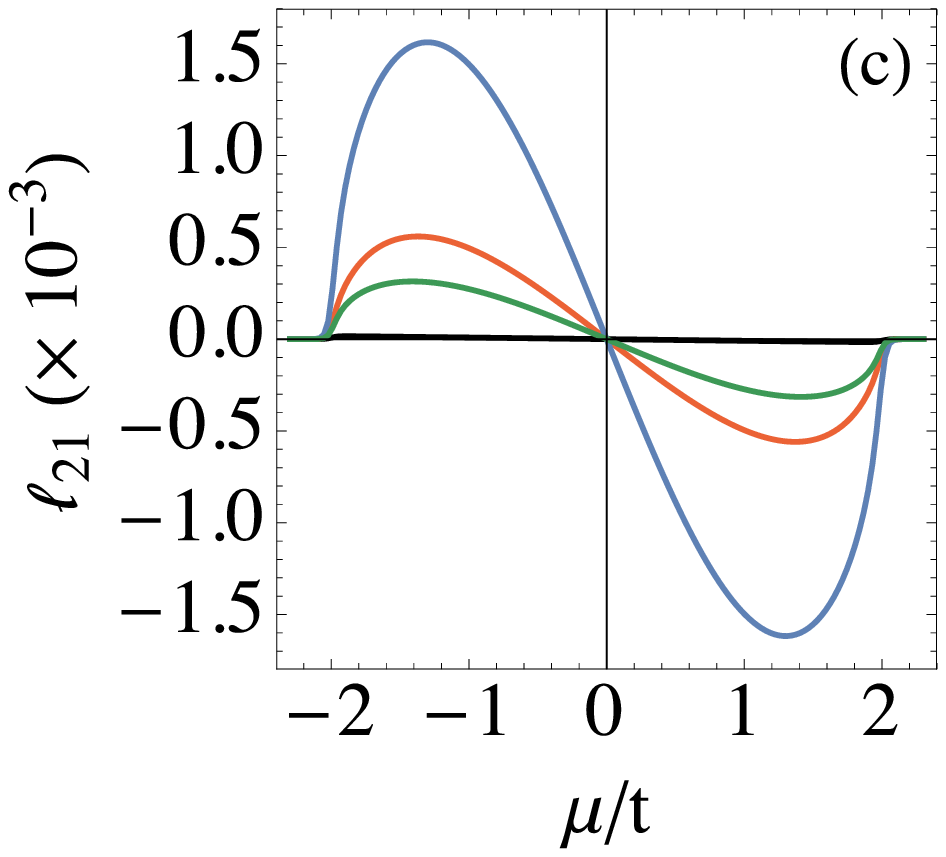}\quad
\includegraphics[width=0.23\textwidth]{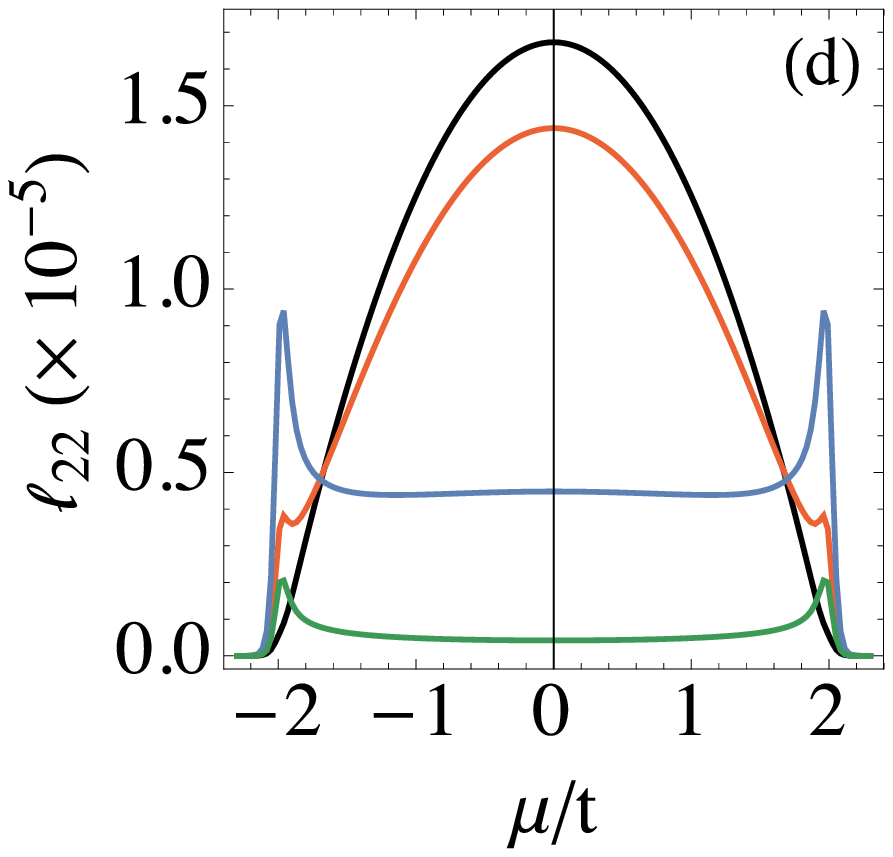}
\caption{\label{fig:Onsager}Onsager coefficients computed from Eqs.~(\ref{onsager_1}) and (\ref{onsager_2}), in the thermodynamic limit, with $T = 0.02 t$ and different values of $\gamma/t$, as shown in image (b).
}
\end{figure}

\begin{figure}
\centering
\includegraphics[width=0.23\textwidth]{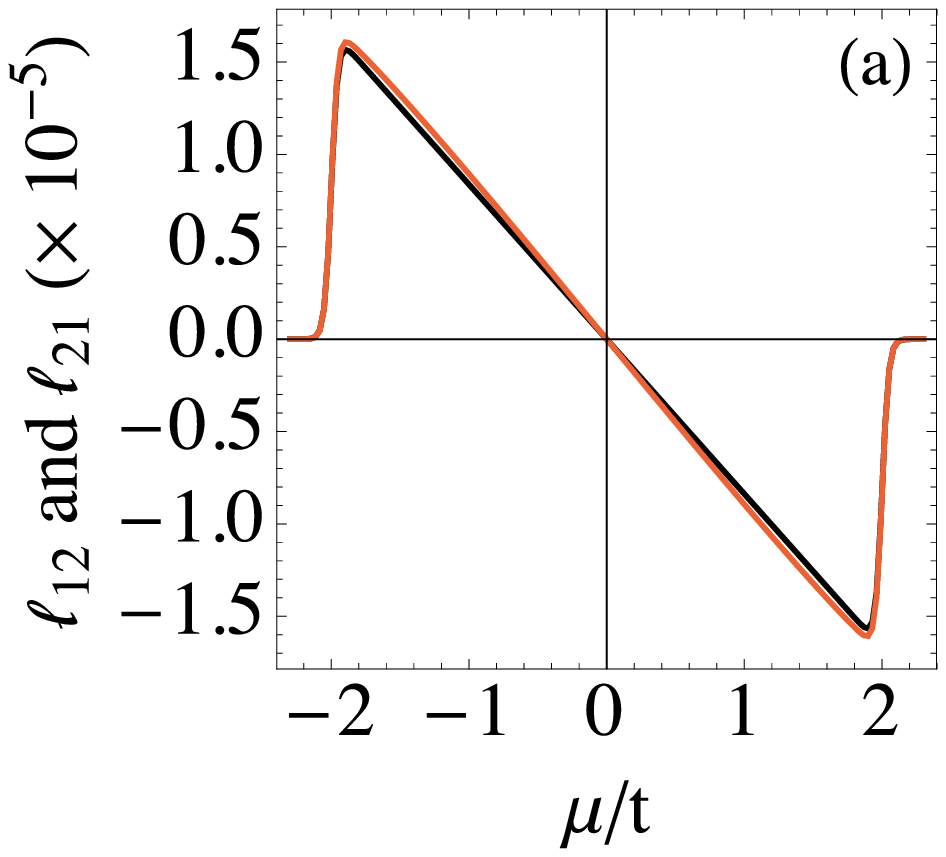}\quad
\includegraphics[width=0.23\textwidth]{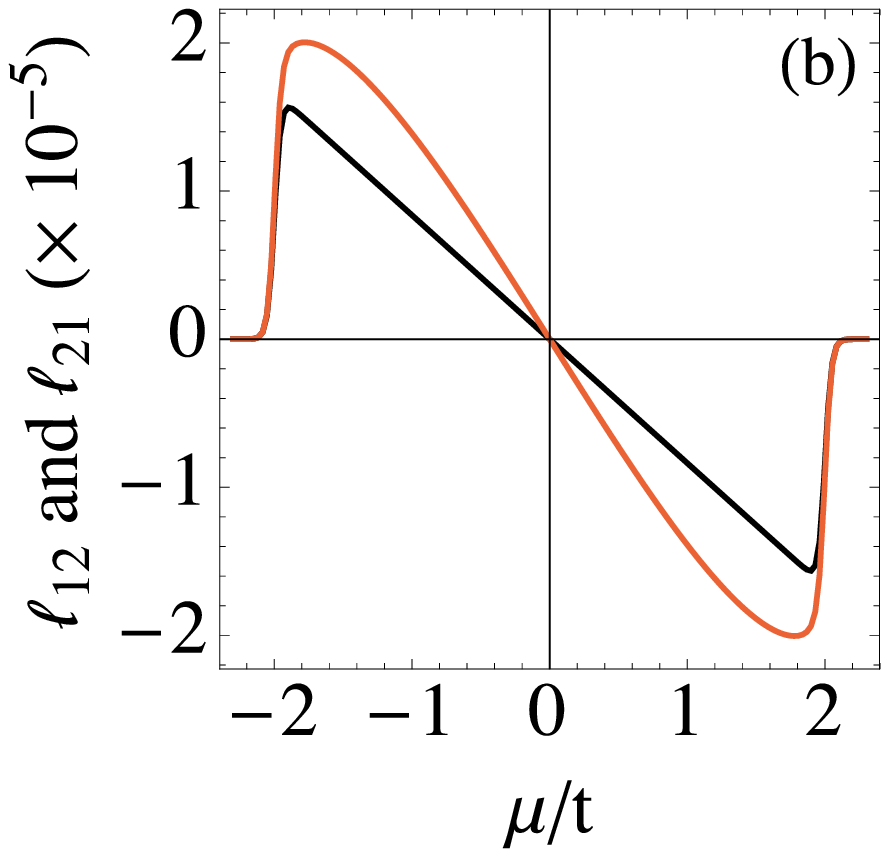}
\caption{\label{fig:Onsager_cross}
Comparison between the cross Onsager coefficients $\ell_{12}$ and $\ell_{21}$ for $\gamma/t = 0.0001$ and $0.001$, with $T = 0.02 t$. 
}
\end{figure}

Examples of the Onsager coefficients, computed in the thermodynamic limit, are shown in Fig.~\ref{fig:Onsager} for different values of $\gamma$. 
As can be seen, the direct coefficients $\ell_{11}$ and $\ell_{22}$ are always positive, as expected. 
The corresponding determinant is also always positive, thus ensuring a positive entropy production.  
However, the cross coefficients $\ell_{12}$ and $\ell_{21}$ only coincide for small values of $\gamma$. 
This is illustrated specifically in Fig.~\ref{fig:Onsager_cross}, where we compare $\ell_{12}$ and $\ell_{21}$ for $\gamma/t = 0.0001$ and $\gamma/t = 0.001$. 
As can be seen, only for the smallest value of $\gamma/t$ does the two quantities coincide. 

This fact can actually be demonstrated analytically, using Eqs.~(\ref{Ftl}) and (\ref{G_TL}).
The coefficient $\ell_{12}$ in Eq.~(\ref{onsager_1}) reads
\begin{equation}
\ell_{12} = \frac{4 g^2 \gamma}{\pi} \int\limits_0^\pi \ud k \; \mathcal{I}(k) T^2 \frac{\partial \bar{n}}{\partial T}
\end{equation}
which holds for any value of $\gamma$. 
However, for the coefficient $\ell_{21}$ in Eq.~(\ref{onsager_2}) we must distinguish between the different $\gamma$ regimes. 
From Eq.~(\ref{G_TL}), if $\gamma/t \ll 1$, then we will have 
\begin{equation}
\ell_{21} = \frac{4 g^2 \gamma}{\pi} \int\limits_0^\pi \ud k \; \mathcal{I}(k) T(\epsilon_k - \mu)\frac{\partial \bar{n}}{\partial \mu}
\end{equation}
Since [cf. Eq.~(\ref{bar_n})]
\[
\frac{\partial \bar{n}_k}{\partial T} = \frac{\epsilon_k-\mu}{T} \frac{\partial \bar{n}_k}{\partial \mu}
\]
we conclude that when $\gamma/t \ll 1$, $\ell_{12} = \ell_{21}$. 
Conversely, in the case when $\gamma/t \gg 1$, no such equality holds due to the factor of 1/2 in the second line of Eq.~(\ref{G_TL}). 
This therefore explains the results in Fig.~\ref{fig:Onsager_cross}. 

The reciprocity relation $\ell_{12} = \ell_{21}$ is a direct consequence of detailed balance in the system \cite{Onsager1931,*Onsager1931a}. 
These results therefore indicate that in our multi-site setup, the system should only satisfy detailed balance when $\gamma/t \ll 1$. 
This is further corroborated by the results in Fig.~\ref{fig:occupation}, where we found that only for low $\gamma/t$ did chain B correctly thermalize locally, something expected from a system satisfying detailed balance.

%
%
%
%

\section{Conclusions}

The non-equilibrium properties of open quantum chains is known to be extremely sensitive to the type of dissipator employed. 
In addition, unless one has detailed experimental knowledge of the system-bath coupling,  the structure of the dissipator is not unique.
Hence the importance of understanding the properties of the NESS under the influence of different dissipators. 
In this paper we have discussed in detail the properties of multi-site baths, where the Lindblad dissipator acts on groups of spins and is such that the entire group, were it isolated, is correctly thermalized. 
For our system, which is quadratic (in the language of second quantization), this type of dissipator is readily constructed by coupling to the normal modes of the Hamiltonian. 
Indeed, it is worth mentioning that this approach can be used for any quadratic Hamiltonian, fermionic or bosonic. 
Hence, together with the perturbative solution presented here, this multi-site bath structure opens avenues to the research of many other systems in statistical mechanics and condensed matter in general.

We have shown that the multi-site baths introduce physical properties which are substantially richer from those of a single-site bath. 
Using a perturbative method we have shown that the particle and energy currents  have the structure of Landauer's formula, and we have been able to find analytical formulas for the Onsager coefficients. 
In all results, we have observed a sensitive dependence on the coupling constant $\gamma$. 
When $\gamma/t \ll 1$, which is the situation expected experimentally, the system obeys the Onsager reciprocal relations and the middle chain correctly thermalizes to its Fermi-Dirac distribution. 
Conversely, when $\gamma/t \gg 1$ we find that all modes tend to contribute equally, leading to substantial modifications in the properties of the system.

\begin{acknowledgements} 

The authors would like to thank Prof. Dragi Karevski for fruitful discussions.
For their financial support, the authors would like to acknowledge the Brazilian funding agencies CNPq and FAPESP  (2014/01218-2).

\end{acknowledgements}

\appendix
\section{\label{sec:app}Microscopic derivation of the Lindblad dissipator}

In this appendix we will show how to derive  the dissipator~(\ref{D}). 
The basic idea will be to assume that, since the Hamiltonian~(\ref{H3}) factors into a sum of commuting terms for each normal mode, we may treat each  mode individually. 
We therefore only need to consider a Hamiltonian~$H = \epsilon \eta^\dagger \eta$ for one normal mode [here $\eta$ is a simplified notation for each of the $\eta_k$ defined in Eq.~(\ref{ak_an}) and not the original $\eta_n$ of Eq.~(\ref{JW})]. 
The total dissipator will then be a sum of the dissipators of each mode. 

We will further assume that this normal mode is coupled to an infinite number of bosonic degrees of freedom with Hamiltonian~$H_B = \sum_\ell \Omega_\ell b_\ell^\dagger b_\ell$, where $b_\ell$ are bosonic operators satisfying $[b_\ell, b_{\ell'}^\dagger] = \delta_{\ell,\ell'}$. 
The interaction Hamiltonian is assumed to 
\begin{equation}\label{app:HI}
H_I = \sum\limits_\ell f_\ell (\eta + \eta^\dagger) (b_\ell + b_\ell^\dagger)
\end{equation}
where $f_\ell$ are certain coupling constants. The only assumption here is that the bath couples linearly in the $\eta$ (or, more precisely, in the $\eta_k$). 
Notice that since the normal modes $\eta_k$ are linearly related to the original operators $\eta_n$ [cf. Eq.~(\ref{H0})], it does not matter if the bath is coupled to the normal modes $\eta_k$ or to the $\eta_n$. This will only change the constants $f_\ell$. 

Under the assumption of weak-coupling and in the rotating wave approximation we may trace out the bath and write a corresponding Lindblad dissipator. 
This is most readily done using the method of eigenoperators, which is discussed in detail in Ref.~\cite{Breuer2007}. 
An arbitrary operator $\mathcal{O}(\omega)$ is termed an eigenoperator of $H$ when 
\[
[H,\mathcal{O}(\omega)] = - \omega \mathcal{O}(\omega)
\]
for some given frequency $\omega$. 
According to the derivation in~\cite{Breuer2007}, we must construct the eigenoperator associated to $\mathcal{O} = (\eta + \eta^\dagger)$, which is the operator coupling to the bath. 
Due to the diagonal structure of $H = \epsilon \eta^\dagger \eta$, it follows that this eigenoperator will be
\begin{equation}\label{app:O}
\mathcal{O}(\omega) = \eta\; \delta_{\omega, \epsilon} + \eta^\dagger \; \delta_{\omega,-\epsilon}
\end{equation}
Intuitively speaking, the coupling $(\eta+\eta^\dagger)$ to the bath induces transitions in the system and $\omega$ represents the allowed energy transitions due to this coupling. For our case the only allowed transitions have energy differences $\epsilon$ and $-\epsilon$.
 
In terms of the eigenoperators $\mathcal{O}(\omega)$, the Lindblad dissipator corresponding to the bath coupling~(\ref{app:HI}) will be \cite{Breuer2007}:
\begin{equation}\label{app:D1}
D(\rho) = \sum\limits_\omega \Gamma(\omega) \bigg[\mathcal{O}(\omega) \rho \mathcal{O}^\dagger (\omega) - \frac{1}{2}\{\mathcal{O}^\dagger(\omega) \mathcal{O}(\omega),\rho\}\bigg]
\end{equation}
where
\[
\Gamma(\omega) =  \int\limits_{-\infty}^\infty \ud t e^{i \omega t} \tr\Bigg\{ (e^{i H_B t} B e^{-i H_B t} ) B \frac{e^{-H_B/T}}{\tr(e^{-H_B/T})}\Bigg\}
\]
is the Fourier transform of bath correlation functions, with $B = \sum\limits_\ell f_\ell (b_\ell + b_\ell^\dagger)$ [see Eq.~(\ref{app:HI})].
Substituting Eq.~(\ref{app:O})  into Eq.~(\ref{app:D1}) we get 
\begin{IEEEeqnarray}{rCl}
\label{app:D2}
D(\rho) &=& \Gamma(\epsilon) \bigg[ \eta \rho \eta^\dagger - \frac{1}{2} \{\eta^\dagger \eta,\rho\} \bigg]	\\[0.2cm]
&&+ \Gamma(-\epsilon) \bigg[ \eta^\dagger \rho \eta - \frac{1}{2} \{\eta \eta^\dagger, \rho\} \bigg] 	\nonumber
\end{IEEEeqnarray}

The quantities $\Gamma(\omega)$ may be resolved further by computing the Fourier transform and using the integral representation of the $\delta$-function. 
As a result we get 
\[
\Gamma(\omega) =2\pi \sum\limits_\ell f_\ell^2 \bigg[ \delta(\omega - \Omega_\ell) [1 + \bar{n}_B(\Omega_\ell)] + \delta(\omega + \Omega_\ell) \bar{n}_B(\Omega_\ell)\bigg]
\]
where $\bar{n}_B(x) = 1/(e^{x/T}-1)$ is the Bose-Einstein occupation number for the bath frequencies. 
Next we assume that the bath frequencies $\Omega_\ell$ cover a continuum of values (as expected from photonic or phononic baths) so that we may convert the $\ell$-sum into an integral over $\Omega$. 
We define the spectral density 
\[
\gamma(\Omega) = \sum\limits_\ell 2\pi f_\ell^2 \delta(\Omega - \Omega_\ell)
\]
in terms of which we may write
\[
\Gamma(\omega) = \int\limits_0^\infty \ud \Omega \; \gamma(\Omega) \bigg[ \delta(\omega - \Omega) [1 + \bar{n}_B(\Omega)] + \delta(\omega + \Omega) \bar{n}_B(\Omega)\bigg]
\]
This can be further simplified  to 
\begin{equation}\label{app:Gamma}
\Gamma(\omega) = \begin{cases} 
\gamma(\omega) [1 + \bar{n}_B(\omega)], & \text{ if } \omega > 0 \\[0.2cm]
\gamma(-\omega) \bar{n}_B(-\omega), & \text{ if } \omega < 0
\end{cases}
\end{equation}
The appearance of the Bose-Einstein occupation numbers in a fermionic problem may at first seem strange. 
But that is indeed correct, since they appear due to the effect of the bath, which is bosonic. 
Notwithstanding, the Fermi-Dirac occupation numbers $\bar{n}_F(x) = 1/(e^{x/T} + 1)$ may be introduced naturally as follows. 

From Eq.~(\ref{app:D2}), we must now compute $\Gamma(\pm\epsilon)$. 
In doing so we must differentiate between $\epsilon >0$ and $\epsilon <0$. 
Suppose first that $\epsilon > 0$. 
Then we use the identities 
\begin{equation}\label{app:ID}
\frac{\bar{n}_B(\epsilon)}{2 \bar{n}_B (\epsilon) + 1} = \bar{n}_F(\epsilon),\qquad 
\frac{1+ \bar{n}_B(\epsilon)}{2\bar{n}_B(\epsilon) + 1} = 1- \bar{n}_F(\epsilon)
\end{equation}
to write  Eq.~(\ref{app:Gamma}) as
\begin{IEEEeqnarray*}{rCl}
\Gamma(\epsilon) &=&  \gamma(\epsilon) [2 \bar{n}_B(\epsilon) + 1] [1 - \bar{n}_F(\epsilon)]	\\[0.2cm]
\Gamma(-\epsilon) &=&  \gamma(\epsilon)  [2 \bar{n}_B(\epsilon) + 1]  \bar{n}_F(\epsilon)
\end{IEEEeqnarray*}

To simplify the problem we will restrict the discussion to the case where $2 n_B(\epsilon) + 1 = \coth(\epsilon/2T) \simeq 1$.
This will generally be true for Fermionic systems. The relevant energies here are  $\epsilon_k = -h - 2t \cos k$ [Eq.~(\ref{H3})] so this approximation will in general be reasonable, except for those momentum values where $\epsilon_k \sim 0$. 
Notwithstanding, with simplicity in mind, we will continue to assume this to hold.
As a result, we get 
\[
\Gamma(\epsilon) = \gamma(\epsilon) [1 - \bar{n}_F(\epsilon)],\qquad
\Gamma(-\epsilon) = \gamma(\epsilon) \bar{n}_F(\epsilon)
\]
Hence, Eq.~(\ref{app:D2}) finally becomes 
\begin{IEEEeqnarray}{rCl}
\label{app:D3}
D(\rho) &=& \gamma(\epsilon) [1 - \bar{n}_F(\epsilon)] \bigg[ \eta \rho \eta^\dagger - \frac{1}{2} \{\eta^\dagger \eta,\rho\} \bigg]	\\[0.2cm]
&&+ \gamma(\epsilon) \bar{n}_F(\epsilon) \bigg[ \eta^\dagger \rho \eta - \frac{1}{2} \{\eta \eta^\dagger, \rho\} \bigg] 	\nonumber
\end{IEEEeqnarray}
This dissipator has precisely the structure of each of the terms in Eq.~(\ref{D}). 

Next we consider the case $\epsilon<0$. 
In this case we  use the identity $\bar{n}_B(-x) = - [1 + \bar{n}_B(x)]$ to write 
 Eq.~(\ref{app:Gamma}) as
\begin{IEEEeqnarray*}{rCl}
\Gamma(\epsilon) &=&  \gamma(-\epsilon) \bar{n}_B(-\epsilon) = -\gamma(-\epsilon) [1 + \bar{n}_B(\epsilon)] 	\\[0.2cm]
\Gamma(-\epsilon) &=&  \gamma(-\epsilon) [1 + \bar{n}_B(-\epsilon)] = - \gamma(-\epsilon) \bar{n}_B(\epsilon)
\end{IEEEeqnarray*}
Next we use Eq.~(\ref{app:ID}) once again but, this time, we note that since $\epsilon <0$, $\coth(\epsilon/2T) \simeq -1$.
Consequently, we will get
\begin{IEEEeqnarray*}{rCl}
\Gamma(\epsilon) &\simeq &  \gamma(-\epsilon) [1 - \bar{n}_F(\epsilon)] 	\\[0.2cm]
\Gamma(-\epsilon) &\simeq&  \gamma(-\epsilon) \bar{n}_F(\epsilon)
\end{IEEEeqnarray*}
As a result we will get a dissipator which is essentially the same as Eq.~(\ref{app:D3}), but with $\gamma(\epsilon)$ replaced by $\gamma(-\epsilon)$. 

We may write both cases in a unified way as 
\begin{IEEEeqnarray}{rCl}
\label{app:D4}
D(\rho) &=& \gamma(|\epsilon|) [1 - \bar{n}_F(\epsilon)] \bigg[ \eta \rho \eta^\dagger - \frac{1}{2} \{\eta^\dagger \eta,\rho\} \bigg]	\\[0.2cm]
&&+ \gamma(|\epsilon|) \bar{n}_F(\epsilon) \bigg[ \eta^\dagger \rho \eta - \frac{1}{2} \{\eta \eta^\dagger, \rho\} \bigg] 	\nonumber
\end{IEEEeqnarray}
which is valid for arbitrary $\epsilon$. 
This concludes our derivation of the dissipator~(\ref{D}). 
The total dissipator for all modes $\eta_k$ will be a sum of dissipators with the structure~(\ref{app:D4}), each with its own Fermi-Dirac occupation number $\bar{n}_k$ and coupling constants $\gamma_k$. 
Since we have no direct physical model for the oscillator bath, it is not possible to determine the functional form of the coupling constants $\gamma_k$.
It is also important to notice that the terms $2 \bar{n}_B (\epsilon) + 1$, which we have approximated to unity, may be included inside the definition of the $\gamma_k$ if one wishes. This would merely introduce a temperature dependence on the $\gamma_k$.

\bibliography{/Users/gtlandi/Documents/library}
\end{document}